\documentclass{aastex}
\usepackage{emulateapj5}
\usepackage[]{graphicx}

\newcommand{\ap}{\approx}
\newcommand{\ds}[1]{\ensuremath{\sigma = #1}}
\newcommand{\br}{\hbox{$B\!-\!R$}}

\newcommand{\vi}{\hbox{$V\!-\!I$}}

\newcommand{\kms}{km~s$^{-1}$}
\newcommand{\hii}{\ion{H}{2}}
\newcommand{\hi}{\ion{H}{1}}
\newcommand{\ha}{H$\alpha$}
\newcommand{\nii}{[\ion{N}{2}]}
\newcommand{\oiii}{[\ion{O}{3}]}
\newcommand{\Msun}{M_{\sun}}

\newcommand{\rone}{R\hbox{$^{\prime}_{1}$}}  

\newcommand{\emax}{\hbox{$e_{\rm max}$}}

\newcommand{\amax}{\hbox{$a_{\rm max}$}}
\newcommand{\amin}{\hbox{$a_{\rm min}$}}
\newcommand{\aten}{\hbox{$a_{10}$}}

\slugcomment{To appear in \textit{The Astrophysical Journal Supplement}}

\shorttitle{Early-Type Barred Galaxies}
\shortauthors{Erwin \& Sparke}

\begin{document}

\title{An Imaging Survey of Early-Type Barred Galaxies}

\author{Peter Erwin}
\affil{Instituto de Astrofisica de Canarias, C/ Via L\'{a}ctea s/n, 
38200 La Laguna, Tenerife, Spain}
\email{erwin@ll.iac.es}
\and
\author{Linda S. Sparke}
\affil{University of Wisconsin-Madison, 475 North Charter Street, 
Madison, WI 53706}
\email{sparke@astro.wisc.edu}

\begin{abstract}
This paper presents the results of a high-resolution imaging survey,
using both ground-based and \textit{Hubble Space Telescope} images, of
a complete sample of nearby barred S0--Sa galaxies in the field, with
a particular emphasis on identifying and measuring central structures
within the bars: secondary bars, inner disks, nuclear rings and
spirals, and off-plane dust.  A discussion of the frequency and
statistical properties of the various types of inner structures has
already been published.  Here, we present the data for the individual
galaxies and measurements of their bars and inner structures.  We set
out the methods we use to find and measure these structures, and how
we discriminate between them.  In particular, we discuss some of the
deficiencies of ellipse fitting of the isophotes, which by itself
cannot always distinguish between bars, rings, spirals, and dust, and
which can produce erroneous measurements of bar sizes and
orientations.

\end{abstract}

\keywords{galaxies: structure --- galaxies: active --- galaxies: 
elliptical and lenticular, cD --- galaxies: spiral}

\section{Introduction}

In the past decade, high-resolution images of the centers of disk
galaxies have shown that many contain distinct, small-scale
structures, such as inner bars, nuclear spirals, and nuclear rings. 
\citet{buta93} present a compilation of data on nuclear rings, while
\citet{friedli96b} reviews double-barred galaxies: those with a
smaller, concentric bar inside the main bar.  Infrared imaging of
bars-within-bars \citep[e.g.,][]{shaw93,shaw95,friedli96a,jung97}
has shown that the structures are not composed purely of dusty gas and
young stars, but are also seen in the light of the old red stars that
make up most of the disk mass.

As Buta \& Crocker point out, these features can be used to probe the
galaxy dynamics; they argue that nuclear rings and spirals are linked
to the inner Lindblad resonance of the large-scale bar or spiral
pattern.  The prevalence of small-scale stellar features tells us that
the centers of many disk galaxies are dynamically ``cool'' enough to
be responsive to gravitational effects on such small scales: most of
the kinetic energy is in rotation, rather than random motions.  A
dynamically ``hot'' bulge, where the velocity dispersion $\sigma$ is
of the same order as the rotation speed $V$, would not form bars or
stellar rings.  As \citet{kormendy93} suggested, the bright inner
regions of many galaxies may be dominated by ``cool'' disks, rather
than ``hot'' bulges.  When the inner features themselves contain
significant mass, their non-axisymmetric gravitational pull affects
the motions of stars and gas.  For example, \citet{shlosman89}
suggested that inner bars may be important in removing angular
momentum from gas, feeding it into the center where it may fuel a
starburst or active nucleus.  \citet{witold02} and \citet{shlosman02}
have presented simulations of this complex flow.

To conduct a census of central structures, we observed a complete
sample of nearby disk galaxies which were already known to contain a
kiloparsec-scale main bar.  Selecting S0 and Sa systems to minimize
the effect of dust, we used the 3.5 m WIYN telescope, and archival
optical and near-IR images from the \textit{Hubble Space Telescope}
(HST), to search for multiple bars and other central
features\footnote{The WIYN Observatory is a joint facility of the
University of Wisconsin-Madison, Indiana University, Yale University,
and the National Optical Astronomy Observatories.  Observations with
the NASA/ESA Hubble Space Telescope obtained at the Space Telescope
Science Institute, operated by the Association of Universities for
Research in Astronomy, Inc., under NASA contract NAS5-26555.}.

Our first results for three galaxies were presented in
\citet{erwin99}; \citet[][hereafter Paper~I]{erwin-s02} discusses the
types of structures that we found, their frequencies, and their
relation to other properties of the galaxies.  In this paper, we
discuss our observations and the analysis techniques, and provide
details and measurements for the individual galaxies.  A key advantage
of our approach is that we consider multiple types of inner structures
(bars, disks, rings, spirals, and off-plane dust) and carefully
distinguish between them.  We also show that the common technique of
fitting ellipses to isophotes, while quite useful, can often lead to
erroneous measurements of bar parameters such as orientation and size.

\section{Sample Selection} 

To minimize the potential confusion from dust, we restricted ourselves
to early-type disk galaxies; to maximize our chances of detecting
small circumnuclear structures, we chose large, nearby galaxies that
were not close to edge-on.  We selected all barred S0--Sa galaxies in
the UGC catalog \citep{ugc} which met the following criteria:
declination $> 10\arcdeg$, heliocentric radial velocity $\leq 2000$
\kms, major axis diameter $\ge 2\arcmin$, and ratio of major to minor
axis $a/b \leq 2$ (corresponding to $i \lesssim 60\arcdeg$).  Galaxy
types and axis measurements (at the 25 mag arcsec$^{-2}$ level in $B$)
were taken from \citet[][hereafter RC3]{rc3}; radial velocities are
from the NASA/IPAC Extragalactic Database (NED).  Our use of the UGC
and the restriction to galaxies with $D_{25} \ge 2\arcmin$ means that
the sample is biased in favor of high surface brightness galaxies; we
believe that this is the only significant bias.

We excluded galaxies identified as members of the Virgo Cluster (those
with VCC entries in NED), because the cluster environment may affect
bar properties \citep[e.g.,][]{andersen96}.  In addition, galaxies in
Virgo and the field with the same Hubble type may actually differ
significantly in morphology: \citet{koop98} found that many Virgo Sa
galaxies were more similar to field Sc galaxies than field Sa galaxies
in terms of central light concentration.

Because our galaxies are selected on the basis of being
\textit{optically} barred --- that is, SB or SAB according to RC3 ---
there is the possibility that we are missing some barred galaxies
mistakenly classified as unbarred.  Conflicting claims have been made
about how much the bar fraction increases when galaxies are observed
in the near-infrared, where dust extinction is less of a problem; the
most extensive survey to date is that of \citet{eskridge00}, who
classified 186 disk galaxies using $H$-band images.  They found that
the primary difference between the RC3 optical classifications and
those from the near-IR was the increase in the relative number of
strong (SB) bars; the \textit{total} bar fraction (SB + SAB) increased
by less than 10\%.  This effect is weakest for early type galaxies:
65\% of S0--Sab galaxies in their sample are barred according to the
RC3, while 71\% are barred in the near-IR. Since we restrict ourselves
to S0--Sa galaxies, and select both SB and SAB classes, we are
probably missing only one or two barred galaxies.

There is also the possibility that the barred appearance some of these
galaxies at optical wavelengths is produced by, for example, strong
dust lanes and star formation, rather than by a genuine bar in the
underlying stellar disk.  We restricted ourselves to early-type
galaxies to minimize this problem.  Nonetheless, three of the galaxies
(NGC~2655, NGC~2685, and NGC~3032) show signs that they may not be
barred; we discuss these cases individually in
Section~\ref{sect:ind-galaxies}.

This sample has a total of 38 galaxies; their basic parameters, taken
from the literature and from public databases --- NED and the
Lyon/Meudon Extragalactic Database (LEDA) --- are listed in
Table~\ref{tab:galaxies}.  Twenty of the galaxies are S0, ten are
S0/a, and the remaining eight are Sa; twenty-five are strongly barred
(SB), with the remaining thirteen being SAB.

\section{Observations\label{sect:obs}} 

Table~\ref{tab:obs} summarizes the observations for each galaxy in the
sample.

We observed all but two of the galaxies in the $B$ and $R$ bands at
the 3.5m WIYN telescope in Tucson, Arizona, between 1995 December and
1998 March.  The detector was the S2KB CCD, a thinned $2048 \times
2048$ STIS chip with 21-micron (0.2\arcsec) pixels and an approximate
field of view of $6.8\arcmin \times 6.8\arcmin$.  Seeing --- the mean
FWHM of Moffat-profile fits to stars in $R$-band images --- is given
in Table~\ref{tab:obs}; it ranged from 0.5--1.3\arcsec, with a median
of 0.8\arcsec.  Seeing in $B$-band images tended to be slightly worse
(0.7--1.65\arcsec).  Most of the nights were not photometric, but
since we were primarily interested in morphology and color variations,
relative photometry is adequate for our purposes.  In most cases, we
used exposure times of 60s and 300s in $R$ and 120s and 600s in $B$. 
A few galaxy centers were bright enough to saturate the CCD in the 60s
$R$-band exposures, so we made additional 30s exposures.  For the two
galaxies \textit{not} observed with WIYN --- NGC~936 and NGC~4314 ---
reasonable ground-based images were available in the literature, and
archival HST observations allowed us to carry out our analysis of
their central regions.  For NGC~936, we also obtained ground-based
images of the central region in 2000 December with the 4.2m William
Herschel Telescope (La Palma, Spain).  Finally, we used $B$, $R$, and
$I$ images NGC~3729 from \citet{tully96}, available via the NASA/IPAC
Extragalactic Database (NED), to supplement WIYN images obtained under
poor conditions.

Prior to mid-1996, an unrecognized nonlinearity plagued the S2KB CCD,
dampening the response at levels $\lesssim 200$ counts above the bias
(Ted von Hippel, private communication).  This is a potential problem
only for our $B$-band images, and then only for the outer disk
regions.  Two galaxies (NGC~2273 and NGC~3945) were observed both
before and after the non-linearity was fixed; we found no significant
differences in either the color maps or the ellipse fits.  For images
taken during the non-linear era, we treat color gradients in the
outermost disk with skepticism; affected observations are noted in
Table~\ref{tab:obs}.

Reduction of the images was carried out in standard fashion, using
\textsc{iraf} software.  In a few cases where background sky gradients
were present, we subtracted fitted backgrounds using the
\textsc{imsurfit} task.  Linear fits generally sufficed, though we
were forced to use quadratic fits on one or two occasions.  In images
with no sky gradient, we measured the median sky levels in at least
five regions far from the galaxy, and then subtracted the average of
these values.  To make color maps, $B$ and $R$ images were registered
using stars in both images with the \textsc{geomap} and
\textsc{geotran} tasks, since simple linear shifts were not
sufficient.  We shifted the higher-resolution image to align with the
lower-resolution image, then iteratively smoothed the shifted image
until it matched the other's resolution.  In all cases, we used the
long-exposure images to make color maps, to maximize signal-to-noise. 
When the long $R$-band exposures had resulted in saturated galaxy
centers, we made additional color maps using the shorter exposures.

HST data (see Table~\ref{tab:obs}) were processed with the standard
pipeline, and analyzed using tools in the \textsc{stsdas} package
within \textsc{iraf}.  Multiple exposures were combined with the
\textsc{stsdas} task \textsc{crrej} to eliminate cosmic-ray
contamination; linear shifts were used to register images that were
not aligned.  The images were originally taken for a variety of
programs, so filters and exposure times varied; in some snapshot
survey images, the galaxy was off-center or saturated in the nucleus. 
Linear shifts sufficed to align WFPC2 images for making color maps;
however, because all the galaxies are large enough to extend outside
the WFPC2 field of view, we could not subtract the sky background.  To
make color maps using WFPC2 and NICMOS images, we rebinned the PC2
image to the corresponding NICMOS pixel scale, then shifted it to
align with the NICMOS image.  Fits of Moffat profiles to stars in
WFPC2 and NICMOS images yielded consistent FWHM of $\ap 2.0$ pixels;
pixel scales are 0.0455\arcsec{} pixel$^{-1}$ for the PC2 chip,
0.075\arcsec{} pixel$^{-1}$ for Camera 2 of NICMOS (NICMOS2), and
0.2\arcsec{} pixel$^{-1}$ for Camera 3 (NICMOS3).

\section{Methods of Analysis\label{sect:analysis}}

Our primary goal in analyzing the images was the identification and
measurement of central stellar and gaseous structures --- as well as
identifying galaxies which \textit{lack} such features.  Since we are
interested in the relation of these features to the bars within which
they are found, we also needed to measure the (main) bars in each
galaxy.  In this section, we discuss the techniques we used, what
guidelines we used to discriminate among the various features we find,
and how we measured them.  Our measured values are listed in 
Table~\ref{tab:features}.

\subsection{Techniques}

\subsubsection{Ellipse Fits\label{sect:ellipse}} 

We used ellipse fits to determine the position angle (PA) and
ellipticity of galaxy isophotes as a function of radius.  The specific
algorithm was that of \citet{jz87}, based partly on that of
\citet{carter78} and implemented in the \textsc{iraf} task
\textsc{ellipse} (part of the \textsc{stsdas} package).  For each
specified semi-major axis $a$, measured from the galaxy center, the
program attempts to fit an ellipse to the isophotes of the galaxy; the
center, position angle, and ellipticity $e = 1 - b/a$ (where
$b$ is the semi-minor axis of the ellipse) of the ellipse are allowed
to vary.  The best results came from allowing ellipse centers to vary;
this produced converged fits more often.  We also employed iterative
rejection of pixels with intensity values more than 3-$\sigma$ from
the mean along the ellipse, which minimizes contamination from cosmic
rays, bad pixels, stars, and meteor/satellite tracks.  We used
logarithmic radial sampling with steps of 0.03 in scaled semi-major
axis; sparser sampling can miss small-scale features and undersample
rapid isophote twists, especially at large radii.

A good discussion of the \textsc{ellipse} algorithm and the associated
errors can be found in \citet{rauscher95}; he notes that its error
estimates are likely to be \textit{over}estimates.  Both he and
\citet{jz87} found that fitted ellipses are distorted for semi-major
axis $a < 3$--5 pixels, due to pixelization effects.  Other
distortions can come from limited resolution and atmospheric seeing,
which make isophotes rounder in the center; focus problems with a few
WIYN images produced elliptical distortions.

As an example of the effects of resolution, Figure~\ref{fig:n3412e2}
shows the results of fitting ellipses to both WIYN and WFPC2 images of
the galaxy NGC 3412.  Seeing for the $R$-band image was $\ap
1.2\arcsec$, while the mean FWHM of stellar profiles in the PC2 image
was $\ap 0.1\arcsec$.  There is no evidence for dust lanes in the
central region of this galaxy (see Figure~\ref{fig:n3412}), so the
differences in ellipticity for $a \lesssim 5\arcsec$ for the two
images are due to the dramatic difference in resolution.  At lower
resolution, the point of maximum ellipticity shifts \textit{outward}. 
Thus, estimates of the sizes of small-scale features based on the
location of maximum ellipticity (see Section~\ref{sect:measure} below)
can sometimes be exaggerated by poor seeing.

For a perfect fit to an isophote, the intensity along the ellipse
would be constant.  Variations in intensity along the ellipse can be
expanded as a Fourier sum:
\begin{equation}
I(\theta) = I_{0} \, + \sum_{n = 3}^{4}\, [\tilde{A}_{n} 
\sin n\theta + \tilde{B}_{n} \cos n\theta] \, ;
\end{equation}
the first- and second-order coefficients are zero for a best-fit
ellipse.  The higher coefficients are divided by the local radial
intensity gradient and by the ellipse semi-major axis, to obtain
normalized coefficients of \textit{radial} deviation $\delta r$ from a
perfect ellipse, in a stretched coordinate system where the fitted
ellipse is a circle with radius $r = (ab)^{1/2}$:

\begin{equation}
\frac{\delta r(\theta)}{r} = \sum_{k = 3}^{4} \, [A_{k} \sin k
\theta + B_{k} \cos k \theta].
\end{equation}

The most commonly used coefficient is $B_{4}$, the $\cos 4 \theta$
term, which measures symmetric distortions from pure ellipticity. 
When $B_{4} > 0$, the isophotes are pointed (lemon-shaped) or
``disky''; when $B_{4} < 0$, the isophotes have a more rectangular or
``boxy'' shape.  This term is related to the coefficient $a_{4}/a$ of
\citet{bender88} by $a_{4}/a = \sqrt{b/a} \, B_{4}$.

\begin{figure}
\begin{center}
	\includegraphics[scale=0.8]{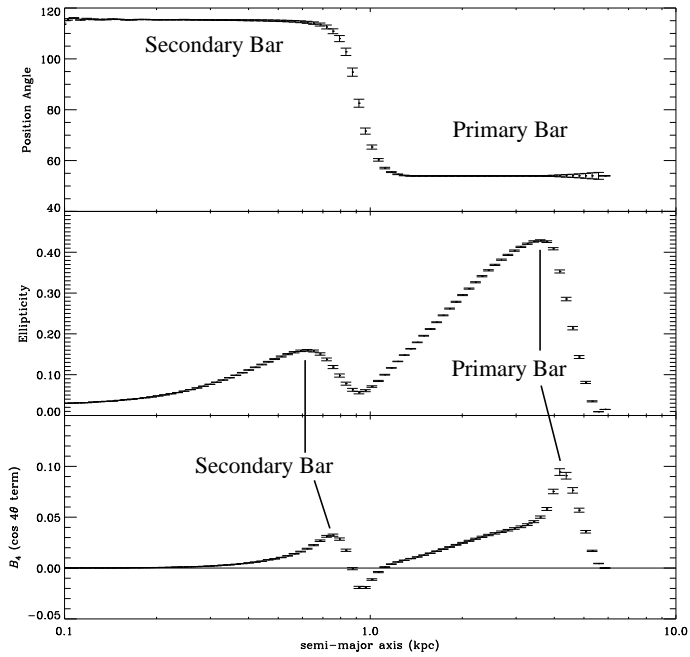}
\end{center}
\caption{Ellipse fits to a face-on model double-bar galaxy,
consisting of disk, bulge, and two Ferrers bars \citep[Model 2
from][]{witold98}.\label{fig:witolde3}}
\end{figure}

In their study comparing SB0 galaxies to ellipticals with twisted
isophotes, \citet{nieto92} argued that boxiness in ellipse fits is
characteristic of bars, since the \textit{ends} of bars in early-type
galaxies often have a rectangular shape --- see \citet{athan90}. 
However, an elongated, bisymmetric structure superimposed on rounder
isophotes can also produce composite isophotes that are disky, with
$B_{4} > 0$.  This effect is particularly strong in their images of
UGC 2534, UGC 2708, and BGP 61; see also Figures~\ref{fig:witolde3}
and \ref{fig:n2950e3} of this paper.  For secondary bars, embedded
more deeply inside bulges than primary bars, we find that the dominant
sign of $B_{4}$ is almost always positive.  For these reasons, $B_{4}$
cannot be used to discriminate bars from other embedded, elongated
structures (such as disks or stellar rings).  However, we \textit{do}
find it useful in some cases as confirmation that some central
structure is present.

\subsubsection{Unsharp Masking} 

Unsharp masking is a simple form of high-pass filtering, suppressing
large-scale, low-frequency variations \citep[e.g.,
][]{malin79,malin83}.  In contrast to the ellipse fits, this technique
is sensitive to the details of low-amplitude, small-scale brightness
variations.  It can, for example, give a clearer view of a small
central bar which may be outshone by a bright bulge, and of narrow
lanes of obscuring dust.  It is also particularly good for
discriminating between structures which produce similar ellipse-fit
features, such as bars and rings.

In origin, unsharp masking is a photographic technique for enhancing
detail and contrast.  For digital images, the procedure is as follows:
make a smoothed copy of the original image, and then either subtract
it from the original, or divide the original by it; the result is the
``unsharp mask.''  A final, sharpened image is generated by either
adding the (subtraction-generated) mask to the original, or
multiplying the original by the (division-generated) mask; the mask
can be scaled beforehand to control the amount of sharpening applied. 
Subtraction-based unsharp masking is the more common technique, since
it is computationally faster.

We are interested in the mask itself, which directly displays the
subtle, small-scale features we want to identify.  In practice,
division-generated masks have smaller dynamic ranges and are thus more
easily viewed on the monitor, though the same features appear in both
types of masks.  (The subtraction method is more useful if the signal
is near zero, as may be the case in the outer edge of a galaxy in a
sky-subtracted image.)  Detail on different spatial scales can be
emphasized by changing the smoothing width: smoothing with a narrow
Gaussian suppresses all but the smallest scales, while using
progressively wider Gaussians allows larger scale variations to show
up (though these then obscure the smaller-scale variations).

Figures~\ref{fig:umask}a and b show how this process highlights
small-scale variations, using a simple 1-D example.  The left-hand
panel is an $r^{1/4}$ intensity profile with a narrow Gaussian added. 
After dividing the original by the smoothed version, the Gaussian peak
stands out as a sharp excess, bordered by shallow deficits, and the
distracting radial gradient is largely gone. 
Figures~\ref{fig:umask}c--f show this process applied to an $R$-band
image of the double-barred galaxy NGC~2950.  Smoothing with a narrow
Gaussian ($\sigma = 5$ pixels) brings out the inner bar; using a wider
Gaussian ($\sigma = 20$ pixels, panel~f) emphasizes the outer bar,
though it makes the small bar much harder to distinguish.  Stars
appear as bright peaks surrounded by dark ``moats.''  In both masks,
the ends of the bars show up as paired bright regions with dark
regions just beyond, showing that the luminosity drops sharply there. 
\nocite{ohta90}Ohta, Hamabe, \& Wakamatsu (1990) noted that the ends
of bars in early-type galaxies generally have sharp cutoffs in their
luminosity profiles; our unsharp masks indicate that inner bars have 
a similar structure.

We generated unsharp masks with a range of smoothing, convolving the
original images with Gaussians of $\sigma = 2$, 5, 10, 20, and 40
pixels (as a reminder, FWHM $= 2.354 \sigma$ for a Gaussian).  The
resulting masks were compared with isophotes, color maps, and the
ellipse fits to help determine the nature and the orientation of
various structures in the galaxies.

\subsubsection{Color Maps} 

Color maps can be used to search for differences in the stellar
population between the inner structures and their surroundings.  They
are also useful in delineating areas affected by dust, which are
reddened, and recent star formation, where young blue stars are
present.  Nuclear rings are traditionally found as distinct red or
blue ringlike features inside bars \citep[e.g., ][]{buta93}.  Because
``raw'' color maps can be dominated by photon noise, we smoothed them
with median filters.  Minimal filtering (using a median box of width
$w=5$ pixels) is effective at cleaning up moderately strong color
contrasts.  Higher levels of smoothing are useful for finding very
smooth, large-scale color features and gradients.

We note that when the dust extinction is high enough to render the
dust opaque at optical wavelengths ($A_{V} > 1$), our $\bv$ color maps
will not probe the true extinction, since only the foreground,
unreddened stars will be visible.

\subsection{Identifying, Naming, and Discriminating Features} 

Our basic approach was to look for distinct
features in ellipse fits --- particularly peaks in the ellipticity. 
This is based on the approach of \citet[][hereafter W95]{w95}, who
looked for bars by finding local ellipticity maxima, accompanied by
stationary position-angle values.

However, there can be cases where a bar or other feature is present
but does \textit{not} appear as a maximum in ellipticity; so we
considered \textit{any} significant deviation in the ellipse fits as a
possible structure.  For example, projection effects cause the bar of
NGC~4143 (Figure~\ref{fig:n4143}) to manifest itself as a strong
twist in position angle \textit{without} an obvious ellipticity peak;
the inner bar of NGC~3945 is another example \citep[see
][Figure~2]{erwin99}.  NGC~2962's inner bar (Figure~\ref{fig:n2962})
is only apparent in the fits as an ellipticity \textit{minimum}, with
a very modest change in position angle; in this case, the $B_{4}$
coefficients clearly indicate disky isophotes, and unsharp masks
confirm the presence of a bar).  Similarly, both \citet{nieto92} and
\citet{busarello96} identified the main bar in NGC~3384 with a minimum
in the ellipticity.

Finally, we used unsharp masks, color maps, and examination of the
original images to evaluate all such features and determine when
they were most likely due only to dust lanes or star formation, to
discriminate between structures such as bars, rings, and spirals,
(which may all have similar ellipse-fit signatures), and to identify 
features not apparent in ellipse fits (such as some nuclear rings and 
off-plane dust lanes).

For galaxies with two (or more) bars, we adopt the terminology of
\citet{friedli93} and \nocite{w95}W95: the larger bar is the
\textit{primary}, \textit{outer}, or \textit{main} bar, and the
smaller, embedded bar is the \textit{secondary}, \textit{inner}, or
\textit{small} bar.  In NGC~2681, which has three bars, the third,
innermost bar is the \textit{tertiary} bar.  In galaxies with only one
bar, we usually refer to it as simply ``the'' bar.

There is some confusion in the literature as to what constitutes a
``nuclear'' ring versus an ``inner'' ring.  We use the scheme of Buta
\citep[e.g., ][]{buta93,buta95}, which uses the bar as a measuring
stick.  \textit{Outer rings} lie well outside the bar, usually at
slightly over twice the bar radius.  \textit{Inner rings} lie just
beyond the ends of the bar, while \textit{nuclear rings} are found
\textit{inside} the bar.  We use the \textit{primary} bar as the
referent, so a nuclear ring in a double-barred galaxy is one inside
the primary bar.  (We found no clear cases of rings inside
\textit{secondary} bars, which lets us avoid the difficult question of
what to call such things.)

Disks residing inside bars are called \textit{inner disks}.  The
alternative, ``nuclear disk,'' has generally been used in the
literature for the \textit{very} small-scale disks (sizes $\lesssim
1\arcsec$ and $\lesssim 100$ pc) revealed by HST in ellipticals and
edge-on S0's \citep[e.g., ][]{scorza98}.  The larger disks found by
\citet{seifert96} have physical and angular sizes roughly the same
as our disks and secondary bars, so we adopt their term.

\begin{figure}
\begin{center}
	\includegraphics[scale=0.8]{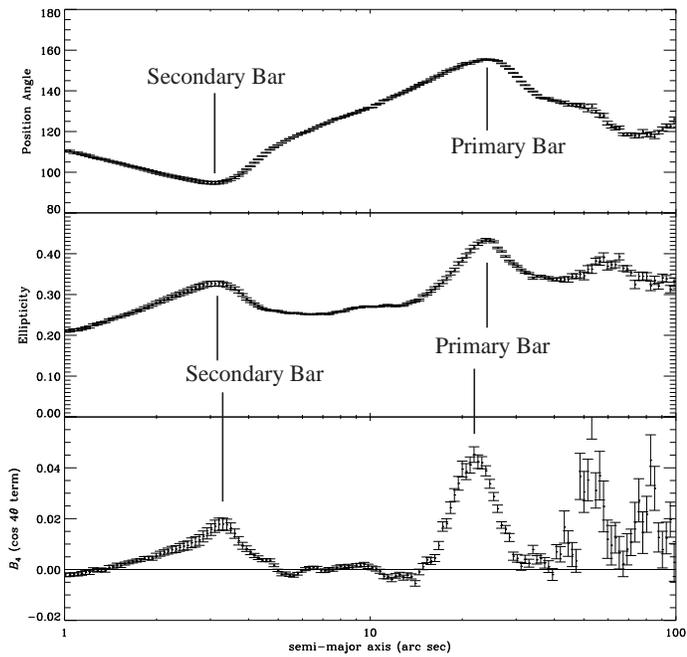}
\end{center}
\caption{Ellipse fits to WIYN $R$-band image of the double-barred
galaxy NGC 2950; see Figure~\ref{fig:n2950} for other data on this
galaxy.\label{fig:n2950e3}}
\end{figure}

\begin{figure}
\begin{center}
	\includegraphics[scale=0.8]{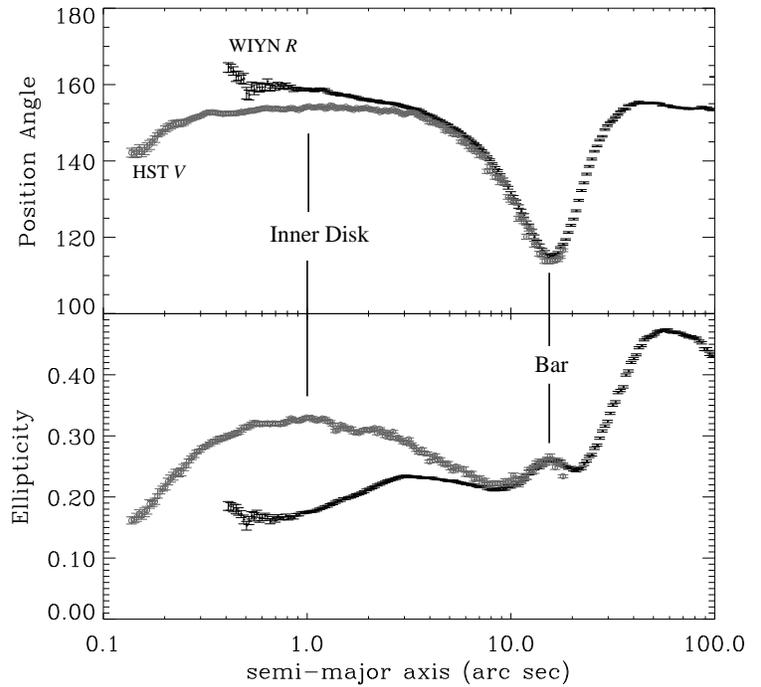}
\end{center}
\caption{Ellipse fits to WIYN $R$-band and WFPC2 F606W (gray)
images of the inner-disk galaxy NGC 3412; see Figure~\ref{fig:n3412}
for other data on this galaxy.\label{fig:n3412e2}}
\end{figure}

\begin{figure}
\begin{center}
	\includegraphics[scale=0.9]{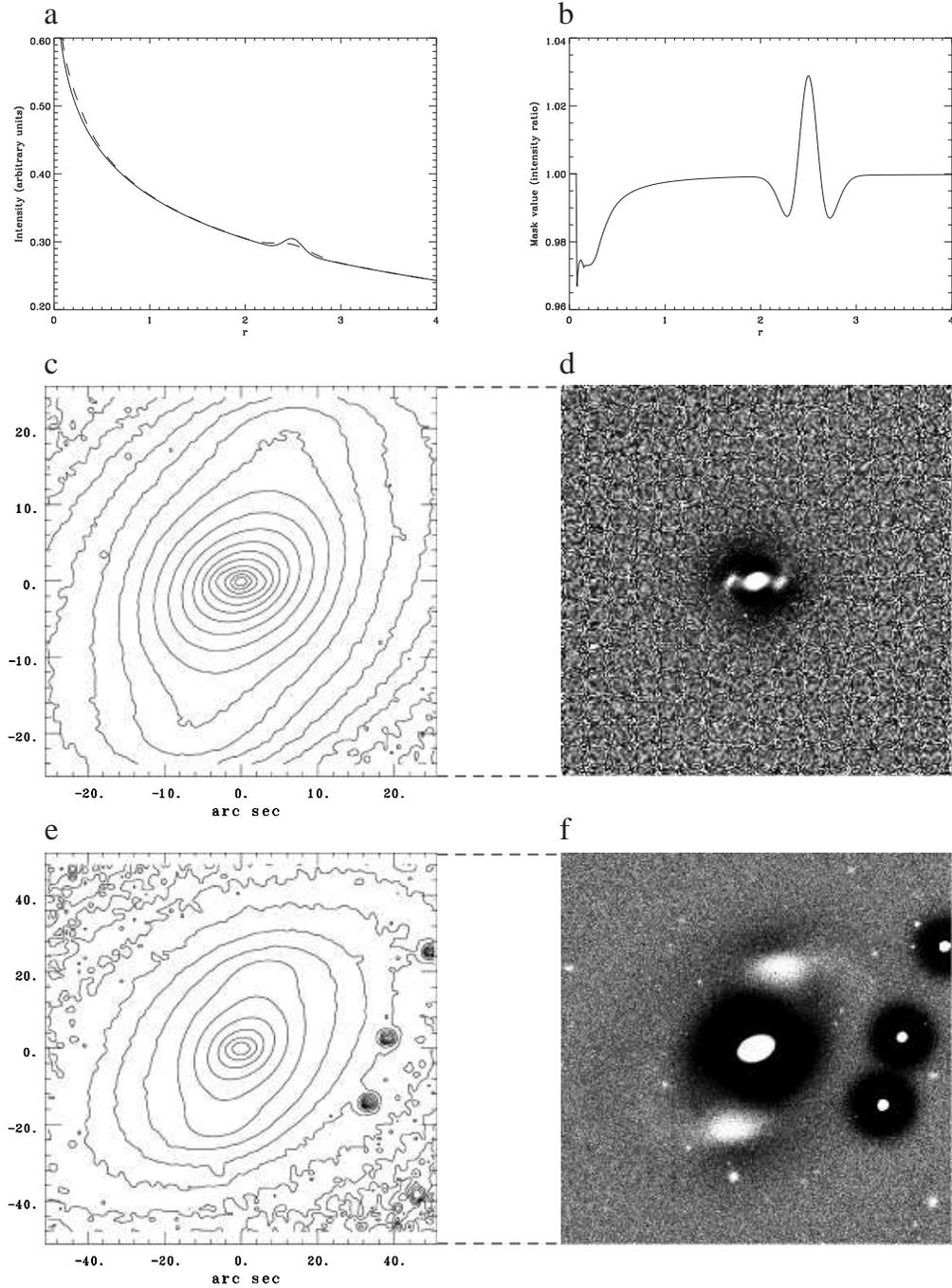}
\end{center}
\caption{Examples illustrating the use of unsharp masks. 
\textbf{a} 1-D $r^{1/4}$ intensity profile with narrow Gaussian added
at $r = 2.5$ (solid) along with Gaussian-smoothed version (dashed);
\textbf{b} 1-D unsharp mask produced by dividing original profile by
smoothed profile.  \textbf{c} $R$-band contours of double-barred
galaxy NGC 2950; \textbf{d} unsharp mask (\ds{5}) of the same image,
showing secondary bar.  \textbf{e} Larger-scale $R$-band contours of
the same galaxy; \textbf{f} unsharp mask (\ds{20}), showing primary
bar.\label{fig:umask}}
\end{figure}

The specific features that we list in Table~\ref{tab:features} are:
\begin{itemize}

  \item \textit{Primary bars.} Since our galaxies are classified as SB
  or SAB, they should in principle all have at least \textit{one} bar. 
  We looked for ellipticity peaks or position angle twists in the
  ellipse fits, and based our measurements on those, but differ from
  \nocite{w95}W95 in that we did \textit{not} require constant PA for
  a bar.  The ellipticity maximum for NGC~3412's bar
  (Figure~\ref{fig:n3412e2}) is a \textit{local} maximum within a
  larger-scale minimum; both here and in Figure~\ref{fig:n2950e3}, the
  position angle twists to a maximum or minimum value without staying
  constant over a large range of semi-major axis.  We inspected color
  maps, unsharp masks, and the original images to check that strong
  dust lanes, star formation, rings, or spiral arms were not
  responsible for features in the ellipse fits, and to ensure that we
  measured the bar itself (as opposed to, e.g., spiral arms outside
  the bar).

  \item \textit{Secondary bars and inner disks.} We usually identified
  these as local ellipticity peaks or distinct PA twists; in certain
  cases, ellipticity \textit{minima} were also an indication.  As with
  primary bars, we examined color maps, unsharp masks, and the
  original images to ensure that strong dust lanes, star formation,
  nuclear rings or spirals were not the cause.  We distinguished inner
  disks by requiring that the PA differ from that of the outer disk by
  less than 10\arcdeg, \textit{and} that the maximum ellipticity be
  less than that of the outer disk.  Figure~\ref{fig:n2950e3} provides
  a clear example of a secondary bar, while Figure~\ref{fig:n3412e2}
  shows an inner disk.  The inner-disk class is somewhat ambiguous: a
  secondary bar fortuitously aligned with the projected outer disk
  would be called an inner disk, unless it appeared more elliptical
  than the outer disk; highly flattened inner bulges could also be
  detected and classed as inner disks.  Our secondary bars almost
  always display a distinct appearance in unsharp masks (e.g.,
  Figure~\ref{fig:umask}), while our inner disks generally do not, so
  we believe that these are distinct populations. 
  \nocite{erwin-s02}Paper~I presents statistical evidence that these
  are indeed distinct classes.
  
  \item \textit{Nuclear rings.} We recognize four general classes of
  nuclear ring.  \textit{Star-forming nuclear rings} are readily
  apparent in images as a ringlike pattern of bright knots, usually
  intermixed with tightly wrapped spiral dust lanes; in color maps,
  the knots are blue.  The well-studied nuclear ring of NGC~4314
  (Figure~\ref{fig:n4314}) is an excellent prototype.  \textit{Blue
  nuclear rings} are smooth, blue features in color maps; they lack
  the uneven, knot-like appearance of star-forming rings, and have
  little or no dust in the ring.  \textit{Dusty} (red) nuclear rings
  are seen in color maps as red features without any blue knots; they
  can also be identified in unsharp masks as dark rings.  Finally,
  \textit{stellar nuclear rings} show up as ellipticity peaks, and
  were distinguished from bars and disks by their appearance in
  unsharp masks; \citet{erwin99} present two clear examples.  They
  seem to have no clumps of dust or young stars, and thus do not stand
  out in color maps.

  \item \textit{Inner and Outer Rings.} We did not always identify or
  measure these, but in some cases it can be unclear from casual
  inspection of the ellipse fits \textit{which} peak corresponds to
  the (primary) bar, so it is helpful to clearly distinguish bars from
  exterior rings.  Sometimes it is difficult to define the end of the
  bar from the ellipse fits; if an inner ring is present, a
  measurement of its size (from color maps or unsharp masks) might
  provide a useful upper bound to the bar size.  Finally, some
  galaxies have strong outer rings which obscure the true orientation
  and ellipticity of the outer disk; the shapes and orientations of
  the inner and outer rings are then useful to help constrain the disk
  inclination and PA (see Section~\ref{sect:inclination}, below).

\end{itemize}

In addition, we found examples of \textit{nuclear spirals} and
\textit{off-plane gas}.  The former are distinct regions of spiral
structure surrounding the nucleus, usually identifiable in unsharp
masks or color maps or both; we typically observed multiple dust lanes
rather than distinct stellar arms.  We use the term ``off-plane gas''
somewhat loosely to identify galaxies where at least some of the gas
appears to be off-plane with respect to the stellar disk.  The ``Helix
Galaxy'' NGC~2685, with dust and stellar rings surrounding the central
regions, is a canonical example.  For an off-plane gas classification,
we also required at least some corroborating kinematic evidence ---
e.g., \hi{} maps from the literature which indicate that some gas is
in orbits inclined with respect to the stellar disk.

We assumed that spiral arms trail, and used the orientation of any
stellar or dusty spiral arms to derive the direction of disk rotation,
as listed in Table~\ref{tab:features}.

\subsubsection{The Problem of Inclination and Deprojection\label{sect:inclination}} 

\begin{figure}
\begin{center}
	\includegraphics[scale=0.8]{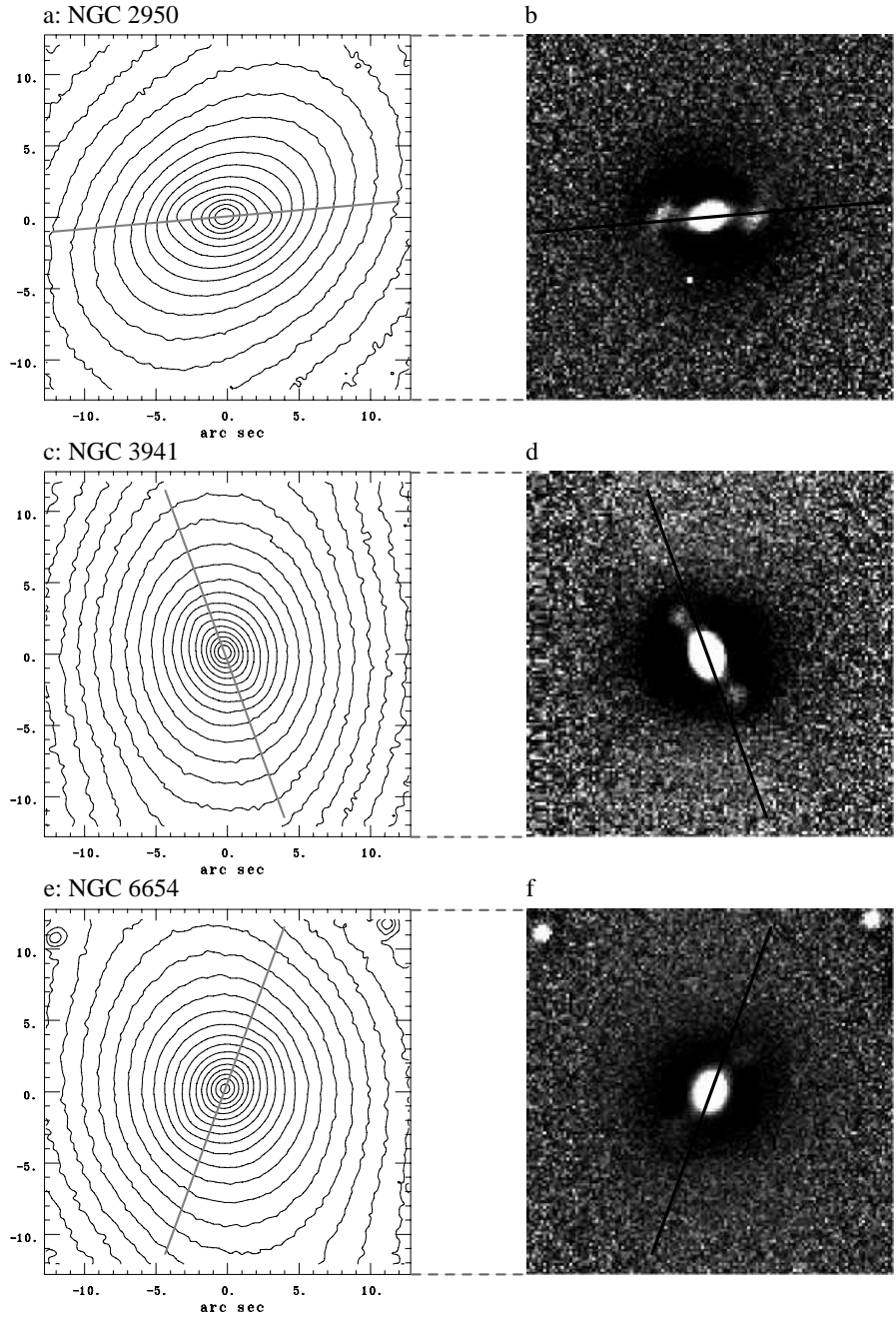}
\end{center}
\caption{Examples of differences between position angles of
secondary bars as given by ellipse fits (diagonal lines) versus actual
isophotes (left) and unsharp masks (\ds{5},
right).\label{fig:umaskpa}}
\end{figure}

\begin{figure}
\begin{center}
	\includegraphics[scale=0.7]{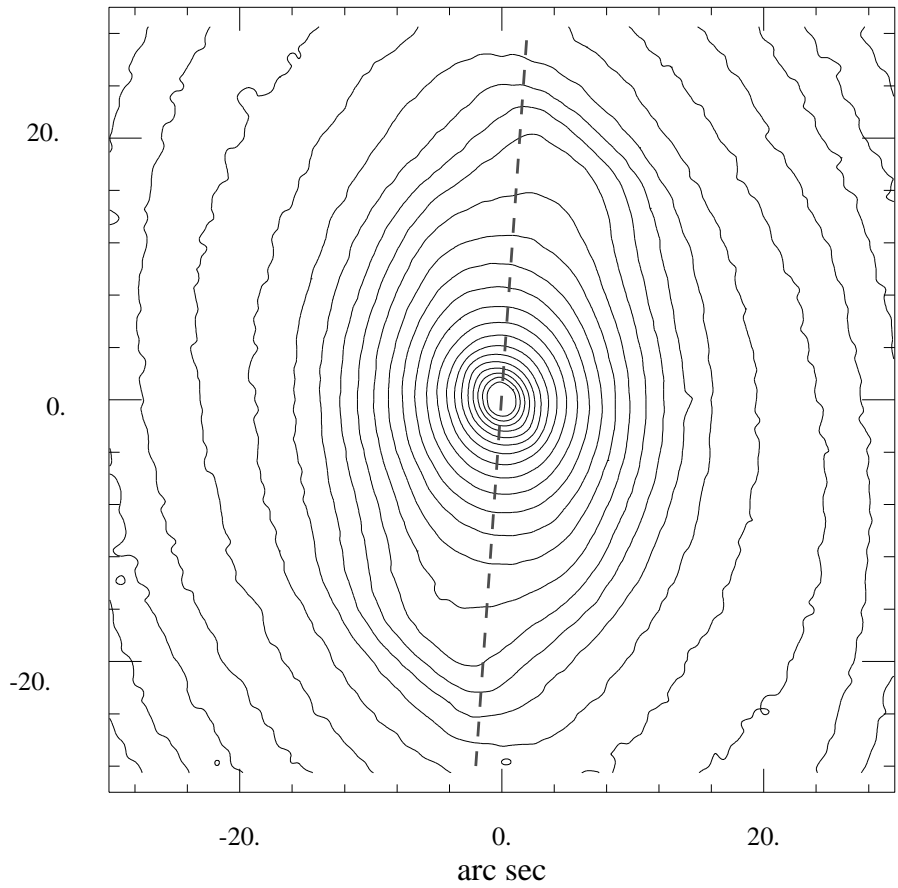}
\end{center}
\caption{$R$-band image of the double-barred galaxy NGC~3941,
showing the outer bar.  The dashed line shows the position angle of
the bar as derived from ellipse fits --- clearly misaligned with the
bar.  As with inner bars (Figure~\ref{fig:umaskpa}), the position
angles derived from ellipse fits must be treated with skepticism.  The
ellipse fits can be found in Figure~\ref{fig:n3941}.
\label{fig:n3941pa}}
\end{figure}

The sizes and orientations of structures reported in
Table~\ref{tab:features} are \textit{measured} values --- i.e.,
without deprojection.  Some authors deproject galaxy images and then
perform isophote fits (e.g., \nocite{w95}W95).  But deprojecting
digital images can create artifacts in the central regions due to the
interpolation of pixel values: tests by \citet{jung97} on artificial
images showed that fitting ellipses to a deprojected image can produce
spurious results, including false signals of secondary bars.  An
alternative approach is to deproject the fitted ellipses analytically,
as done by Jungwiert et al.\ and \citet{laine02}.  While this avoids
interpolation problems, it is vulnerable to the fact that ellipses are
not always good fits to galaxy substructures.  For example, we find
that the position angles of ellipses can be misaligned with respect to
the bars they are ostensibly fitting; this is especially true for
inner bars, where we find deviations of 10--80\arcdeg{}, but it can
also occur for outer bars (see Section~\ref{sect:measure} and
Figures~\ref{fig:umaskpa} and \ref{fig:n3941pa}).  Deprojecting these
ellipses will then give misleading values for both position angles
\textit{and} sizes of bars.  Finally, two-dimensional deprojection,
whether of images or fitted ellipses, only works if all structures in
the galaxy are coplanar and flat.  Since we are interested in
structures with sizes comparable to, or smaller than, the central
bulge, this is not a safe assumption to make.

Nevertheless, we do want to compare the (deprojected) lengths and
orientations of bars and other structures which we presume lie in the
plane of a galaxy's disk, so we \textit{did} attempt to determine the
inclination and orientation of the disk of each galaxy.  This was also
necessary for our inner-disk classification (above).  We generally
assumed that the outermost isophotes that we could measure are those
of an intrinsically circular disk.  The inclination $i$ to the line of
sight ($i = 0\arcdeg$ for a face-on disk, and 90\arcdeg{} for edge-on)
is then
\begin{equation}
\sin i = \sqrt{\frac{2 e - e^{2}}{1 - q_{0}^{2}}},
\end{equation}
where $e$ is the outer disk ellipticity and $q_{0}$ is the intrinsic
(edge-on) disk axis ratio \citep{hubble26}; we assumed an average
value of $q_{0} = 0.2$ \citep{lambas92}.

Several galaxies have strong, ``detached'' outer rings, with no
detectable disk beyond; the ``outer ring'' then is the largest listed
component in Table~\ref{tab:features}.  In these cases, the outer
isophotes may \textit{not} be those of an intrinsically circular
structure: \nocite{buta86}Buta (1986, 1995) has shown that outer rings
have typical axis ratios $\sim0.8$, and are either perpendicular to,
or parallel with, the bar.  The luminous region interior to the ring,
surrounding the bar, may not be circular either, since the bars can be
surrounded by intrinsically elliptical lenses and inner rings, aligned
with the bar \citep{kormendy79,buta86}.  In such cases, we could only
derive a range of possible inclinations, by combining the statistical
results of \nocite{buta95}Buta (1986, 1995) with kinematic information
from the literature.  In particular, we assumed that the circular
velocity $100$ \kms{} $\leq V_{circ} \leq 500$ \kms.  The upper limit
is the highest known value for a disk galaxy \citep[UGC
12591,][]{giovanelli86}, while 100 \kms{} is a lower limit to the
inclination-corrected circular velocities measured for 18 S0 and S0/a
galaxies by \citet{neistein99}.  Details of this process are given in
the discussions of individual galaxies, below.

\subsection{Measuring Orientations and Sizes\label{sect:measure}} 

For each feature identified in Table~\ref{tab:features}, we measured
its semi-major axis, position angle, and ellipticity.  For the outer
disks we list $R_{25}$, taken from RC3.

Position angles (counter-clockwise from north) were measured from the
ellipse fits, \textit{unless} unsharp masking and direct inspection of
the isophotes indicate a significantly different value (see below). 
Using the ellipse fits, we found the point where ellipticity reaches a
local maximum (\emax) at semi-major axis \amax.  The position angle at
\emax{} is often a good indication of the structure's position angle,
though dust lanes and spiral arms can confuse the issue.  In some
cases, the PA curve reaches a local minimum or maximum near \amax;
this extremal value is then a reasonable estimate for the PA of the
bar's major axis.

However, we find that the orientation of the fitted ellipses often
does not match that of the galaxy components involved.  This is a
problem particularly for secondary bars embedded in bright bulges:
unless the galaxy is close to face-on, or the bulge nearly spherical,
the projected isophotes of the bulge will be elliptical, and thus the
combined bulge + bar isophotes may not accurately reflect the bar's
orientation.  Figure~\ref{fig:umaskpa} illustrates cases where the
inner isophotes of double-barred galaxies show pointed (disky)
deviations --- and the unsharp masks show the secondary bar structure
--- at a different position angle from what the ellipse fits report. 
\textit{This is the case for about half the secondary bars in our
sample;} the differences in position angle range from 10\arcdeg{} to
80\arcdeg{}.  In such cases, we measured the bar's PA from the unsharp
mask and the isophotes, and report the PA from the ellipse fits in
parentheses.  We found similar, smaller deviations for twelve primary
bars (e.g., Figure~\ref{fig:n3941pa}); see Table~\ref{tab:features}. 
\citet{busarello96} found a similar discrepancy for NGC~3384, where
the main bar's PA measured with an adaptive Laplacian filter (similar
to an unsharp mask) differed by 60\arcdeg{} from the ellipse-fit
value.

Unfortunately, there are no standard methods for measuring the
\textit{sizes} of bars.  One fairly common measure is \amax{}
\citep[e.g., ][]{w91,regan97,jung97,laine02}.  However,
\nocite{w95}W95 argued that that in at least some galaxies the
isophotes remain elongated, parallel, and ``bar-shaped'' beyond \amax. 
For small features, this measure can also be sensitive to seeing, as
we pointed out in Section~\ref{sect:ellipse}.  For an upper limit to
bar length, \nocite{w95}W95 used the point outside \amax{} where
ellipticity reaches a local \textit{minimum}, at semi-major axis
\amin.  But this can also give misleading results if, e.g., there are
strong dust lanes, spiral arms or a ring starting near the ends of the
bar.  In these cases the isophotal ellipticity can remain high even
though the ellipses are no longer tracing the bar.

Our alternate method is based on the fact that bars in early-type
galaxies are extremely straight, deviating from a constant position
angle by less than $\pm2\arcdeg$ \citep{ohta90}.  A plausible upper
limit for bar length, \aten, is the point where the PA of the ellipses
has changed by $10\arcdeg$ from the bar's PA. This radius is somewhat
sensitive to viewing angle, and the technique fails completely if the
bar happens to be closely aligned with the disk major axis.  So we
used the \textit{smaller} of \amin{} and \aten{} as the length of the
bar.

In Table~\ref{tab:bar-length} we compare measurements of bar length
for four well-studied SB galaxies.  For these galaxies, bar length has
been determined by: decomposition into disk, bulge, and bar components
\citep{kent89}; analysis of how azimuthal bar-interbar contrast varies
with radius \citep{ohta90}; or Fourier decomposition
\nocite{quillen94}(Quillen, Frogel, \& Gonz\'{a}lez 1994). 
Measurements apparently made by eye are also included for two of the
galaxies.  For three of the four galaxies, \aten{} is a better match
to the published bar lengths than either \amax{} or \amin.  However,
for NGC~4314 the published lengths are all closer to \amax. 
Consequently, we list both measurements for bars and disk/bar
candidates in Table~\ref{tab:features}: \amax{} and the smaller of
\aten{} and \amin.

\section{Results for Individual Galaxies\label{sect:ind-galaxies}} 

\begin{figure}
\begin{center}
	\includegraphics[scale=0.85]{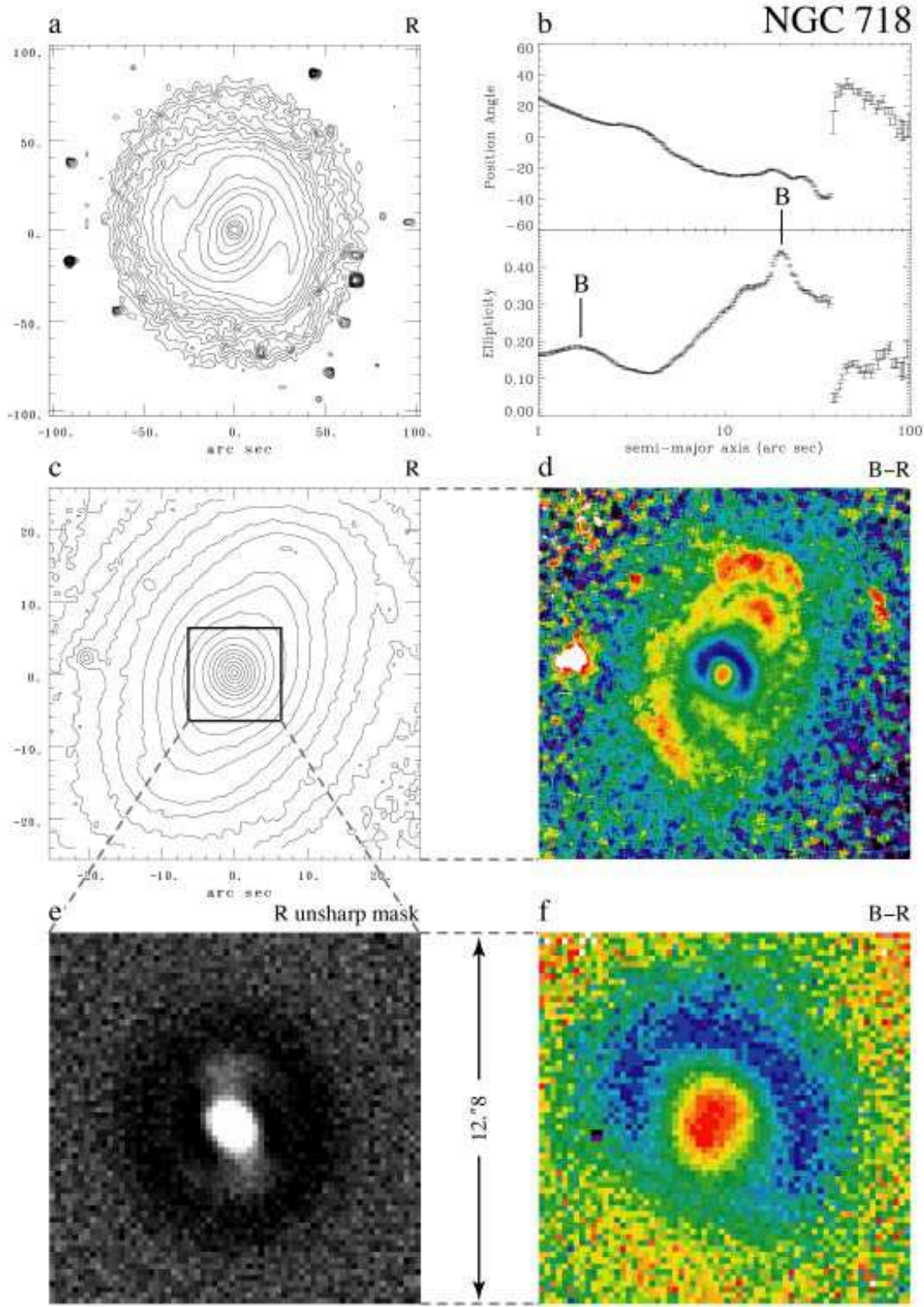}
\end{center}
\caption{NGC 718: $R$-band isophotes on two scales (a, c); ellipse
fits (b, WIYN $R$ only); \br{} color maps (d, median smoothing width
$w = 5$ pixels, color range = 0.85 mag; f, unsmoothed, color range =
0.28 mag between ring and galaxy nucleus); $R$-band unsharp mask (e,
Gaussian \ds{2} pixels).
\label{fig:n718}}
\end{figure}

\begin{figure}
\begin{center}
	\includegraphics[scale=0.85]{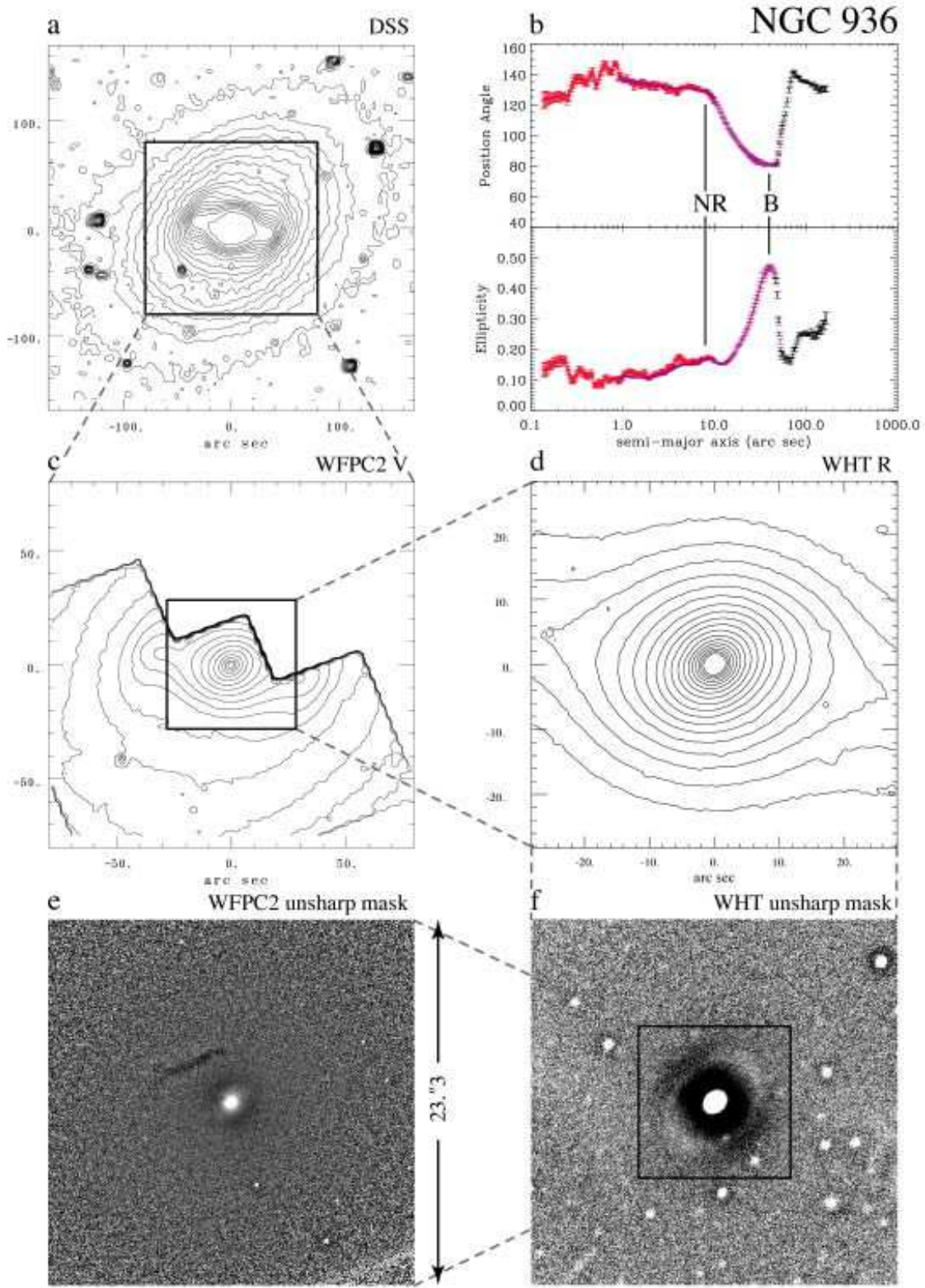}
\end{center}
\caption{NGC 936: large-scale DSS isophotes (a); ellipse fits (b,
PC2 F555W [red], WHT R [purple], DSS [black]); WFPC2 mosaic (c); WHT
$R$-band image (d), unsharp mask of PC2 image (e, \ds{10}); the bright
line at the lower right edge of the unsharp mask image is the edge of
the PC2 chip; unsharp mask of WHT $R$-band image (f,
\ds{10}).\label{fig:n936}}
\end{figure}

\begin{figure}
\begin{center}
	\includegraphics[scale=0.85]{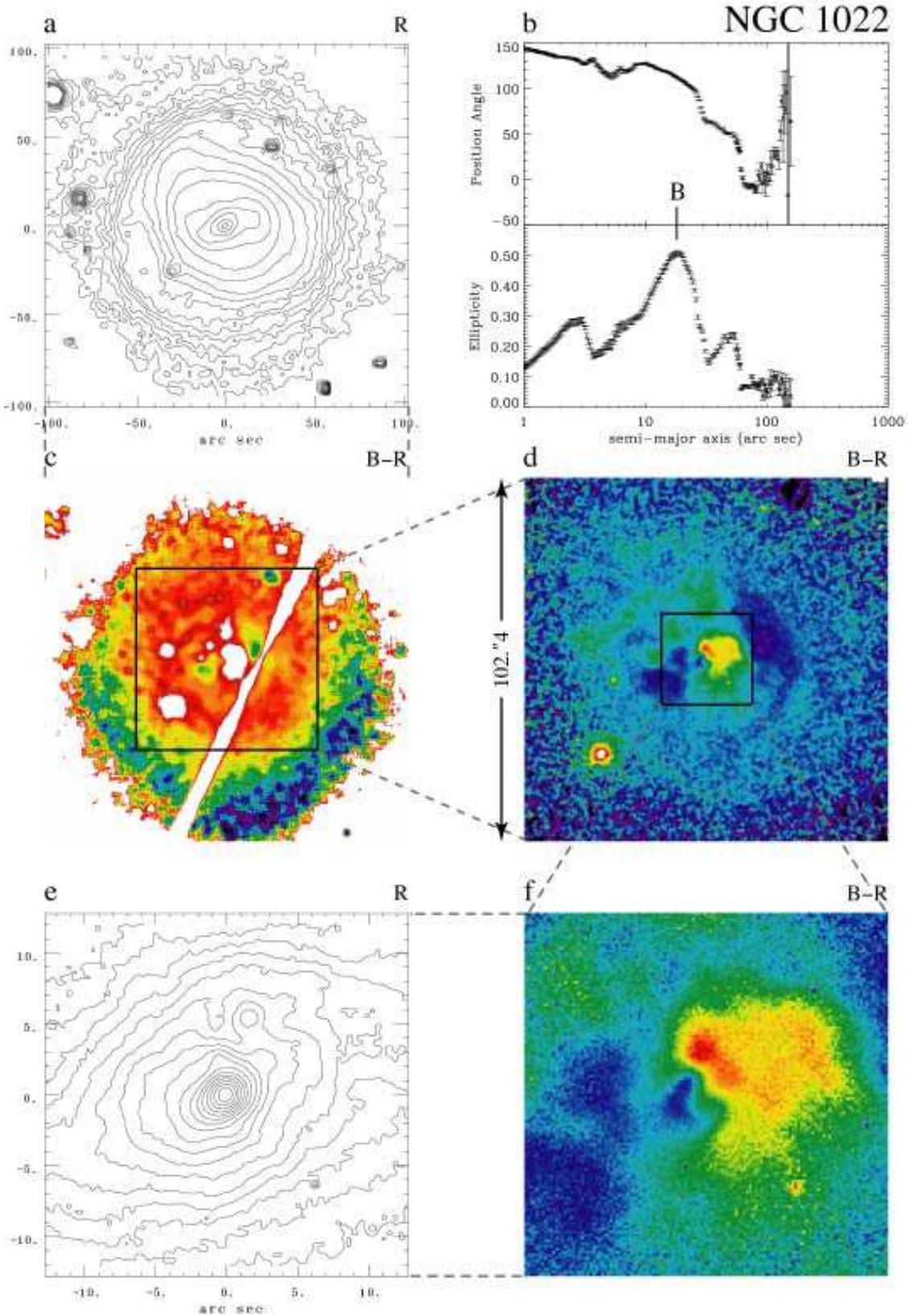}
\end{center}
\caption{NGC 1022: $R$-band isophotes on two scales (a, e); ellipse
fits (b, WIYN $R$ only); \br{} color maps (c, $w = 23$,
color range $\ap 1.40$ mag [excluding satellite.meteor trail); d,
$w = 5$, color range = 1.53 mag; f, unsmoothed, color
range = 0.90 mag).  The red diagonal streak in panel~c is due to a
bright satellite or meteor trail during the long $R$-band exposure;
the color maps in panels~d and f were made with the unmarred short
$R$-band exposure.\label{fig:n1022}}
\end{figure}

\begin{figure}
\begin{center}
	\includegraphics[scale=0.85]{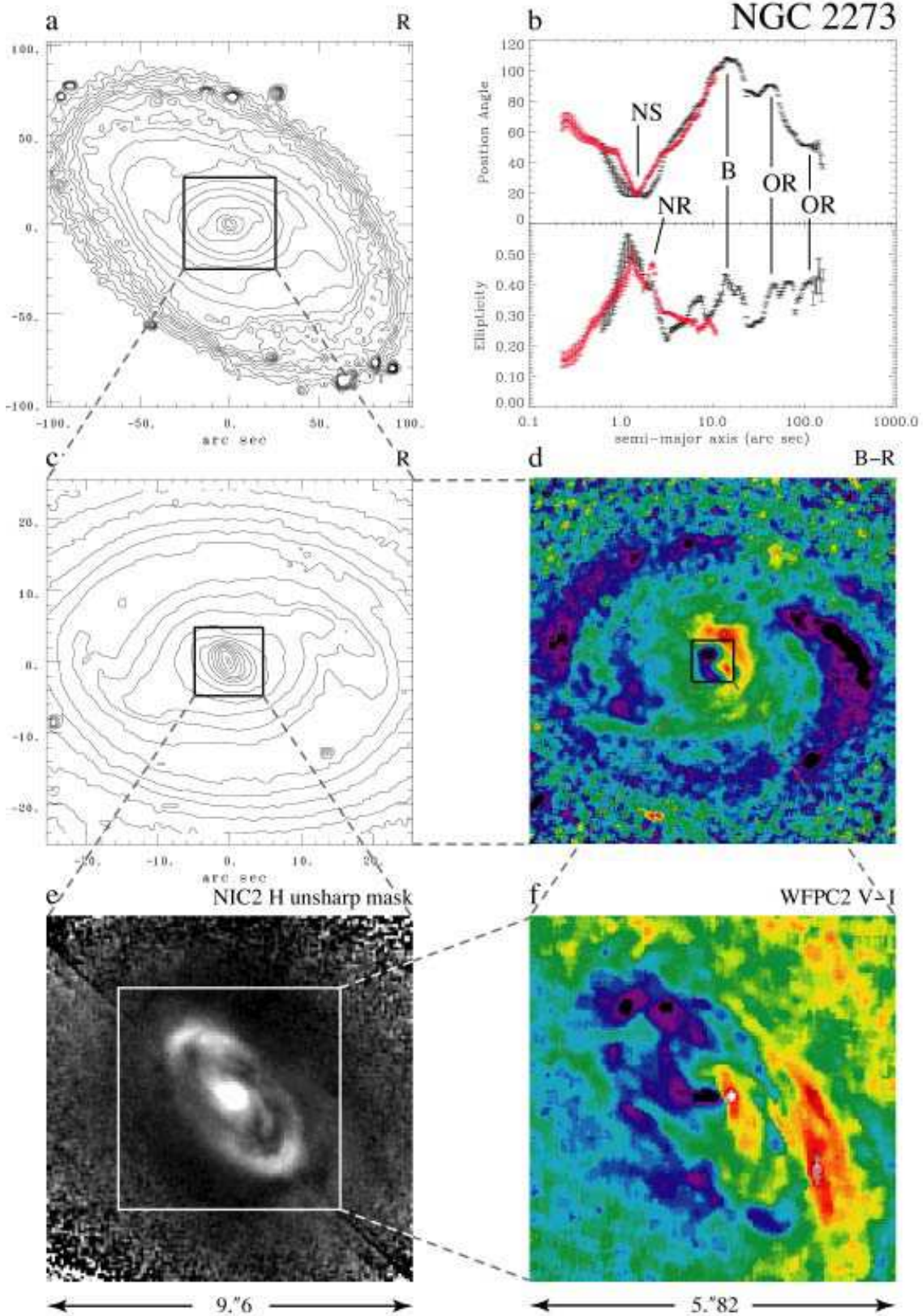}
\end{center}
\caption{NGC 2273: $R$-band isophotes on two scales (a, c); ellipse
fits (b, WIYN $R$ and NICMOS2 F160W [red]); \br{} color map (d, $w =
5$, color range = 0.89 mag); PC2 $V\!-\!I$ [F606W$\,-\,$F791W] color
map (f, $w = 5$, color range = 1.15 mag); unsharp mask of NICMOS2
F160W image (e, \ds{10}).
\label{fig:n2273}}
\end{figure}

\begin{figure}
\begin{center}
	\includegraphics[scale=0.85]{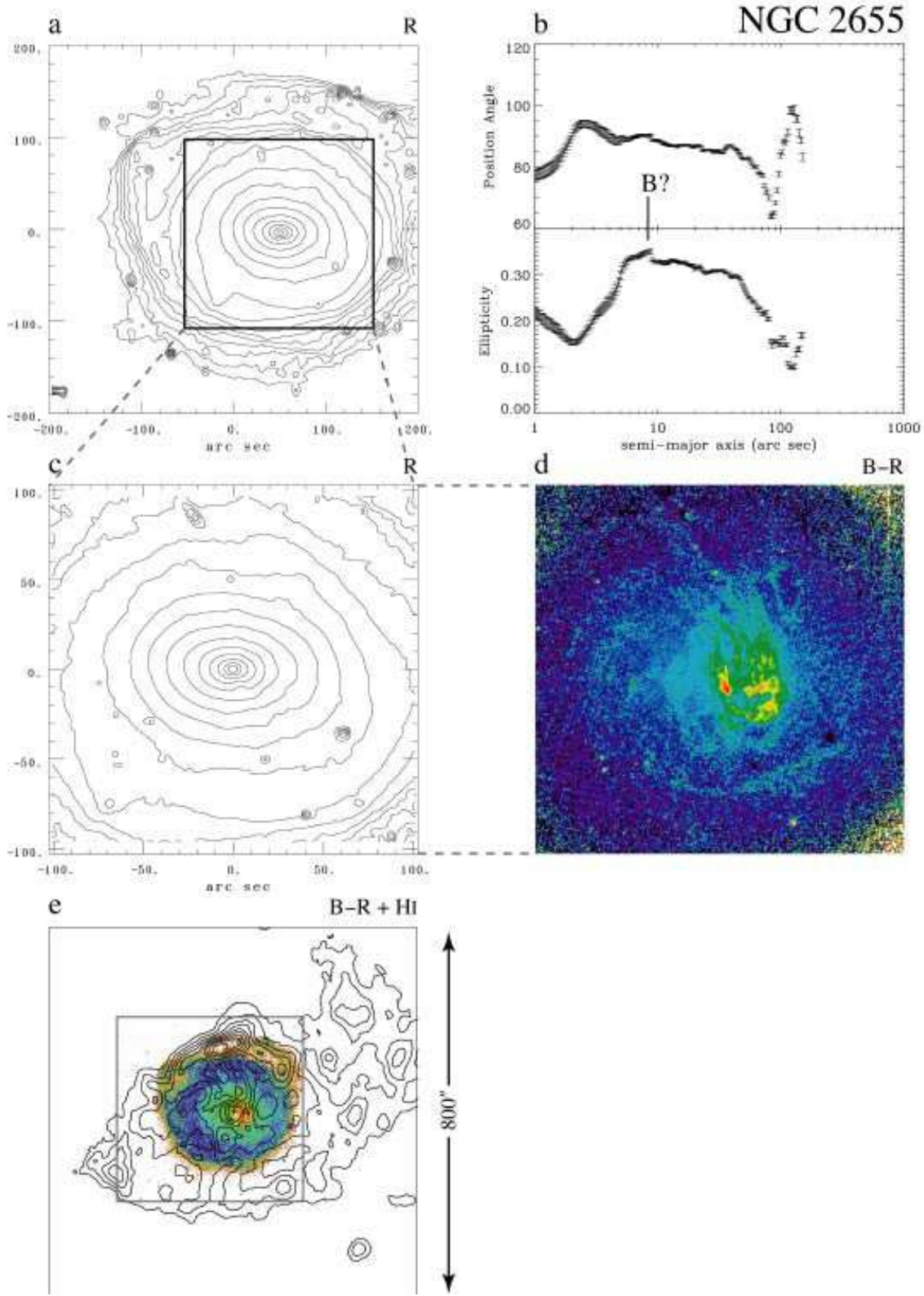}
\end{center}
\caption{NGC 2655: $R$-band isophotes on two scales (a, c); ellipse
fits (b, WIYN $R$); \br{} color map (d, $w = 5$, color range $\ap 1.2$
mag; e, $w = 11$, color range $\ap 1$ mag, with \hi{} radio contours). 
The contour map in panel~a is the entire WIYN CCD; the \hi{} contours
in panel~e is from \citet{vanM02}.
\label{fig:n2655}}
\end{figure}

\begin{figure}
\begin{center}
	\includegraphics[scale=0.85]{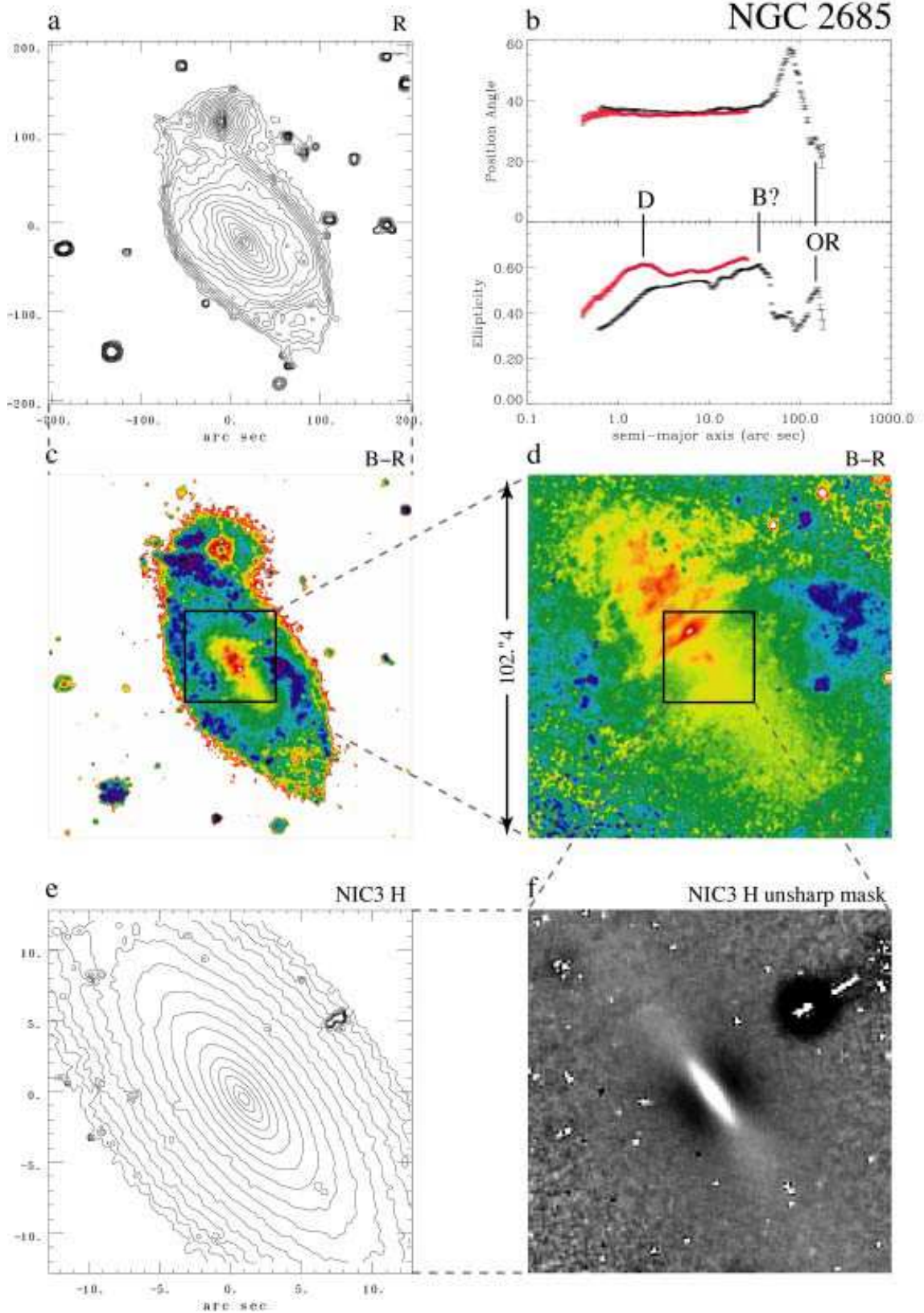}
\end{center}
\caption{NGC 2685: $R$-band isophotes (a); ellipse fits (b, WIYN
$R$ and NICMOS3 F160W [red]); \br{} color maps (c, $w = 23$, color
range $\ap 1.0$ mag; d, unsmoothed, color range = 1.48 mag); NICMOS3
F160W isophotes (e) and unsharp mask (f, \ds{5}).  The contour map in
panel~a and the color map in panel~c are the entire WIYN CCD.
\label{fig:n2685}}
\end{figure}

\begin{figure}
\begin{center}
	\includegraphics[scale=0.85]{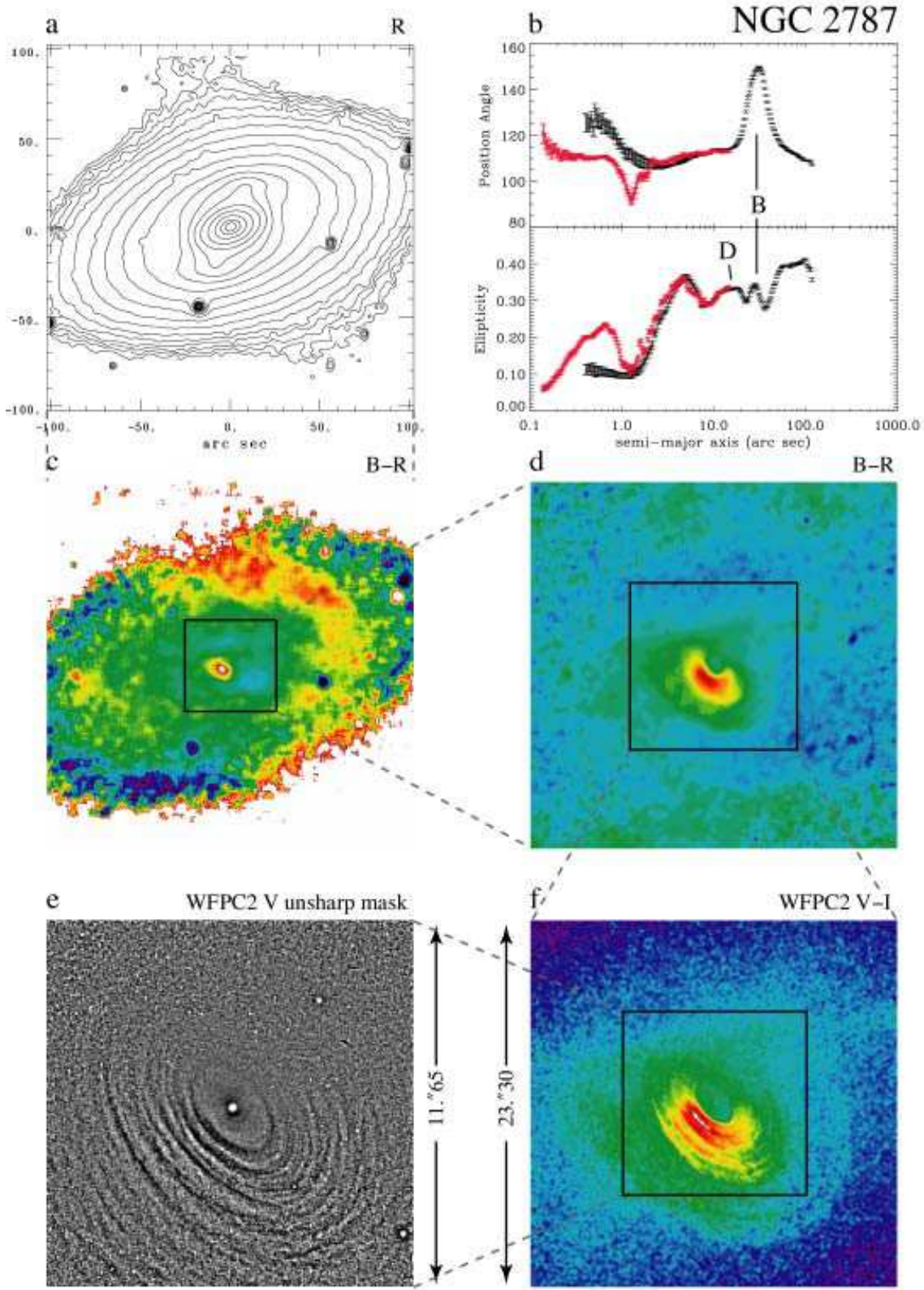}
\end{center}
\caption{NGC 2787: $R$-band isophotes (a); ellipse fits (b, WIYN
$R$ and PC2 F814W [red]); \br{} color map (c, $w = 23$, color range =
0.65 mag); PC2 $V\!-\!I$ [F555W$\,-\,$F814W] color map (f, $w = 5$,
color range = 0.80 mag); unsharp mask of F555W PC2 image (e, \ds{5}). 
The extensions on the NNE and SSW edges of the disk in panel~a are due
to scattered light from an extremely bright star just off the edge of
the CCD.
\label{fig:n2787}}
\end{figure}

\begin{figure}
\begin{center}
	\includegraphics[scale=0.85]{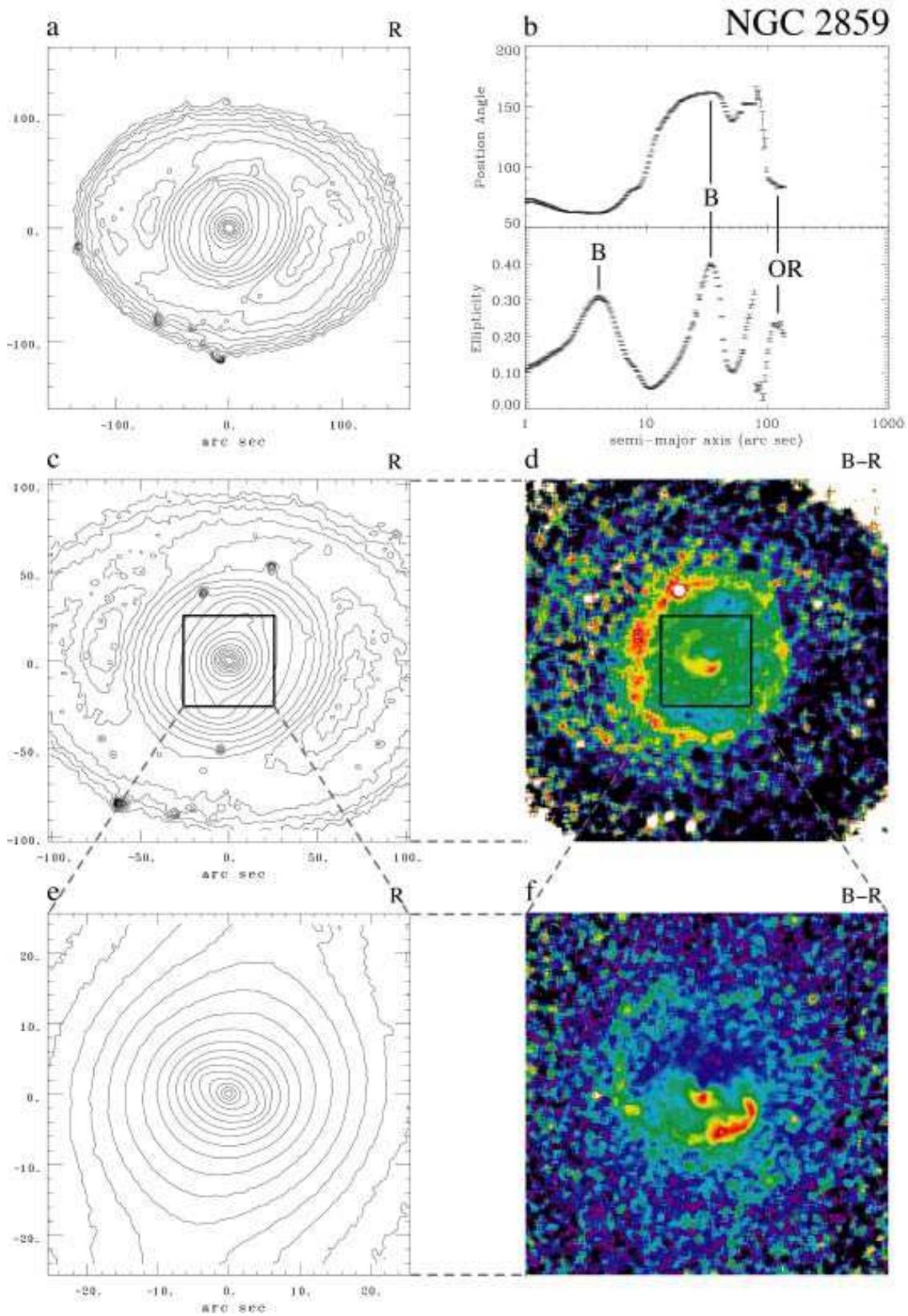}
\end{center}
\caption{NGC 2859: $R$-band isophotes on various scales (a, c, e);
ellipse fits (b, WIYN $R$); \br{} color maps (d, $w = 23$, color range
= 0.54 mag; f, $w = 5$, color range = 0.27 mag).
\label{fig:n2859}}
\end{figure}

\begin{figure}
\begin{center}
	\includegraphics[scale=0.85]{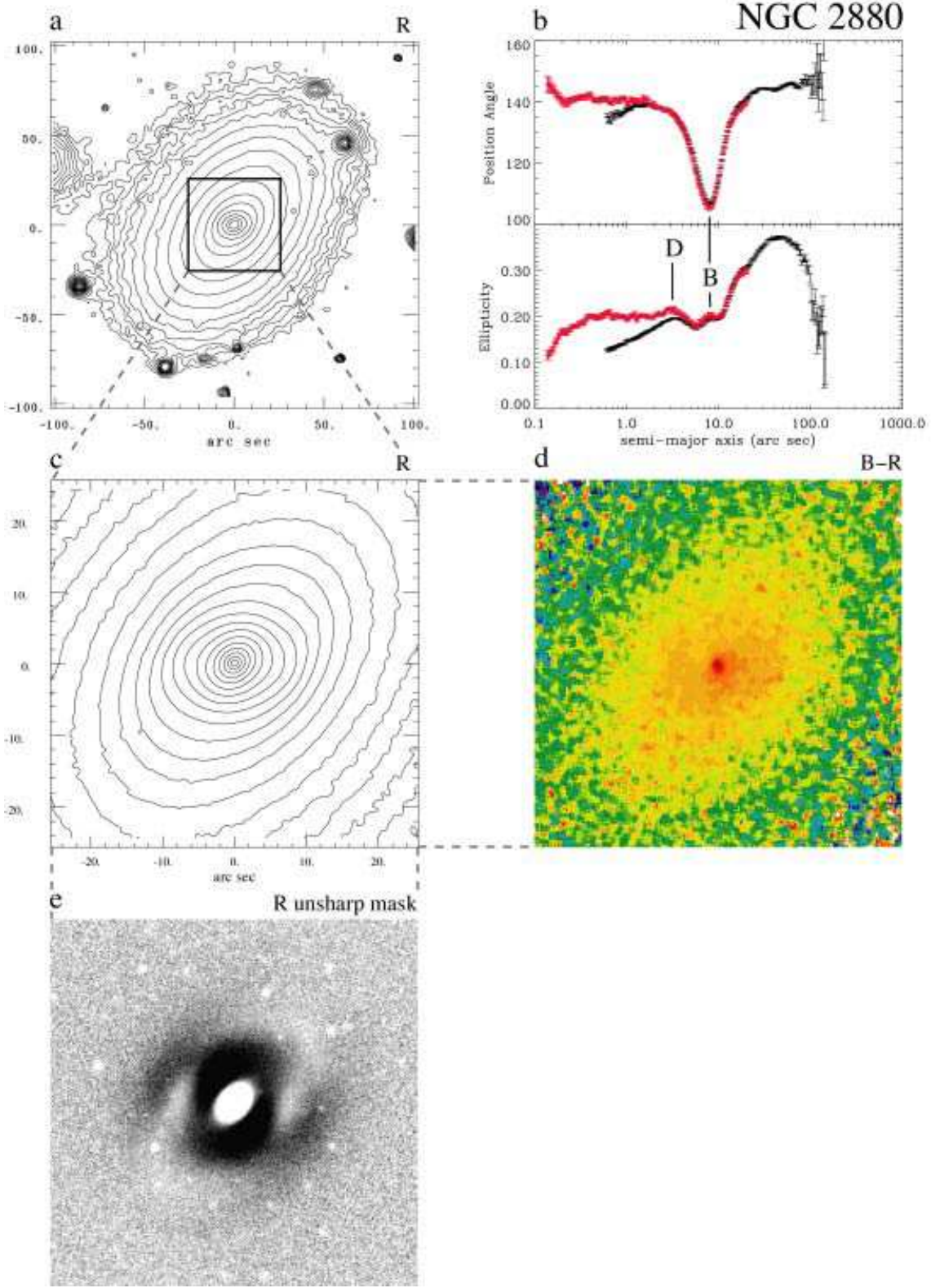}
\end{center}
\caption{NGC 2880: $R$-band isophotes on two scales (a, c); ellipse
fits (b, WIYN $R$ and WFPC2 F814W [red]); \br{} color map (d, $w = 5$,
color range = 0.39 mag); \ds{10} unsharp mask of WIYN $R$-band image
(e), showing the bar and spiral/ring just outside.
\label{fig:n2880}}
\end{figure}

\begin{figure}
\begin{center}
	\includegraphics[scale=0.85]{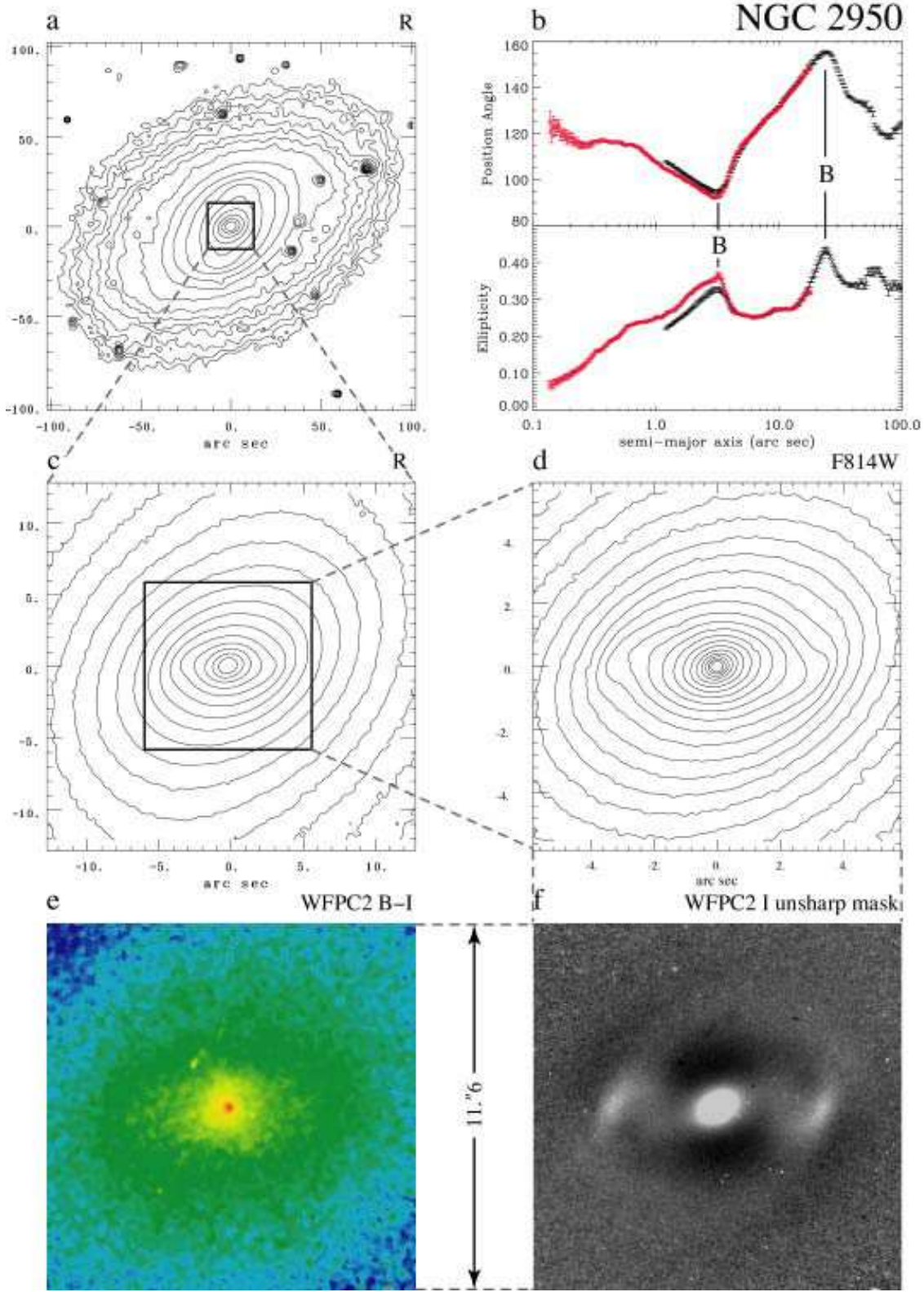}
\end{center}
\caption{NGC 2950: $R$-band isophotes (a, c); ellipse fits (b, WIYN
$R$ and WFPC2 F814W [red]); WFPC2 F814W isophotes (d); WFPC2
F450W$\,-\,$F814W colormap (e, $w = 5$, color range = 0.38 mag);
\ds{15} unsharp mask of the WFPC2 F814W image
\label{fig:n2950}}
\end{figure}

\begin{figure}
\begin{center}
	\includegraphics[scale=0.85]{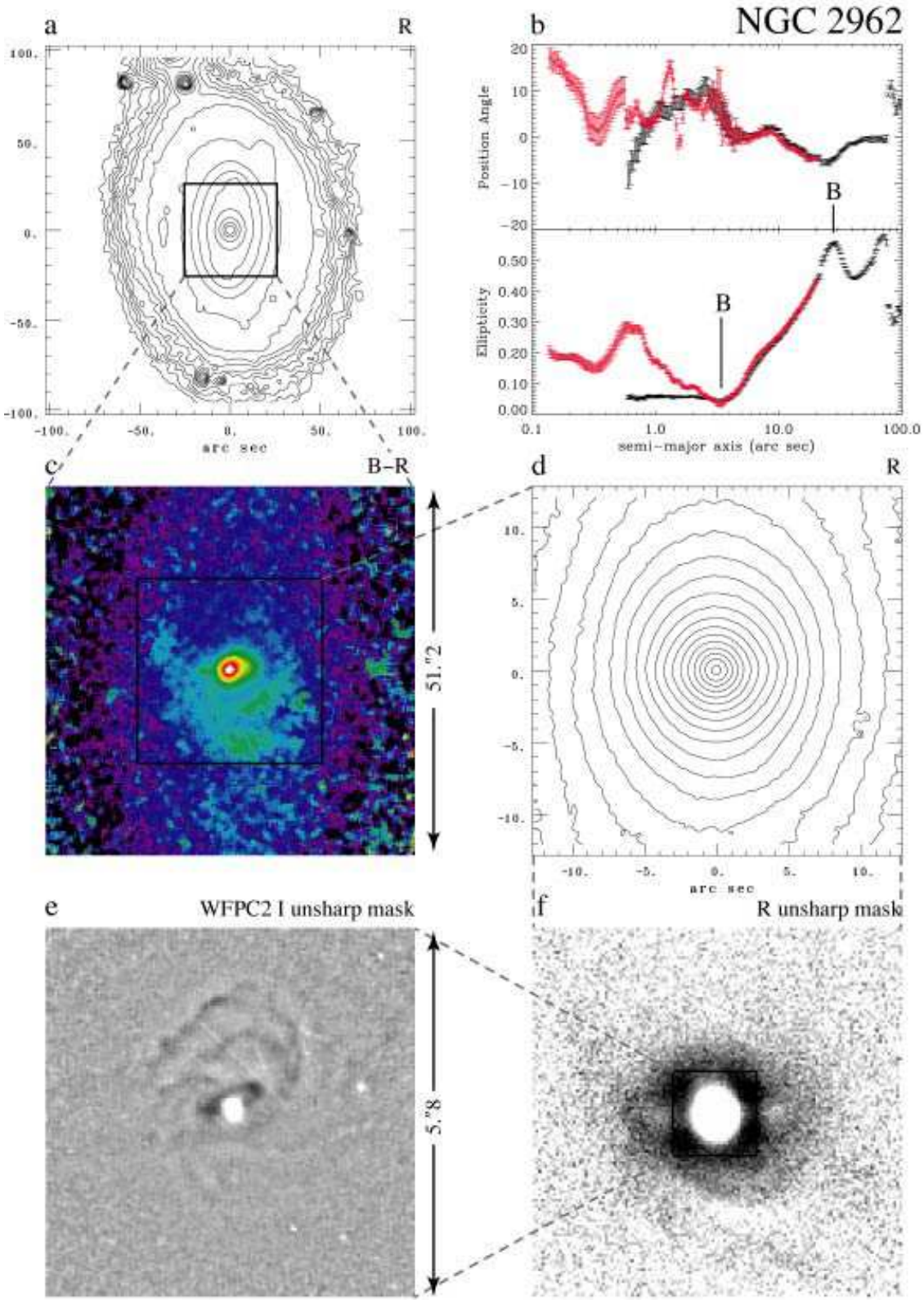}
\end{center}
\caption{NGC 2962: $R$-band isophotes (a, d); ellipse fits (b, WIYN
$R$ and WFPC2 F814W [red]); \br{} color map (c, $w = 5$, color range =
0.52 mag); \ds{5} unsharp mask of the $R$-band image (f), showing the
secondary bar and dust lanes outside it; \ds{3} unsharp mask of the
WFPC2 F814W image.
\label{fig:n2962}}
\end{figure}

\begin{figure}
\begin{center}
	\includegraphics[scale=0.85]{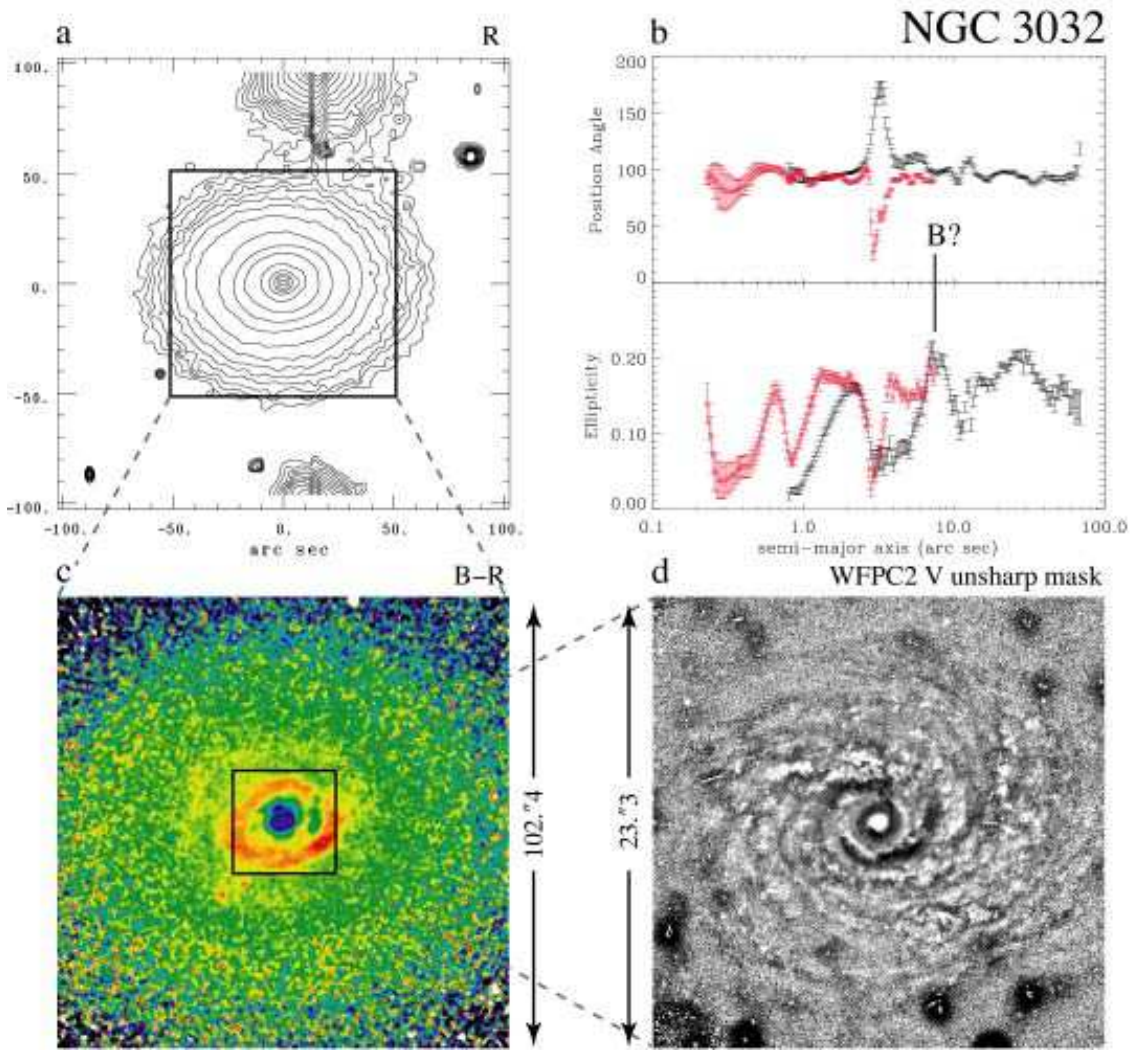}
\end{center}
\caption{NGC 3032: $R$-band isophotes (a); ellipse fits (b,
WIYN $R$ and NICMOS2 F160W [red]); WIYN \br{} color map (c, $w = 5$,
color range between dust ring and nuclear blue spots = 1.07 mag); and
a \ds{10} unsharp mask of the F606W PC2 image (d), showing the nuclear
spiral.  The two objects N and S of the galaxy in (a) are bright,
saturated stars.  Note that pixels in the central 0.6--0.7\arcsec{} of
the PC2 image are saturated.
\label{fig:n3032}}
\end{figure}

\begin{figure}
\begin{center}
	\includegraphics[scale=0.85]{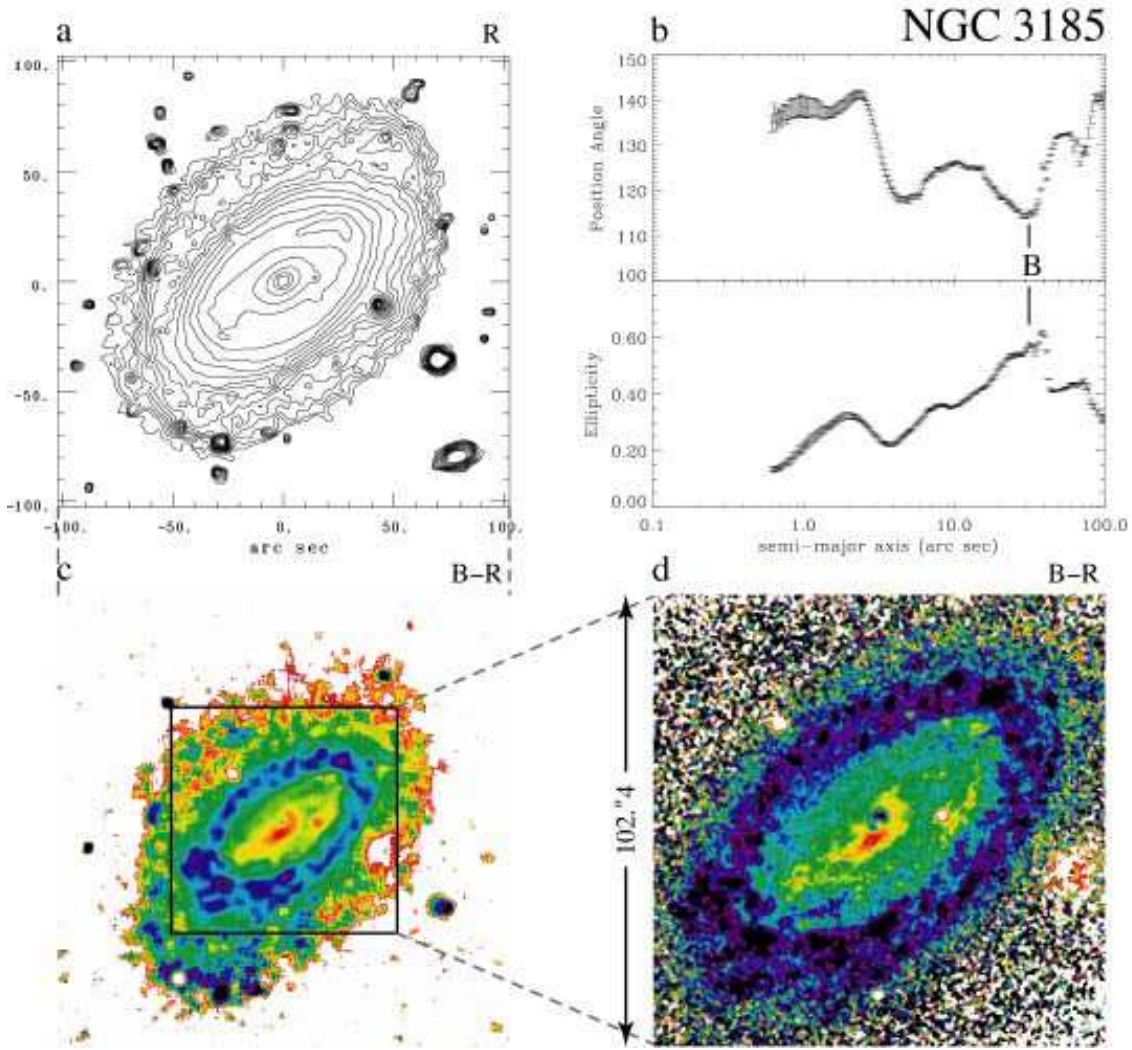}
\end{center}
\caption{NGC 3185: large-scale $R$-band isophotes (a); ellipse fits
(b, WIYN $R$); \br{} colors on the same scale (c, $w = 23$, color
range = 1.1 mag); and a closer view of the \br{} color map showing
dust lanes in the bar and a possible nuclear dust ring (d, $w = 5$,
color range between inner ring and central dust lanes = 1.02 mag).
\label{fig:n3185}}
\end{figure}

\begin{figure}
\begin{center}
	\includegraphics[scale=0.85]{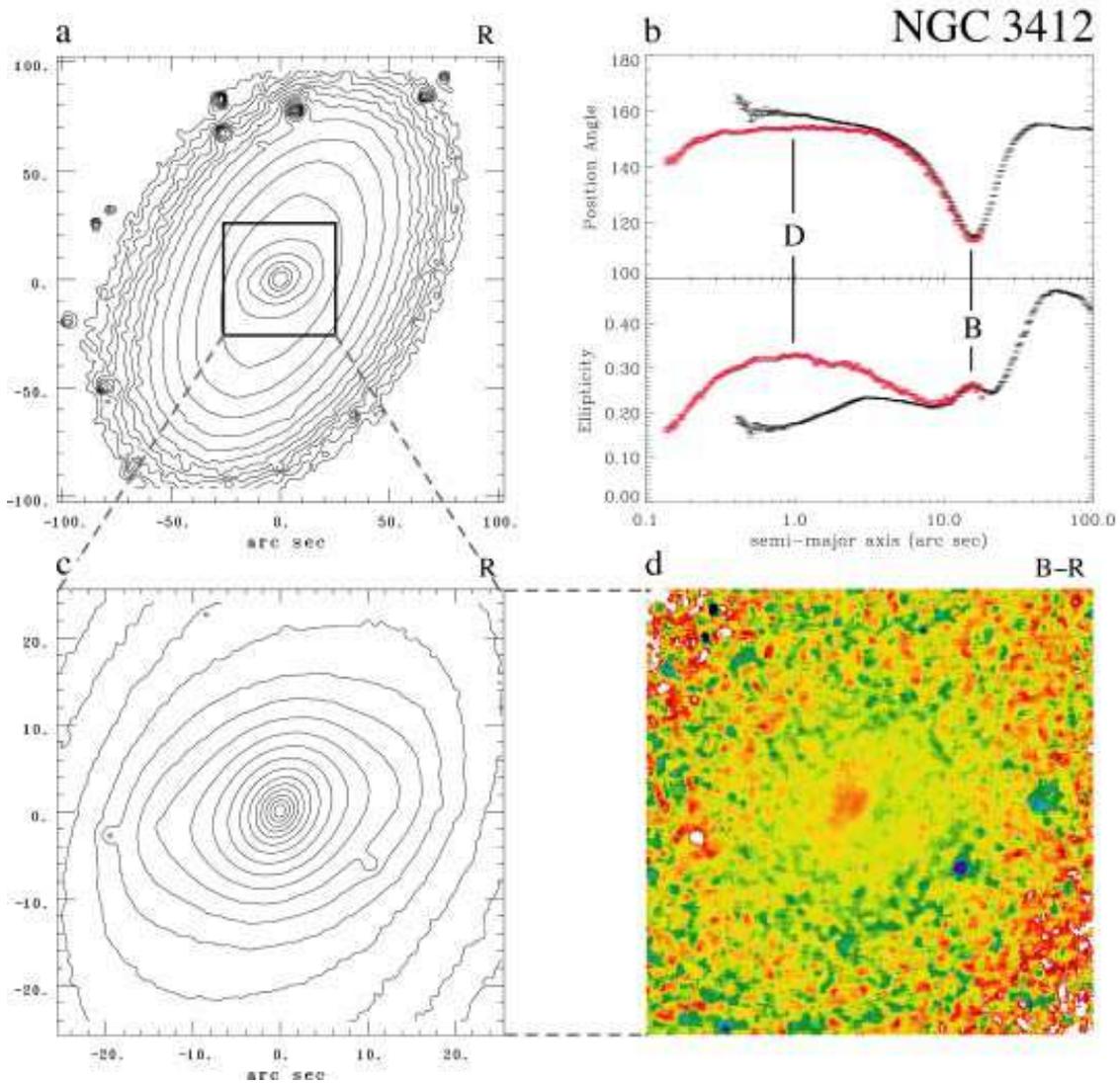}
\end{center}
\caption{NGC 3412: large- and small-scale $R$-band isophotes (a,
c); ellipse fits (b, WIYN $R$ and WFPC2 F606W); along with the \br{}
color map (d, $w = 5$, color range = 0.27 mag).
\label{fig:n3412}}
\end{figure}

\begin{figure}
\begin{center}
	\includegraphics[scale=0.85]{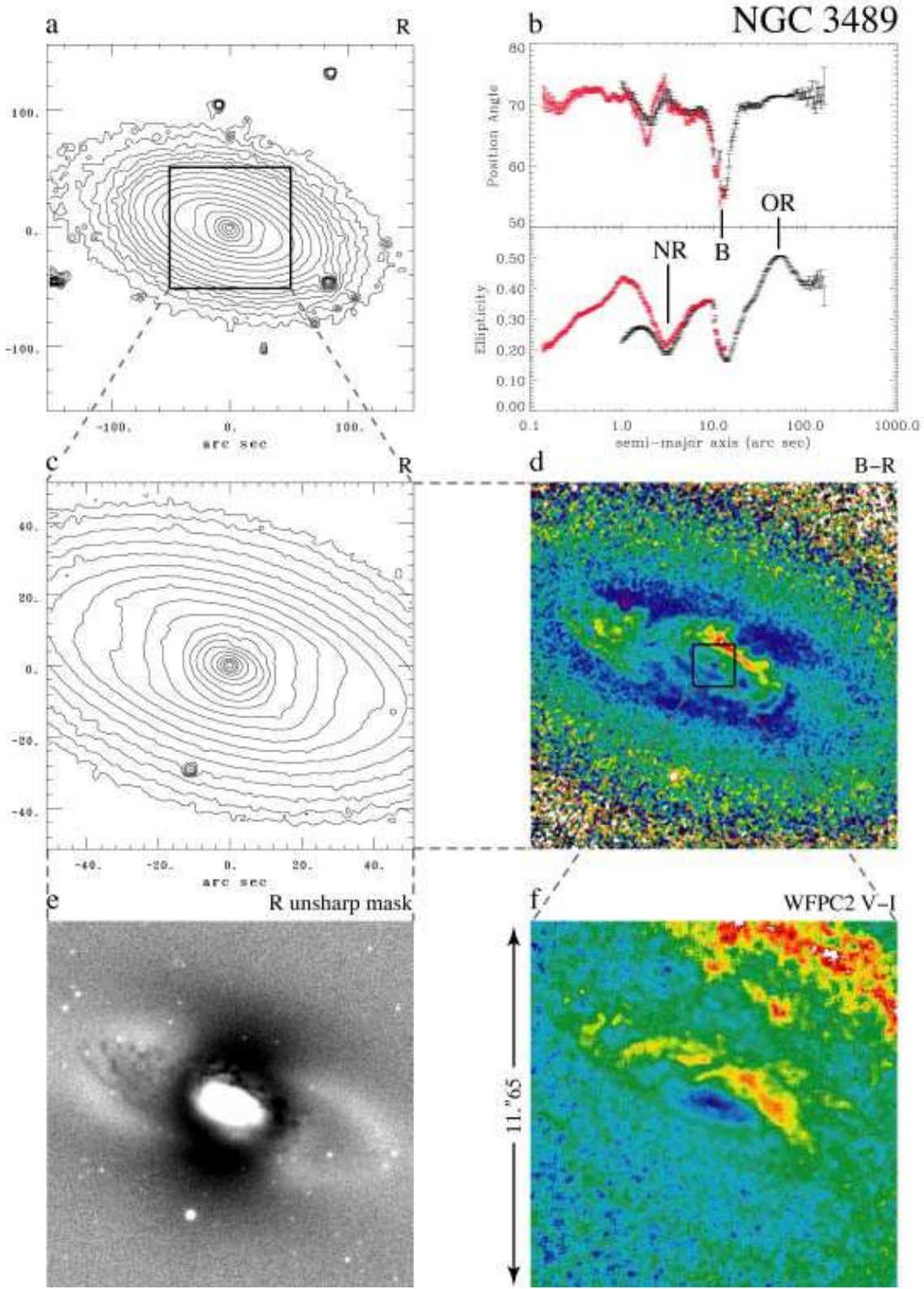}
\end{center}
\caption{NGC 3489: large-scale (a) and medium-scale $R$-band
isophotes (c); ellipse fits (b, WIYN $R$ and WFPC2 F814W [red]); \br{}
color map (d, $w = 5$, color range = 0.57 mag); \ds{40} unsharp mask
of WIYN $R$-band image showing the outer ring, dust lanes, and the bar
(e); HST F555W$\,-\,$F814W color map (f, $w = 5$, color range = 0.80
mag) showing the nuclear dust ring and blue interior.
\label{fig:n3489}}
\end{figure}

\begin{figure}
\begin{center}
	\includegraphics[scale=0.85]{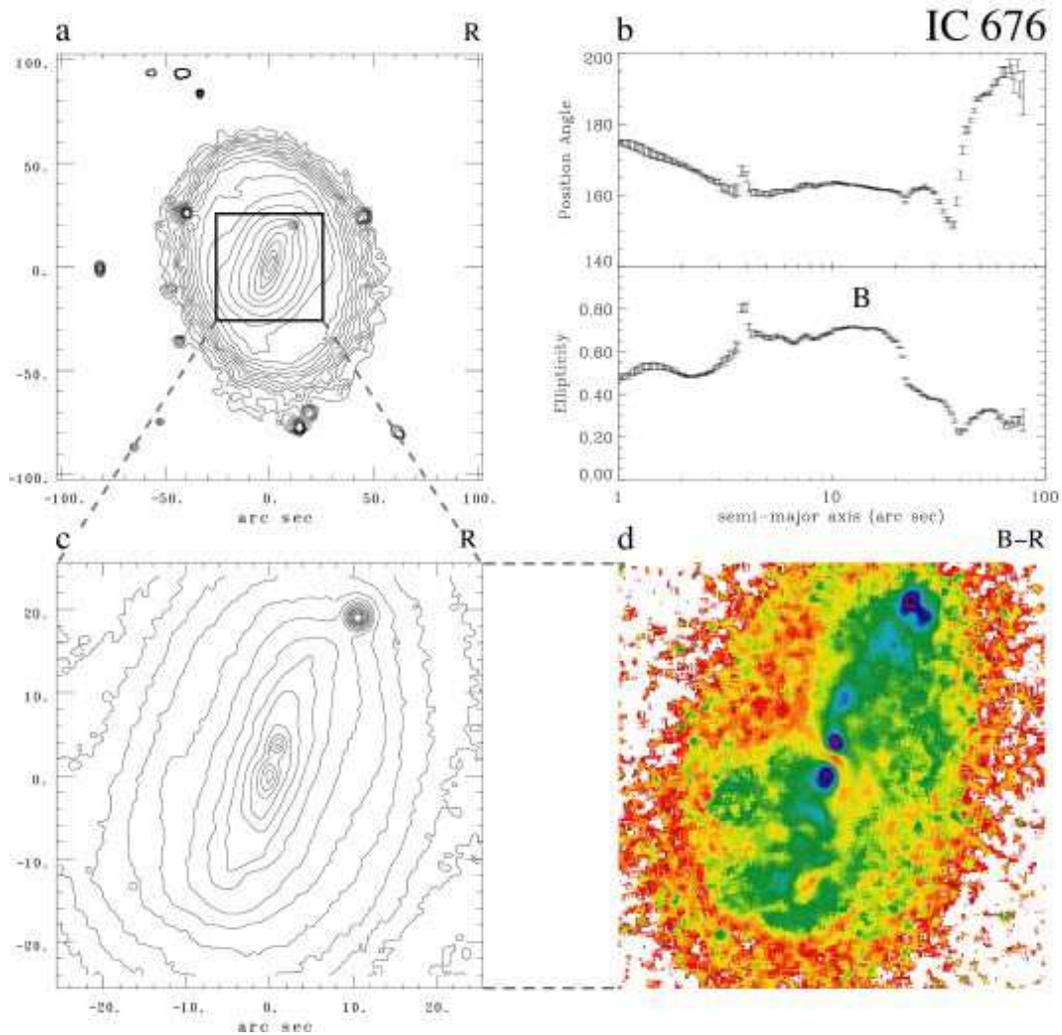}
\end{center}
\caption{IC 676: large-scale $R$-band isophotes (a); ellipse fits
(b, WIYN $R$); a close-up of the bar region (c), which shows the
``binary nuclei''; and a \br{} color map of the same region (d, $w =
5$, color range = 0.92 mag, excluding outer regions).  The bright,
blue object on the NW edge of the bar is a foreground star.
\label{fig:ic676}}
\end{figure}

In this section we discuss the individual galaxies in the sample, and
present figures for each (Figures~\ref{fig:n718}--\ref{fig:n7743}). 
These figures show the ellipse fits, contour maps, and one or more
color maps or unsharp masks.

Isophote maps use the square root of logarithmic (sky-subtracted)
counts for the ground-based $R$-band images.  HST images images are
not sky-subtracted, since most of the galaxies are larger than the WFC
field of view, but use the same scaling; DSS images, which are
taken from photographic plates, are plotted with linear contours.

Ellipse fits to the ground-based $R$-band images are plotted in black;
those for HST images are plotted in gray (red for online figures). 
The longest-wavelength HST image available is used for the ellipse
fits, to minimize confusion due to dust and star formation.  We label
major features in the ellipse fits as follows: ``B'' = bar, ``D'' =
inner disk, ``NR'' = nuclear ring, ``IR'' = inner ring, and ``OR'' =
outer ring.

Color maps are displayed as grayscale images (color in the online
figures).  Lighter values are redder.  (For the online figures, we
use a natural color map, with red = red and blue = blue.)  The
displayed colors are adjusted for each plot to maximize contrast; the
range in the appropriate astronomical color index is given in the
figure captions for each galaxy.  Because our images are uncalibrated,
these are \textit{relative} color maps.  The applied median smoothing
$w$ is given in pixels.  Unsharp masks are displayed as grayscale
images, with excesses light and deficits dark.  The smoothing $\sigma$
is given in pixels.

Two of the galaxies from our sample --- NGC~2681 and NGC~3945
--- have already been discussed in \citet{erwin99}.  The reader is
referred to that paper for detailed analyses and figures of those
galaxies, though we include some additional notes for each below.

\subsection{NGC 718: Double bar, nuclear ring\label{n718}} 

Figure~\ref{fig:n718}: This weakly barred galaxy has two strong,
uneven spiral arms outside the bar, with the western arm both stronger
and bluer than the eastern arm; \citet{kennicut86} concluded that the
western arm was the site of recent star formation.  The two arms may
form a weak outer pseudo-ring, oriented at PA $\ap 45\arcdeg$, which
is the value for the disk given in \nocite{rc3}RC3; our disk
orientation ($PA = 5\arcdeg$) is based on isophotes further out.

The color map shows several strong dust lanes within the bar, only one
of which is a plausible leading-edge lane.  There is also a blue ring
oriented roughly perpendicular to the bar; the southeast side of the
ring is redder, probably due to the nearby dust lane.  This is one of
two examples of a ``blue nuclear ring'' in our sample: it is quite
smooth, with no signs of dust or knots which might indicate current
star formation.  In fact the ring is $\sim 0.1$ mag bluer than most of
the outer disk, but $\sim 0.1$--0.15 mag \textit{redder} than the
strong, blue western arm, so it could be fairly old.  Nonetheless, due
to its distinct color we consider it a ``young'' and presumably
gas-rich nuclear ring.  The ring does not show up in unsharp masks,
except as a possible enhancement surrounding the inner bar in the $R$
band, at the inside edge of the blue ring.  (The lower signal and low
resolution of the $B$ image --- 1.7\arcsec{} --- may explain why the
ring does not appear in $B$-band unsharp masks.)

A secondary bar shows up as an ellipticity peak, and in the \ds{2}
unsharp mask.  The position angle we report for the inner bar is the
value at $\amax$, which agrees with the elongated feature in the
unsharp mask.  Telescope focus problems caused the elliptical shape of
the nucleus in the unsharp mask, and the inward twist in ellipse-fit
PA. Additional images obtained in 2000 December with the William
Herschel Telescope's Auxport camera (see next galaxy) confirm the
presence of the inner bar.

\subsection{NGC 936: Nuclear ring\label{n936}} 

Figure~\ref{fig:n936}: This is a very well studied, prototypical SB0
galaxy \citep[e.g.,][]{kormendy83,kent89,merrifield95}.  To supplement
the archival WFCP2 image (short exposures in the F555W filter), we
obtained high-resolution, ground-based images of the bar region with
the Auxiliary Port Camera of the 4.2m William Herschel Telescope (a
$1024 \times 1024$ CCD with 0.11\arcsec{} pixels).  These show an
elliptical feature with semi-major axis $\approx 8\arcsec$; unsharp
masking (panel~f) reveals this to be a nuclear ring, with orientation
and ellipticity roughly consistent with a circular ring seen in
projection.  Because this appears equally well in $B$, $V$, and $R$
images, and does not appear as a distinct feature in color maps (not
shown), we conclude that this ring is predominantly stellar and
probably similar in population to the surrounding stars of the bulge
and bar.

The ring does \textit{not} show up in unsharp masks of the WFPC2
images (panel~e), although the ellipse fits do show a similar peak. 
By comparing unsharp masks of long (300s) and short (60s) WHT
exposures, we have determined that high signal/noise is necessary to
see the ring clearly in unsharp masks: it is barely detectable in 60s
exposures, but clearly visible in 300s.  Since the combined WFPC2 long
exposures total only 280s, the difference in mirror and pixel sizes
means the 300s WHT images receive $\sim 20$ times the signal per pixel
of the combined HST images (since this is the bright central region, 
the lower sky background of HST images has little effect on S/N).

Both the WHT and WFPC2 images show a weak linear dust feature
approximately 4.5\arcsec{} long, about 3\arcsec{} north-northeast of
the nucleus at closest approach, curving slightly inward at the west
end.  An even fainter straight dust lane extends from the nucleus to
the east-northeast, about 0.8\arcsec{} long; this may be responsible
for the innermost ($a < 1\arcsec$) features in the WFPC2 ellipse fits.

\subsection{NGC 1022: Dust-obscured\label{n1022}} 

Figure~\ref{fig:n1022}: This is a known starburst galaxy
\citep[e.g.,][]{devereux89,usui98}.  The roundness of the outer
isophotes and the low \hi{} width ($W_{50} = 182$ \kms) reported by
\citet{theureau98} both indicate the galaxy is close to face-on; the
position angle for the outer disk is rather uncertain.

Unsharp masks suggest a faint outer ring with \rone{} morphology and
an inner pseudo-ring surrounding the bar.  The latter ring also shows
up as a red feature in the color maps, and as an ellipticity minimum
at $a \sim 30\arcsec$.  The bar itself has notably blue ends,
suggesting recent star formation there.  Within the bar, strong
obscuring dust lanes are visible, especially in the northwest half;
these prevent us from determining what stellar structures might exist
inside the bar.

\subsection{NGC 2273 (Mrk 620): Nuclear spiral, nuclear ring\label{n2273}} 

Figure~\ref{fig:n2273}: This is a rather spectacular
\textit{four}-ringed spiral galaxy, with two outer rings, an inner
ring, and a bright star-forming nuclear ring with a luminous blue
nuclear spiral inside.  The latter structure masquerades as a
secondary bar in ellipse fits and ground-based images.

The two outer rings make determining the disk inclination and position
angle somewhat difficult.  Fortunately, the \hi{} mapping performed by
\citet{vanD91} yields plausible values of 50\arcdeg{} for both
inclination and position angle for the two outer rings, where most of
the \hi{} is concentrated.  We assume these rings to be coplanar with 
the stellar disk.

The outer of the two outer rings is blue, as is the inner ring which
wraps around the bar.  Blue regions at the end of the bar are probably
associated with the \hii{} regions found by \citet{gonzalez97}.  The
color maps show leading-edge dust lanes, and strong curving lanes in
between the nuclear region and the ends of the bar that are
particularly strong examples of what \citet{sheth00} term ``dust
spurs.''  The combination of leading-edge lanes, which are strongest
to the north and south of the galaxy center, and the spurs are
probably responsible for the $r \sim 7\arcsec$ red ring seen in the
low-resolution optical color maps of \citet{yankulova99}.

\citet{mulchaey97b} suggested there was a secondary bar, based on
ellipse fits to their ground-based near-IR images and the appearance
of the images.  We disagree.  Our ellipse fits do agree with theirs
--- in fact, the NICMOS2 (F160W) fits show a distinct ellipticity
peak, accompanied by a sharp twist in position angle to a minimum
value which is not that of the outer disk.  But the WFPC2 and NICMOS
\textit{images} show an $r \approx 2\arcsec$ nuclear ring with a
spiral inside, and \textit{not} a secondary bar.  This combined
structure can also be seen in the WFPC2 images presented by
\nocite{ferruit00}Ferruit, Wilson, \& Mulchaey (2000).  The ring is
luminous and blue, and its irregular, patchy character suggests recent
star formation.  Ferruit et al.'s \nii{} + \ha{} image shows emission
from the ring, with \ha{}:\oiii{} ratios characteristic of \hii{}
regions, so we class this as a star-forming ring.  The spiral, which
appears to be the cause of the innermost ellipticity peak, consists of
spiral dust lanes \textit{and} a luminous, two-armed blue spiral
(compare panels~e and f); its size and position angle of 20\arcdeg{}
also matches the extended amorphous 20 cm radio emission
\citep{ulvestad84}.  The molecular ``bar'' seen in the intermediate
resolution ($3.0\arcsec \times 2.5\arcsec$) CO maps of
\citet{petitpas02} is probably the gas counterpart of this ring +
spiral structure.

A blue spiral arm \textit{outside} the ring on the west side (at the
right-hand edge of panels~f; see also the F547M image in
\nocite{ferruit00}Ferruit et al.)  is reminiscent of the blue spiral
arms outside the nuclear ring in NGC~4314 (see below).  A fainter
counterpart may be present on the east side.

The galaxy has a Seyfert 2 nuclear spectrum, with broad \ha{} emission
detected in polarized light by \citet{moran00}.  In the HST images,
the nucleus is blue and luminous, elongated east-west, parallel to the
east-west double-nucleus radio structure seen by Ulvestad \& Wilson
and \citet{nagar99}.  We may be seeing the strong \oiii{} 5007
emission noted by \citet{ferruit00} in their WFPC2 images, since the
F606W filter we used for our color maps is wide enough to let in
\oiii{} emission.

\subsection{NGC 2655: Dust-obscured, off-plane gas\label{n2655}} 

Figure~\ref{fig:n2655}: This is the largest, most luminous galaxy in
the sample, with both an active nucleus and substantial off-plane gas. 
The isophotes are nearly round, though the outer disk is irregular. 
We measure the PA and inclination at the ellipticity \textit{minimum}
at $a \sim 110$--130\arcsec.  The central elliptical region bears
little resemblance to any of the other galaxies in our sample; this
galaxy may not, in fact, be barred.

We find two sets of dust features in this galaxy: faint, outer,
counter-clockwise spiral arms, which appear on all sides of the galaxy
for $r \gtrsim 50\arcsec$, and much stronger, disorganized lanes
closer to the center, almost entirely on the west side of the galaxy. 
The kinematics of the extended, irregular \hi{} gas
\citep{huchtmeier82,shane83} indicates significant off-plane gas.  We
show our \hi{} map \citep{vanM02} in panel~e of the figure.

WFPC2 images \citep{barth98} show that the strong, one-sided dust
lanes continue into the nuclear regions, so we do not plot ellipse
fits to the HST data.

\subsection{NGC 2681: Triple bar, nuclear ring\label{n2681}} 

This triple-barred galaxy is discussed by \citet{erwin99}; updated
measurements for the secondary and tertiary bar can be found in
Table~\ref{tab:features}.  The WIYN \br{} color map and unsharp
masking of the PC1 image show curved dust lanes within the secondary
bar, along the leading edges.  The innermost visible lanes appear to
form a spiral pattern just \textit{outside} the tertiary bar; the
strong diffraction pattern in the PC1 image from the (saturated)
nucleus prevents us from determining whether there are dust lanes
within the tertiary bar.  Unsharp masking at the lowest level (\ds{2})
of the NICMOS3 F160W image shows that the tertiary bar \textit{may} be
bounded by a circular, stellar nuclear ring.  Since this is at the
edge of detection, we do not list it in Table~\ref{tab:features}.

\subsection{NGC 2685: Inner disk, off-plane gas\label{n2685}} 

Figure~\ref{fig:n2685}: This the famous ``Helix'' or ``Spindle''
Galaxy, a canonical polar ring system.  The classification ---
(R)SB$0^{+}$ --- is apparently based on the combination of the outer
optical ring and the inner elliptical structure (the ``spindle''). 
However, stellar velocity curves along the spindle
\citep{schechter78,silchenko98} show strong rotation, and these
authors conclude that the spindle is an S0 \textit{disk}, seen nearly
edge-on.  If that is so, then the outer ring lies in a different
plane, though its \hi{} has the same sense of rotation as the stars
\citep{schechter78,mahon92}.  Because of the extreme uncertainty about
the galaxy's orientation, we cannot determine its inclination.  We
take the position angle for the line-of-nodes as that of the outer
ring and the spindle; since the inner disk (see below) has the same
alignment, deprojection would not change its size.

The spindle appears more elliptical in the near-IR than it is in the
optical (panel~b of the figure), as noted by \citet{peletier93}.  This
is undoubtedly because the strong polar dust lanes crossing over the
NE side of the spindle make the optical isophotes rounder.  The
NICMOS3 image shows an inner ellipticity peak, aligned with the
spindle, which also appears in the ground-based images of Peletier \&
Christodoulou.  Unsharp masks (e.g., panel~f) show an elliptical
feature with no bar-like characteristics, so we are fairly confident
this is an inner disk; \citet{ravindranath01} noted the presence of
this disk after subtracting a 2-D Nuker-law fit to the inner regions
of the NICMOS image.  The inner disk shows up in the color map as an
elliptical red feature with $a \sim 2\arcsec$, parallel with the
spindle.  There is a color gradient within the disk, with the center
about 0.2 mag redder in \br{} than the edge.

In the color maps, the spindle is dominated by the polar rings, seen
as red arcs crossing over the brightest part of the spindle and blue
outside (e.g., the blue arc near the western edge of panel~d).  These
luminous rings have been mapped in both \hi{} and CO
\citep{shane80,mahon92,watson94} and are sites of star formation
\citep[e.g., ]{eskridge97}.  Faint red and blue lanes in the region
between the spindle and the outer ring (panel~c) do not appear
connected with the polar rings.

\subsection{NGC 2787: Dust-obscured, inner disk, off-plane gas\label{n2787}} 

Figure~\ref{fig:n2787}: This galaxy is notable for both a large inner
disk and a dramatic, off-plane dust disk in the central regions. 
Although the latter obscures the inner $r \lesssim 10\arcsec$, the
inner disk is large enough to be seen outside the obscuration. 
(Unfortunately, the NICMOS3 images are severely off-center and thus
not usable.)

\hi{} mapping by \citet{shostak87} showed that the neutral gas in this
galaxy is distributed in a ring with diameter $= 6.4\arcmin$, inclined
(at either 25\arcdeg{} or 84\arcdeg) with respect to the stellar disk;
the gas distribution's line of nodes is at a position angle of $\ap
140\arcdeg$, in contrast to the position angle of 109\arcdeg{} which
we measure for the stellar disk.

The ellipse fits show peaks in ellipticity due to the outer disk, the
bar, the bar, and a large inner disk (the latter is also apparent in
\ds{40} unsharp masks, not shown here).  There are additional,
stronger ellipticity peaks at $a \ap 5\arcsec$ and (in the HST ellipse
fits) $a \ap 6.5\arcsec$; these are both due to the strong dust lanes
near the center (see below).

The ground-based \br{} color maps show disturbed, asymmetric patches
of dust.  A large lane seems to cross over the northwest side of the
disk (panel~c), and might be associated with some of the tilted \hi. 
There are fainter, fairly linear lanes in the bar--inner disk region,
none of which appear to be leading-edge lanes of the bar.  Finally,
there is a broad arc of dust surrounding the center (panel~d).

In the HST images, this near-nuclear dust is clearly resolved into a
spectacular set of concentric, elliptical dust rings, covering a
radial range of $\sim 1.5$--10\arcsec{} \citep[previously noted
by][]{sarzi01}.  The rings have a common position angles of $\ap
40\arcdeg$ and an ellipticity $\ap 0.43$ (best measured from the
innermost rings, panel~e), indicating an inclination to the line of
sight of about 55\arcdeg{} if they are intrinsically circular. 
Tracing the degree of extinction suggests a line-of-nodes of $\sim
70\arcdeg$ for the rings' intersection with the stellar disk of the
galaxy.  Careful inspection of the isophotes shows that essentially
\textit{all} of the variation in the ellipse fits for $r \lesssim
10\arcsec$ is due to extinction by this dust.

There is a central region almost free of dust, with a distinctly blue 
nucleus; this nucleus is the bluest part of the PC image (about 0.2 
magnitudes bluer in \vi{} than the region outside the dust lanes).  
This is probably the active nucleus \citep[broad \ha{} emission was 
seen by ][]{ho97d}; \citet{sarzi01} recently found evidence for a 
central black hole with mass $\sim 10^{8} \Msun$.

\subsection{NGC 2859: Double bar, nuclear ring\label{n2859}} 

Figure~\ref{fig:n2859}: This is our strongest example of a double
barred galaxy; the secondary bar is readily apparent in the isophotes
(e.g., panel~e of the figure), and was noted by \citet{kormendy79} and
\nocite{w95}W95.

The \br{} color map shows a faint leading-edge dust lane in the
eastern half of the primary bar, curving in toward the central
regions.  There, it appears to join a dusty nuclear ring which
surrounds and is aligned with the secondary bar; unsharp masking
(\ds{5}, not shown here) indicates faint spiral dust lanes just
outside of the nuclear ring and merging with it, on both the north and
south sides.  The secondary bar shows up as an elongated red feature
in the color maps, but is clearly asymmetric in color.  This suggests
that the color difference is caused by dust in the secondary bar,
rather than by a different stellar population.

NGC~2859 also has a very clear example of a detached outer ring, which
makes determining the galaxy's inclination and PA difficult.  The
\hi{} profile width of $W_{50} \ap 170$ \kms{} \citep{wk86,burstein87}
permits $10\arcdeg < i < 58\arcdeg$ for $100 < V_{circ} < 500$ \kms. 
Since the maximum rotation velocity in \nocite{kormendy82}Kormendy's
(1982) optical spectrum at PA = 90\arcdeg{} (86 \kms) is very close to
half of $W_{50}$, the galaxy's line of nodes is probably not far from
90\arcdeg.  From here, we can use the shapes of the outer ring and
lens to further constrain the inclination.  An intrinsically round
outer ring would give $i = 40\arcdeg$; however, if it has a typical
axis ratio of 0.8 and is perpendicular to the bar, then $i =
15\arcdeg$.  The lens, outlined by a dusty inner ring in the \br{}
color map (panel~d), is seen with $e \ap 0.10$ at $a \ap 50\arcsec$,
elongated parallel with the bar; this implies $i < 39\arcdeg$, for an
upper limit of 0.7 on the lens's intrinsic axis ratio, and $i =
27\arcdeg$ for a typical lens axis ratio of 0.8.  We adopt $i =
25\arcdeg$, and PA = 90\arcdeg{} for the line-of-nodes.

\subsection{NGC 2880: Inner disk\label{n2880}} 

Figure~\ref{fig:n2880}: This relatively featureless galaxy was the
subject of extensive long-slit spectral analysis by \citet{munn92},
who found a moderate metallicity gradient and no evidence for emission
lines.

The weak bar shows up mainly as a pronounced twist in position angle;
the position angle we measure, based on the isophotes and unsharp
masks, differs by more than 20\arcdeg{} from that given by the ellipse
fits.  Unsharp masking (panel~d) shows evidence for a partial ring
just outside the bar.  Interior to the bar is a small ellipticity peak
with a position angle very close to the outer disk; this is presumably
the twist noted by \citet{elmegreen96} in a $B$-band atlas image. 
Because its position angle differs by only 6\arcdeg{} from that of the
outer disk, we consider this an inner disk; given the low amplitude of
the ellipticity peak, it is the weakest case for an inner disk in our
sample.  The color map shows a slight inward reddening trend which
appears to follow the isophotes; this may be associated with the
metallicity gradient.

The WIYN $R$-band ellipse fits also show a strong \textit{outer}
ellipticity peak at $a \sim 45\arcsec$; unsharp masking with high
smoothing (\ds{40}) shows barely perceptible hints of a bar signature
at this radius.  While it is tempting to suggest that this is a
(rather round) bar --- and thus that this galaxy is double-barred ---
the ellipticity peak could instead represent the outer disk, with
rounder isophotes outside due to a luminous halo
\citep[e.g.,][]{michard93}.  Lacking a clear indication of a bar, we
adopt the outer disk interpretation; deeper images are clearly
desirable.

\subsection{NGC 2950: Double bar, nuclear ring\label{n2950}} 

Figure~\ref{fig:n2950}: This is the second most strongly double-barred
galaxy in the sample, with the secondary bar visibly distorting the
isophotes.  Although \citet{kormendy82} called this the ``prototypical
SB0 galaxy with a triaxial bulge,'' we agree with \nocite{w95}W95 that
the distortion is in fact a bar.

The ellipse fits yield a PA for the outer disk of 120\arcdeg, or
25\arcdeg{} less than given by \nocite{rc3}RC3.  There is a shoulder
in the position angles (PA $\ap 135\arcdeg$) corresponding to the
lens, and very clear features due to the primary and secondary bars. 
Figure~\ref{fig:umask} shows unsharp masks of this galaxy,
highlighting both bars; Figure~\ref{fig:umaskpa} shows how the PA for
the secondary bar derived from the disky isophotes and the unsharp
mask differs from that of the ellipse fits.  \nocite{w95}W95 called
attention to a weak ellipticity peak between the primary and secondary
bars at $a \sim 10\arcsec$.  This peak occurs at a position angle
close to that of the outer disk, coincident with a weak $B_{4}$ peak
which suggests disky isophotes, so it might indicate a faint inner
disk rather than a triaxial bulge.

Unsharp masking of the WFPC2 image (panel~f) shows the secondary bar
quite clearly; the parallel straight ends of the bar seem to suggest
that this bar, at least, has the square-ended isophotes characteristic
of large-scale bars in early-type galaxies \citep[as in the outer bar
of this galaxy; see also][]{athan90}.  The unsharp masks also show a
faint nuclear ring just outside the inner bar; this ring is smooth and
does not appear in any of the color maps.  It is a weaker feature than
either of the other two stellar nuclear rings (NGC~936 and NGC~3945),
and thus rather hard to measure; the values in
Table~\ref{tab:features} are crude estimates.  The orientation and
observed ellipticity \textit{are} consistent with a circular ring in
the galaxy plane, though we cannot rule out a moderately elliptical
ring; in addition, unsharp masking with low smoothing (\ds{5}, not
shown) hints at a possible diamond-shaped, four-armed structure, which
argues against a perfectly circular ring.  The HST color maps
(panel~e) show only a smooth inward reddening trend.  Unsharp masking
does seem to show a couple of faint, very thin dust lanes inside the
nuclear ring; the stronger of these matches a slight asymmetry in the
color maps (about 2\arcsec{} north-northeast of the nucleus in 
panels~e and f).

\subsection{NGC 2962: Double bar\label{n2962}} 

Figure~\ref{fig:n2962}: The ellipse fits to the WFPC2 F814W image of
this galaxy show a strong ellipticity peak at $a \sim 0.7\arcsec$,
which could be mistaken for a secondary bar.  Careful inspection of
the unsharp masks (of both WIYN and HST images) reveals that this peak
is almost certainly due to strong circumnuclear dust lanes (see
panel~e).  The WIYN and HST isophotes and unsharp masks \textit{do}
reveal the existence of a genuine secondary bar, whose presence in the
ellipse fits is marked by the \textit{minimum} at $a \approx
3.4\arcsec$, and by a strong positive $B_{4}$ signature peaking at $a
\approx 3.6\arcsec$; we use the mean of these two measurements for the
secondary bar's ``maximum ellipticity'' radius.  The secondary bar
lies at a position angle of $\ap 90\arcdeg$ --- approximately
80\arcdeg{} different from the best-fitting ellipse!  We suspect this
difference results from an oblate central bulge --- or possibly an
inner disk --- which dominates the inner light of the galaxy.  The
isophotes just outside the secondary bar in panel~f are clearly
elongated north-south, parallel with the outer disk, as would be
expected from such a bulge.  Since the distortions introduced by the
secondary bar are almost at right angles to the isophotes outside,
they force the fitted ellipses to be rounder --- and introduce some
slight uncertainty into their PA --- but do not substantially change
their orientation.

The color maps show dust lanes both outside and inside the primary
bar.  The inner lanes visible in the WIYN images (panel~c) are almost
all in the south half of the primary bar, which is somewhat odd given
that the disk major axis runs north-south.  This
\textit{might}indicate off-plane gas in the central regions, but,
lacking any other evidence, we do not consider this galaxy to be a
polar ring candidate.  (Comparison of the major-axis stellar velocity
curve in \nocite{simien00}Simien \& Prugniel 2000 and the Arecibo
\hi{} mapping by \nocite{burstein87}Burstein et al.\ 1987 indicates
that the outer \hi{} and the stars are corotating.)  The secondary bar
appears to show up in the color map as an asymmetric red feature
elongated at PA $\ap 90\arcdeg$; unsharp masking of the WFPC2 image
shows numerous strong dust lanes inside the secondary bar, mostly on
the north side, and what may be a very narrow dust ring (semi-major
axis $\ap 0.4\arcsec$, ellipticity $\ap 0.7$, PA $\ap 110\arcdeg$).

\subsection{NGC 3032: Nuclear spiral\label{n3032}} 

Figure~\ref{fig:n3032}: It is difficult to identify even one bar in
this galaxy.  The weak ellipticity peak at $r \ap 8\arcsec$ appears to
be the bar referred to by \citet{deV57}; in the \br{} color map, it is
surrounded by a strong dust ring.  While this ring might be considered
an inner ring surrounding a bar, it is unusually asymmetric for such a
ring, and the ``bar'' may simply be a side effect of dust extinction.

In the NICMOS2 image, the ellipticity variations appear to be due to
spiral arms and dust lanes.  \citet{regan99} classified the nuclear
regions as ``spiral,'' using a colormap constructed from WFPC2 F606W
and NICMOS2 F160W images.  Our unsharp mask of the F606W image
(panel~d) shows that the entire region inside the \br{} dust ring is
indeed a nuclear spiral, with luminous patches and arms in addition to
dust lanes; the brightest regions in our unsharp mask generally match
the bluest regions in Regan \& Mulchaey's color map.

\citet{schweizer88} found interleaved ``ripples'' throughout this
galaxy; these are faintly visible in our unsharp masks, suggesting a
past merger.  The innermost ``ripple'' (8\arcsec{} W of the nucleus)
is blue and may be part of the nuclear spiral.  There is no indication
of off-plane gas or dust; the \hi{} profile width $W_{50} = 161$
\kms{} \citep{duprie96} is consistent with the galaxy's low
inclination.

\subsection{NGC 3185: Dust-obscured\label{n3185}} 

Figure~\ref{fig:n3185}: This is an actively star-forming galaxy in a
Hickson Compact Group \citep[][galaxy 44c]{hickson89}, with abundant
molecular gas \citep{leon98}.

Unsharp masking shows that the bar of this galaxy is surrounded by an
inner ring composed of two tightly wrapped spiral arms, coinciding
with the ring of \hii{} regions noted by \citet{gonzalez97}.  The
color map also shows dust lanes on the inner edges of the ring,
leading-edge lanes in the bar (strongest in the SE half of the bar),
and a possible dusty nuclear ring.  The nuclear regions are asymmetric
and quite blue, perhaps indicating star formation; Gonz\'{a}lez
et al.\ found extended \ha{} in the nuclear region.

The ellipse fits show considerable position angle changes within the
bar; this is almost certainly due to the strong dust lanes.  The
innermost ellipticity peak is suggestive of a secondary bar, but the
strong nuclear dust lanes make it impossible to be certain without
infrared observations.

\subsection{NGC 3412: Inner disk\label{n3412}} 

Figure~\ref{fig:n3412}: The bar manifests itself as a strong position
angle twist and a weak ellipticity peak.  Much more pronounced is an
inner ellipticity peak with the same position angle as the outer disk:
one of the stronger inner disks in our sample.  (The twisting in
position angle between the bar and the inner disk was noted by
\nocite{shaw95}Shaw et al.\ 1995 in the IR.) Unsharp masking shows a
smooth elliptical feature, with no signs of bar-like structure.

The \br{} color map is almost featureless.  \citet{fisher96} noted a
high central H$\beta$ absorption-line strength with a declining
outward gradient in their optical spectra, suggestive of a younger
stellar population.  Though this would seem to agree with the blue
nucleus and extended blue structure reported in the near-IR by
\citet{shaw95}, we can find no sign of either in our color map. 
Instead, there is a suggestion of a red asymmetry in the region of the
inner disk, possibly a sign of dust.

The peak in ellipticity at $a \sim 55\arcsec$ is apparently due to a
large lens component \citep[noted by][]{kormendy79}; we measure a
lower ellipticity of $\ap 0.37$ further out, at the limits of reliable
isophote fits, and base the disk inclination estimate on that.

\subsection{NGC 3489: Dust-obscured, nuclear ring\label{n3489}} 

Figure~\ref{fig:n3489}: Although the dust lanes in this galaxy are
complex and asymmetric, \citet{bertola95} reported similar kinematics
for both ionized gas and stars, so the dusty gas is probably in the
disk.  Dust lanes distort the inner isophotes, making the bar
difficult to distinguish; we rely on the isophotes and the unsharp
mask to derive its position angle.  In the ellipse fits, it shows up
most clearly as the sharp twist in position angle between $a \ap
10\arcsec$ and 20\arcsec.  Unsharp masking shows that the strong
ellipticity peak outside the bar is due to an outer ring (see
panel~e).

The color map of panel~f shows a dusty nuclear ring, with a position
angle similar to that of the outer disk.  The interior of the nuclear
ring is rather blue, smooth and uniform , with no signs of a nuclear
spiral or other dust features.  The strong dust lanes so close to the
nucleus could be masking inner features, but the innermost ellipticity
peak appears to be due to the dust lanes of the nuclear ring.

\subsection{IC 676 (Mrk 731): Dust-obscured\label{ic676}} 

Figure~\ref{fig:ic676}: This is a ``double-nucleus'' Markarian galaxy,
with an extremely narrow bar and peculiar dust lanes.  Both of the two
bright, blue ``nuclei'' are emission-line sources
\citep{nordgren95,contini98}.  The CO spectra of \citet{contini97} ---
$W_{50} = 100$ \kms, $W_{20} = 128$ \kms --- are consistent with the
low inclination that we derive from the outer isophotes.

The bar is bordered by red regions in the color map; there is a
suggestion of two dust lanes, parallel with the bar.  The stellar
spiral arms leading off the ends of the bar (best seen in unsharp
masks) suggest clockwise rotation, though this would put the straight
dust lanes on the trailing edges of the bar.  The dust lanes are
joined by a narrow lane which crosses the bar between the two blue
nuclei.  Several other blue regions strung along the bar are
symmetrically spaced with respect to the central dust lane.  The 
spiral arms suggest that the upper limit to the bar's radius is closer 
to 17\arcsec{} or 18\arcsec{} than the \amin{} or \aten{} values from 
the ellipse fits.

\subsection{NGC 3729: Dust-obscured\label{n3729}} 

\begin{figure}
\begin{center}
	\includegraphics[scale=0.85]{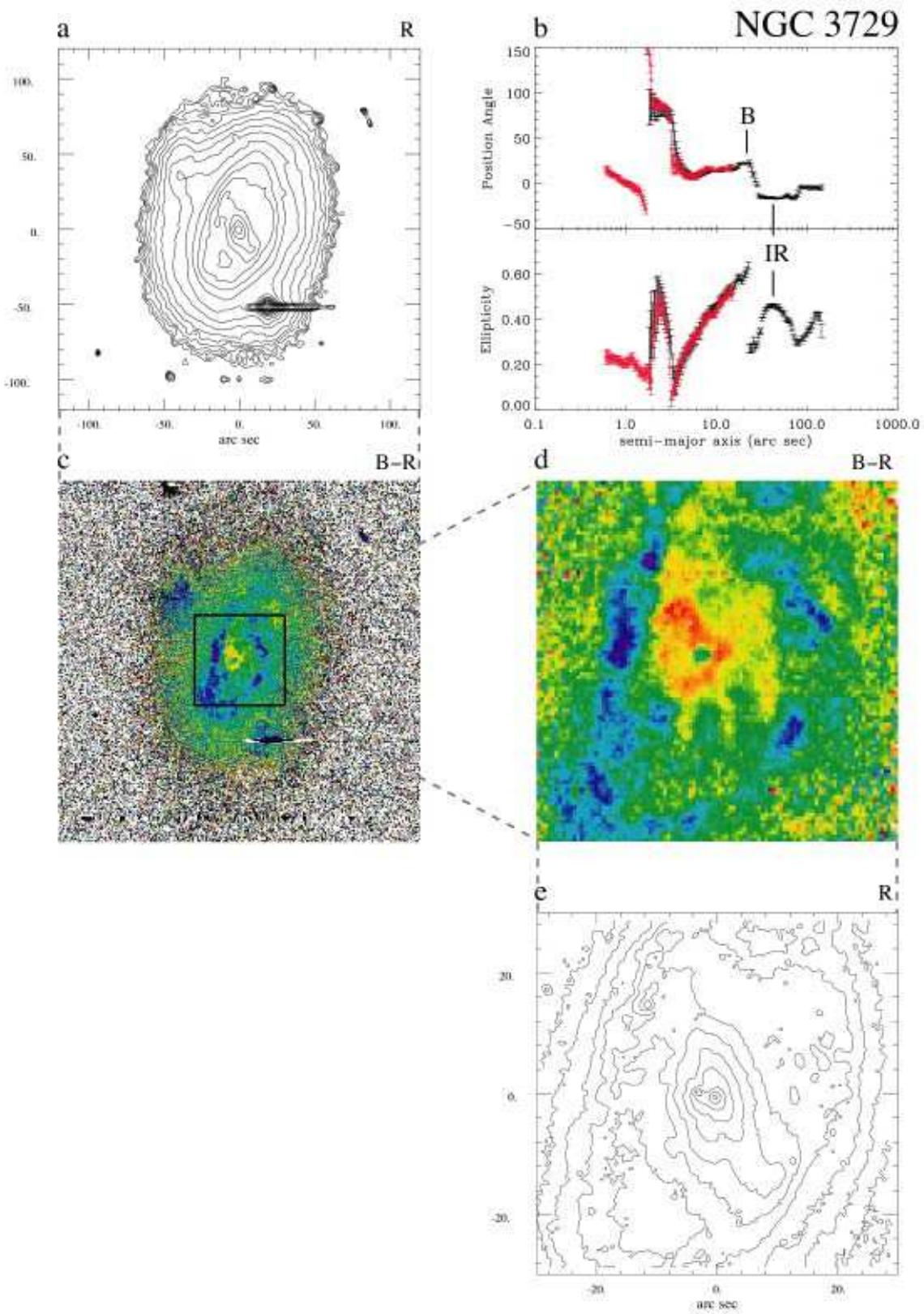}
\end{center}
\caption{NGC 3729: large-scale Mauna Kea $R$-band contours (a);
ellipse fits (b, Mauna Kea $R$ and WIYN $R$ [red]); \br{} color maps
(c, no smoothing, color range $\approx 1.1$ mag; d, no smoothing,
color $\ap 0.8$ mag); and a close-up of the bar region (e, $R$-band
contours from WIYN image).\label{fig:n3729}}
\end{figure}

\begin{figure}
\begin{center}
	\includegraphics[scale=0.85]{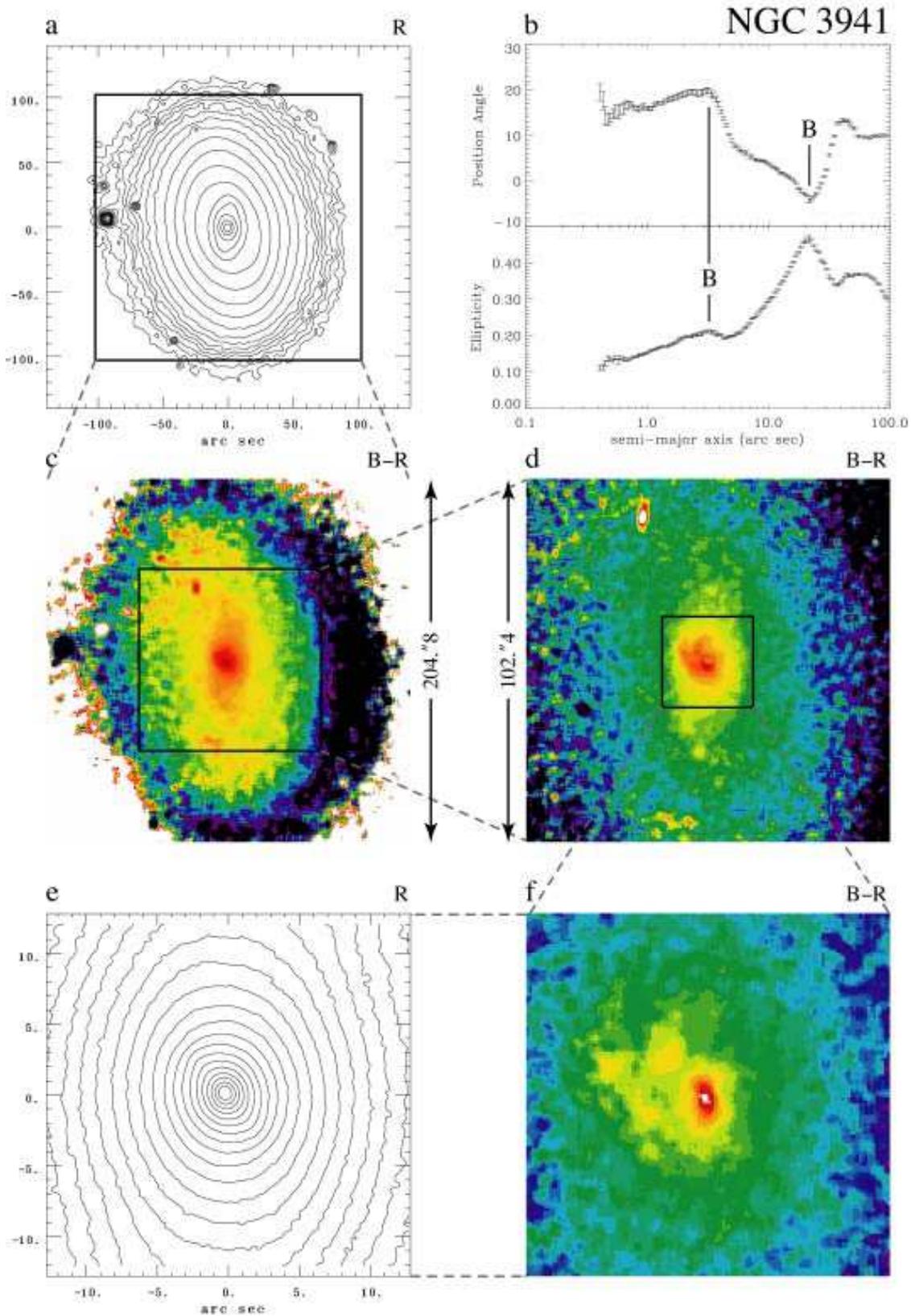}
\end{center}
\caption{NGC 3941: large-scale $R$-band isophotes (a); ellipse fits
(b, WIYN $R$); \br{} color maps at two scales (c, $w = 23$, color
range = 0.51 mag; d, $w = 11$, color range = 0.39 mag); small-scale
isophotes (e); and \br{} color map at the same scale (f, $w = 5$, 
color
range = 0.51 mag).  The apparent blue nucleus in (d) is an artifact of
saturation in the $R$-band image; the shorter-exposure $R$-band image
used in (f) was not saturated.
\label{fig:n3941}}
\end{figure}

\begin{figure}
\begin{center}
	\includegraphics[scale=0.85]{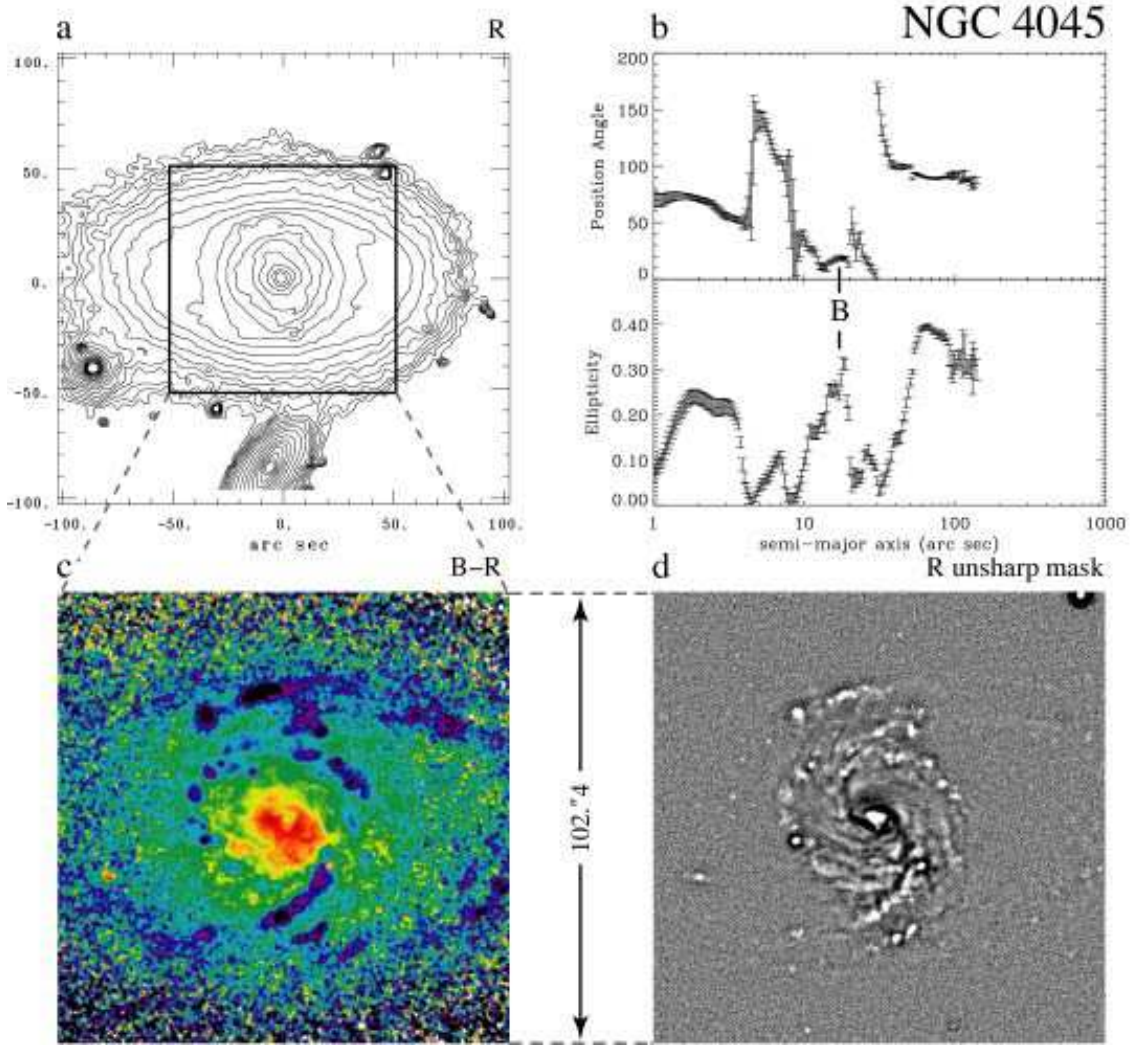}
\end{center}
\caption{NGC 4045: $R$-band isophotes (a); ellipse fits (b, WIYN
R); \br{} color map (c, $w = 5$, color range = 1.13 mag); and \ds{5}
unsharp mask of the $R$-band image, showing dust lanes and blue spiral
arms.  The disk galaxy immediately to the south, NGC 4045A, is at a
redshift of 5040 \kms{} and so is not physically associated.
\label{fig:n4045}}
\end{figure}

\begin{figure}
\begin{center}
	\includegraphics[scale=0.85]{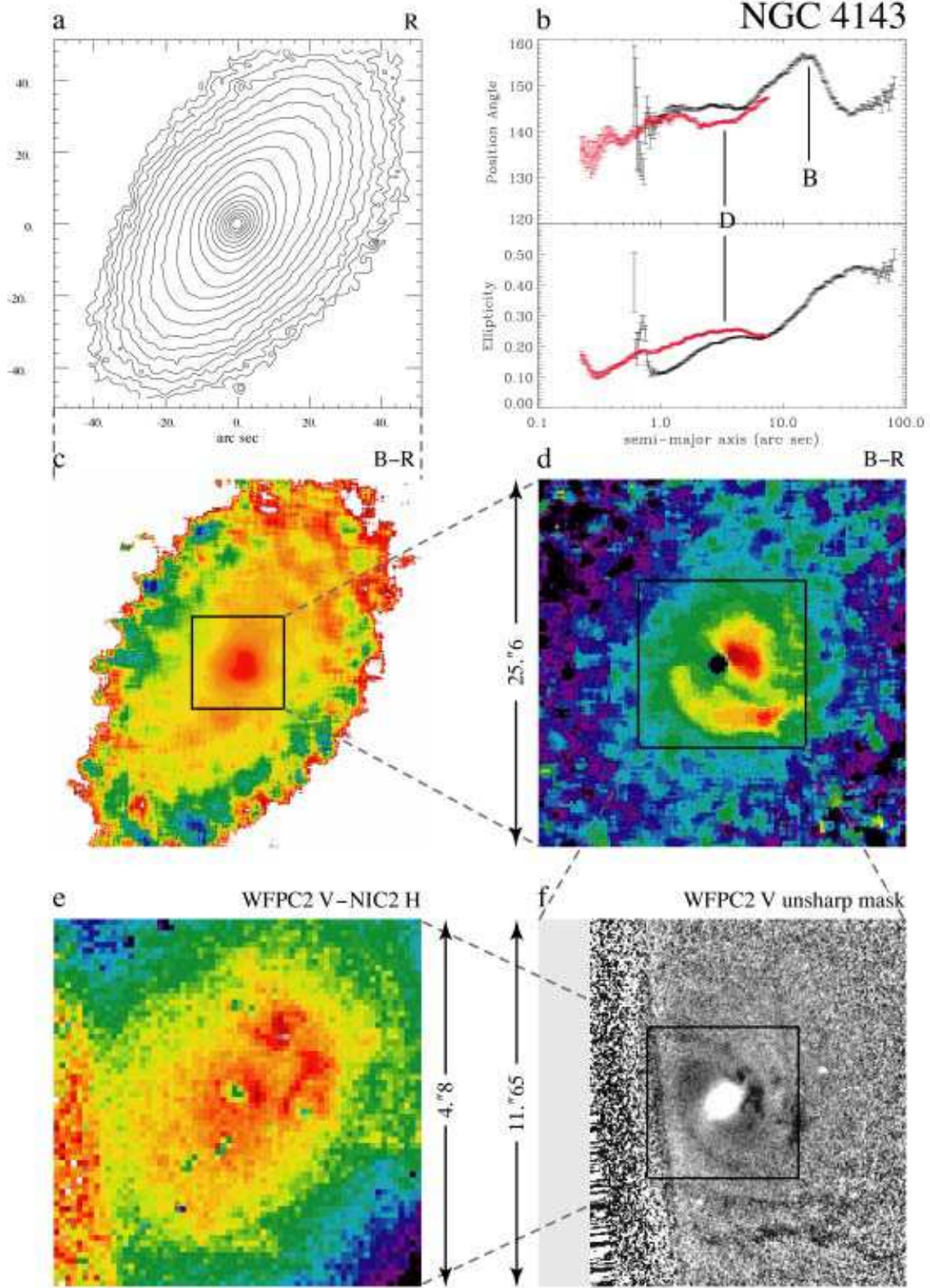}
\end{center}
\caption{NGC 4143: large-scale $R$-band isophotes (a); ellipse fits
(b, WIYN $R$ and NICMOS2 F160W [red]); \br{} color map on the same
scale as panel~a (c, $w = 23$, color range within the disk = 0.44
mag); small-scale \br{} color map (d, $w = 5$, color range = 0.31 mag,
excluding the saturated center); WFPC2 + NICMOS2 F606W$\,-\,$F160W
color map (e, color range in central region = 0.34 mag); and \ds{10}
unsharp mask of the WFPC2 F606W image (f).  Note that the galaxy
nucleus is near the edge of the PC2 chip, which affects both (e) and
(f).  The blue nucleus in (d) is an artifact of a saturated $R$-band
image.
\label{fig:n4143}}
\end{figure}

\begin{figure}
\begin{center}
	\includegraphics[scale=0.85]{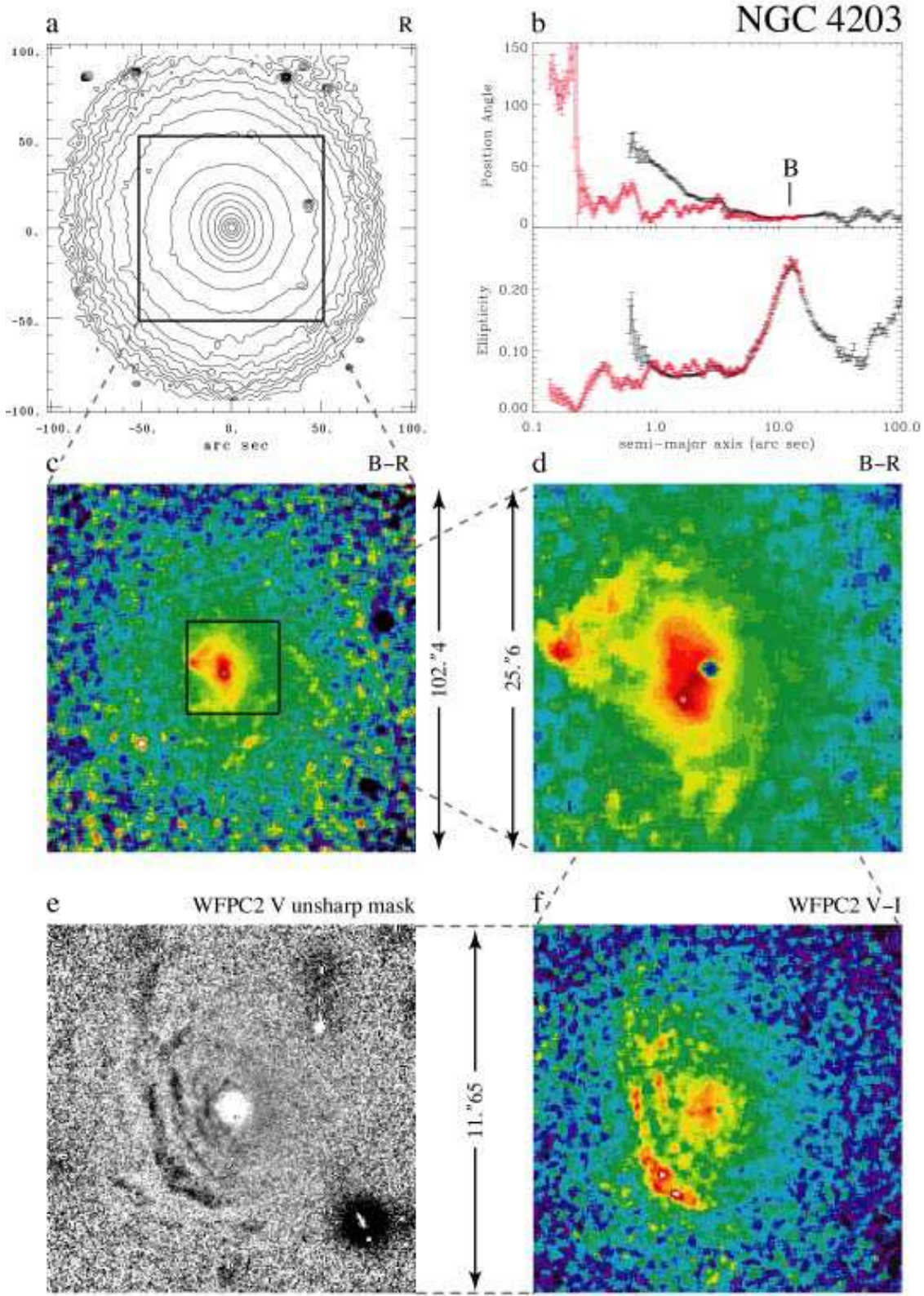}
\end{center}
\caption{NGC 4203: large-scale $R$-band isophotes (a); ellipse fits
(b, WIYN $R$ and WFPC2 F814W [red]); \br{} color map on two scales (c,
$w = 11$, color range = 0.76 mag; d, $w = 5$, color range = 0.32 mag);
\ds{10} unsharp mask of the WFPC2 F555W image (e); and WFPC2
F555W$\,-\,$F814W color map (f, $w = 5$, color range = 1.34 mag). 
Note that the blue nucleus in (d) is an artifact of saturation in the
$R$-band image; the blue nucleus in the WFPC2 color map (f) is real.
\label{fig:n4203}}
\end{figure}

\begin{figure}
\begin{center}
	\includegraphics[scale=0.85]{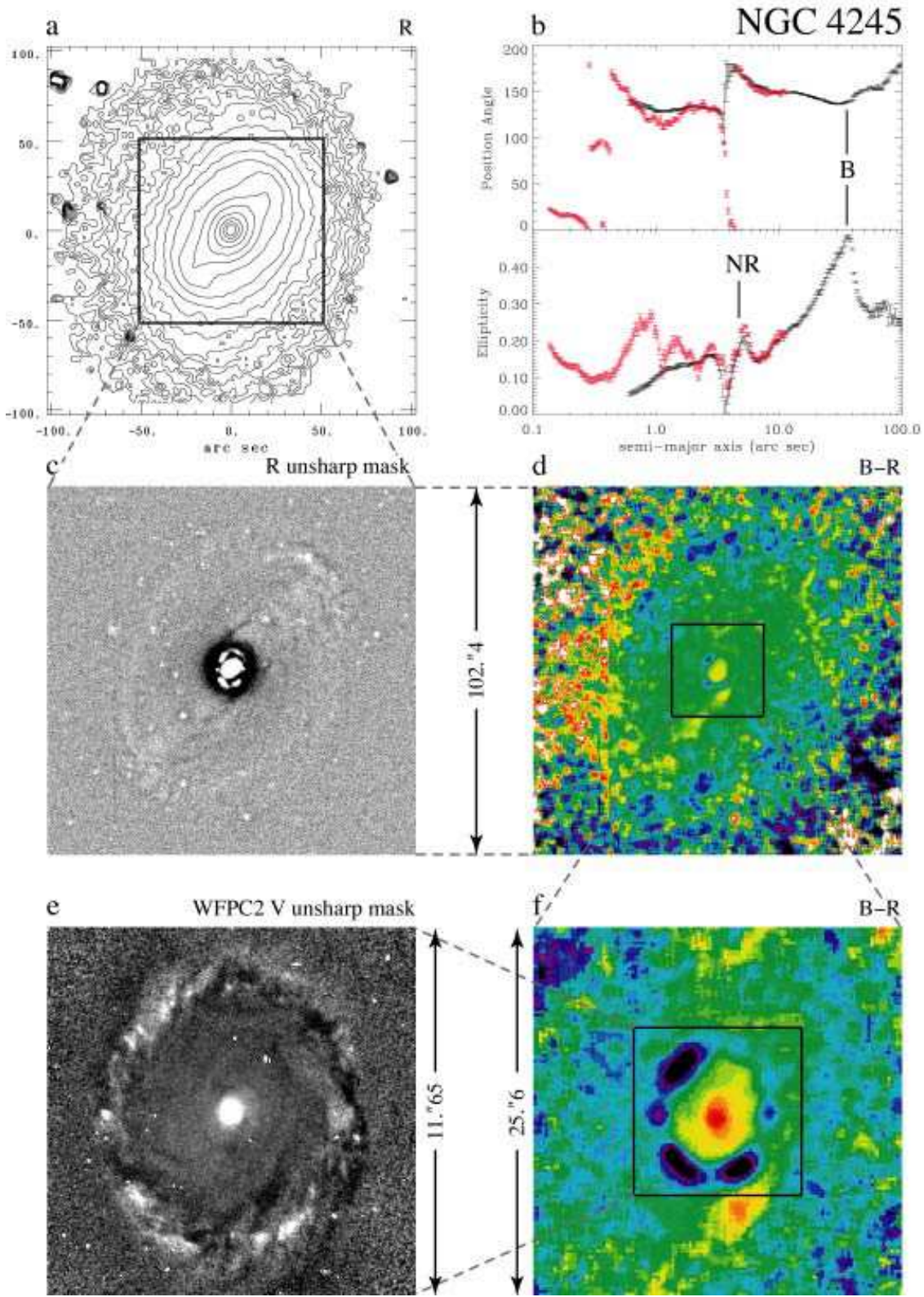}
\end{center}
\caption{NGC 4245: $R$-band isophotes (a); ellipse fits (b, WIYN
$R$ and WFPC2 F606W [red]); unsharp mask (\ds{10}) of $R$-band image
(c); \br{} color map on two scales (d, $w = 11$, color range = 1.26
mag; f, $w = 5$, color range = 0.53 mag); \ds{10} unsharp mask of the
WFPC2 F606W image (e).
\label{fig:n4245}}
\end{figure}

\begin{figure}
\begin{center}
	\includegraphics[scale=0.85]{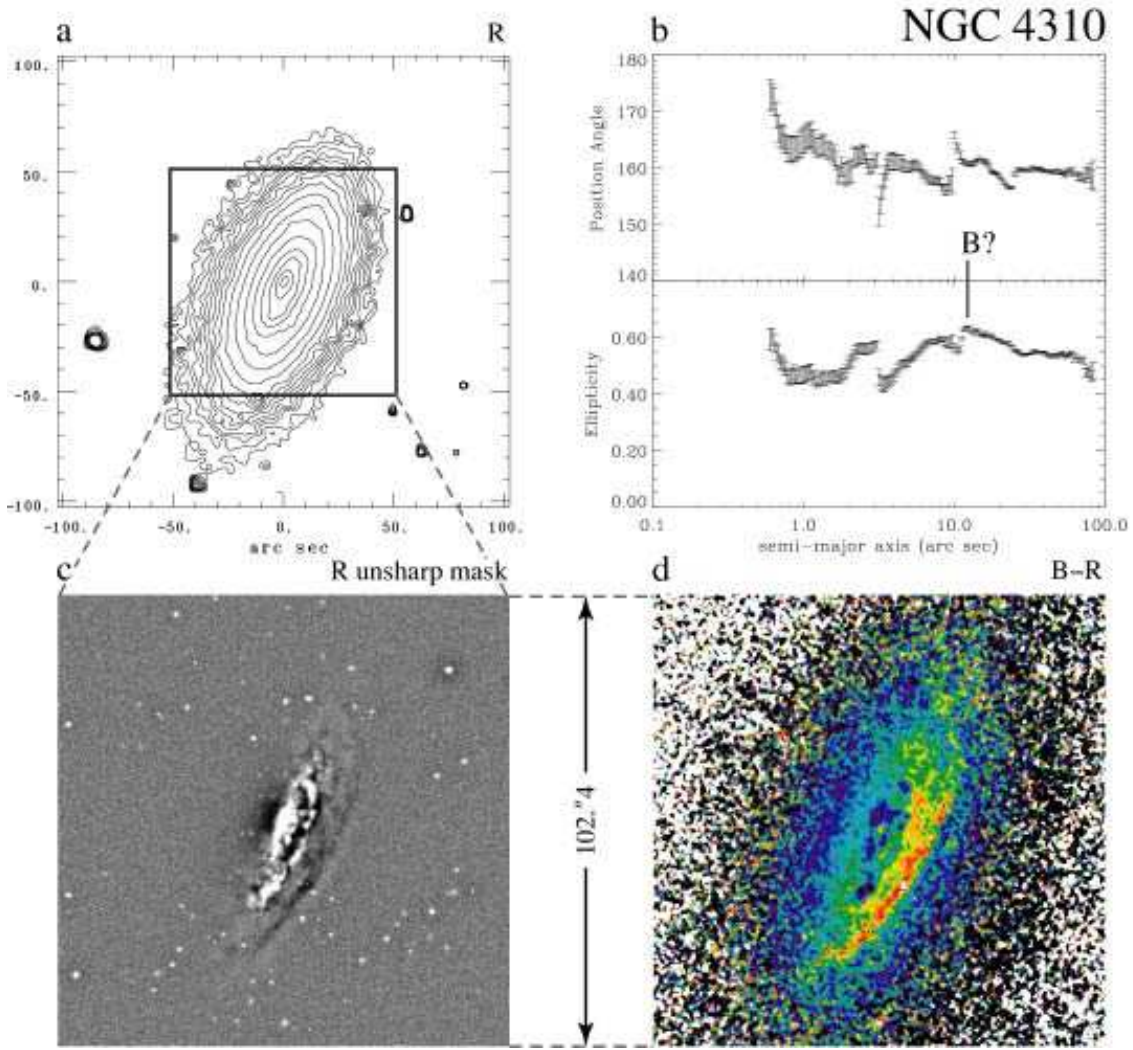}
\end{center}
\caption{NGC 4310: $R$-band isophotes (a); ellipse fits (b, WIYN
$R$); unsharp mask of $R$-band image (c, \ds{10}); \br{} color map (d,
$w = 5$, color range $\ap 1.0$ mag).
\label{fig:n4310}}
\end{figure}

\begin{figure}
\begin{center}
	\includegraphics[scale=0.85]{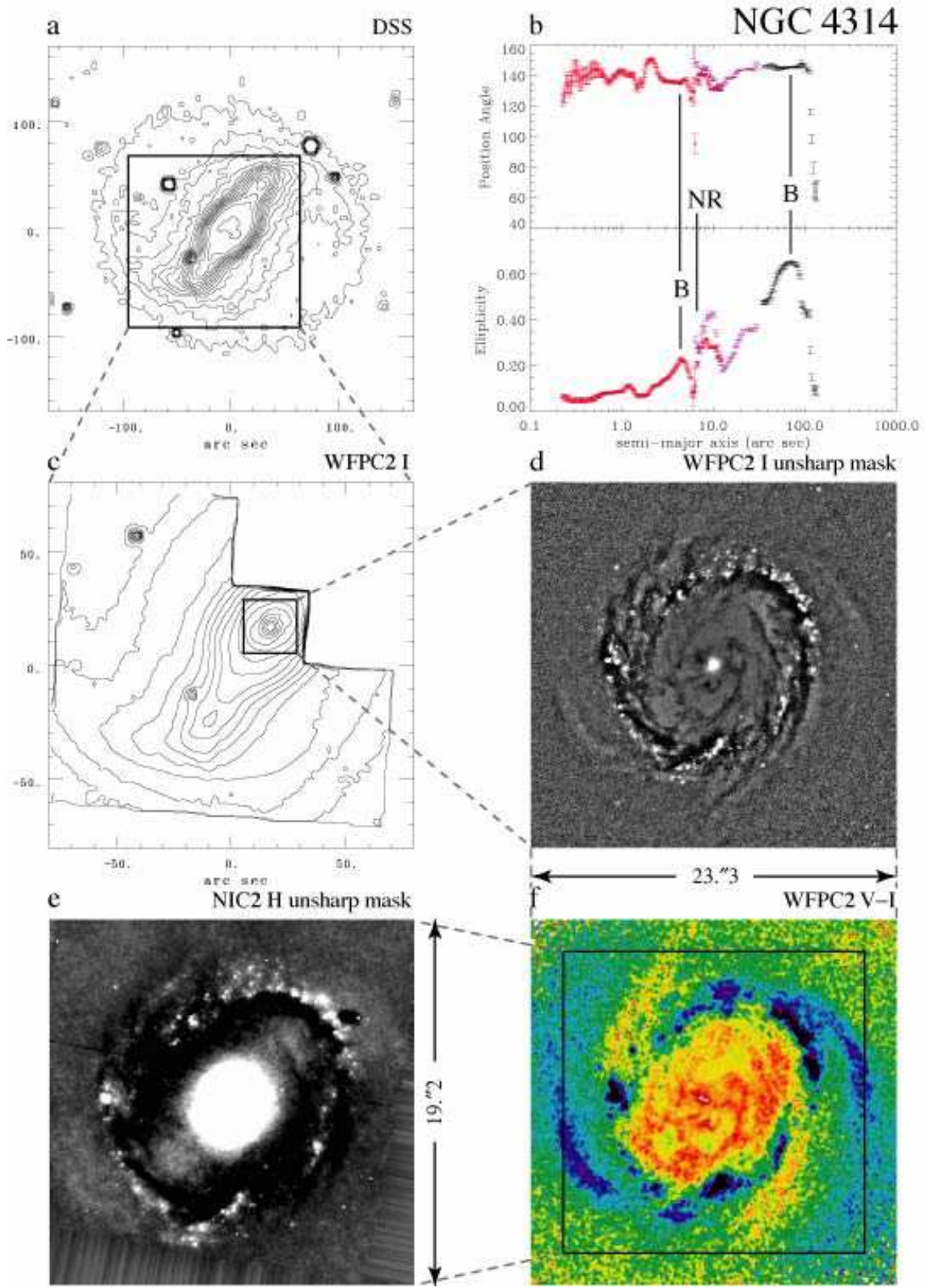}
\end{center}
\caption{NGC 4314: DSS contours (a); ellipse fits (b, DSS [black],
WFPC2 F814W mosaic [purple], and NICMOS2 F160W [red]); WFPC2 F814W
mosaic contours (c); \ds{} unsharp mask of PC2 F814W image (d);
\ds{30} unsharp mask of NICMOS2 F160W image; PC2 F439W$\,-\,$F814W
color map (f, $w = 5$, color range = 2.06 mag).\label{fig:n4314}}
\end{figure}

\begin{figure}
\begin{center}
	\includegraphics[scale=0.85]{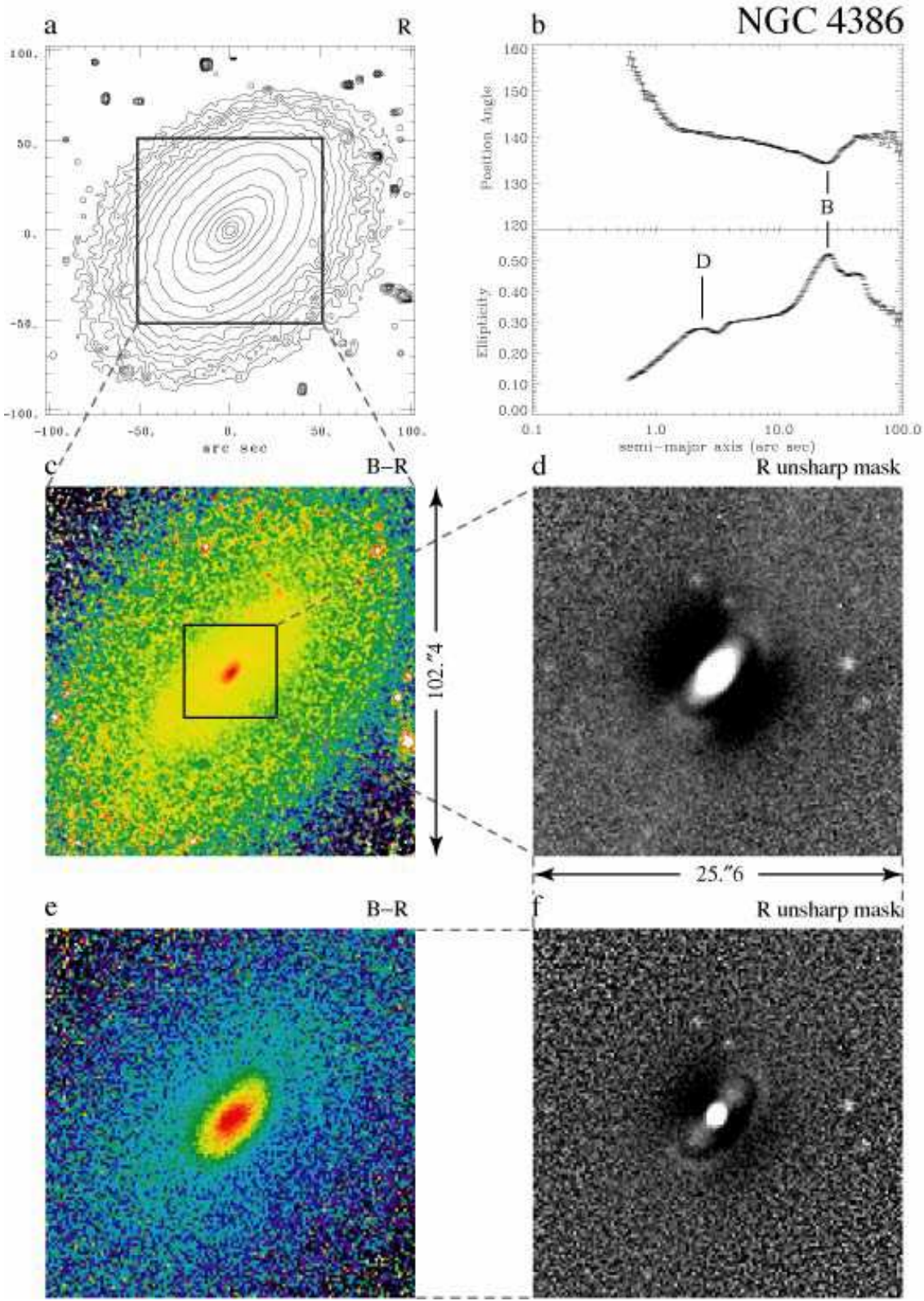}
\end{center}
\caption{NGC 4386: $R$-band isophotes (a); ellipse fits (b, WIYN
$R$); \br{} color maps (c, $w = 5$, color range $\ap 2.25$ mag between
outer disk and central red disk; e, unsmoothed, color range $\ap 0.71$
mag); unsharp masks of $R$-band image (d, \ds{5}; f, \ds{2}).
\label{fig:n4386}}
\end{figure}

\begin{figure}
\begin{center}
	\includegraphics[scale=0.85]{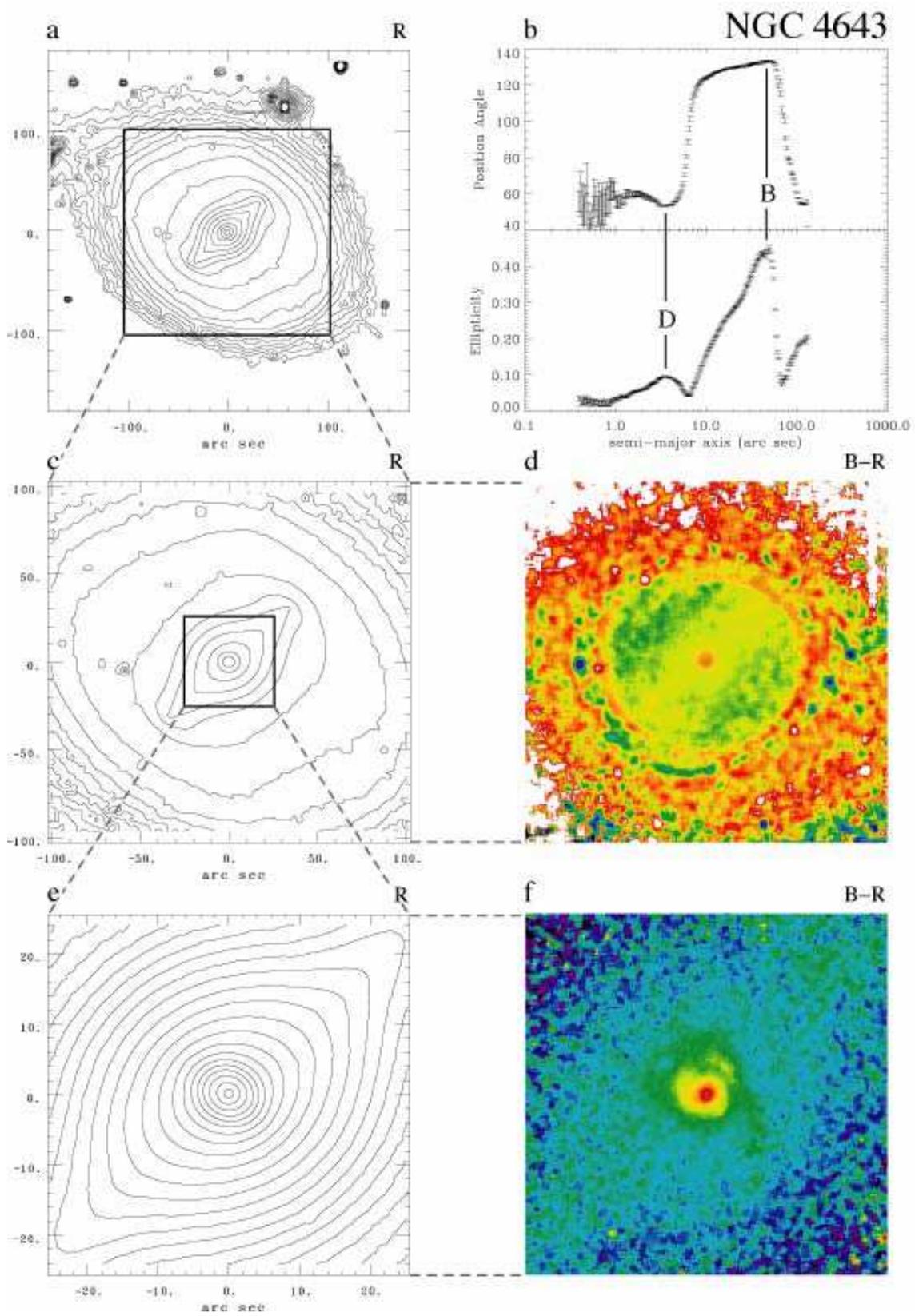}
\end{center}
\caption{NGC 4643: $R$-band isophotes on three scales (a, c, e);
ellipse fits (b, WIYN $R$); \br{} color map on two scales (d, $w =
23$, color range $\ap 0.9$ mag; f, $w = 5$, color range = 0.30 mag).
\label{fig:n4643}}
\end{figure}

\begin{figure}
\begin{center}
	\includegraphics[scale=0.85]{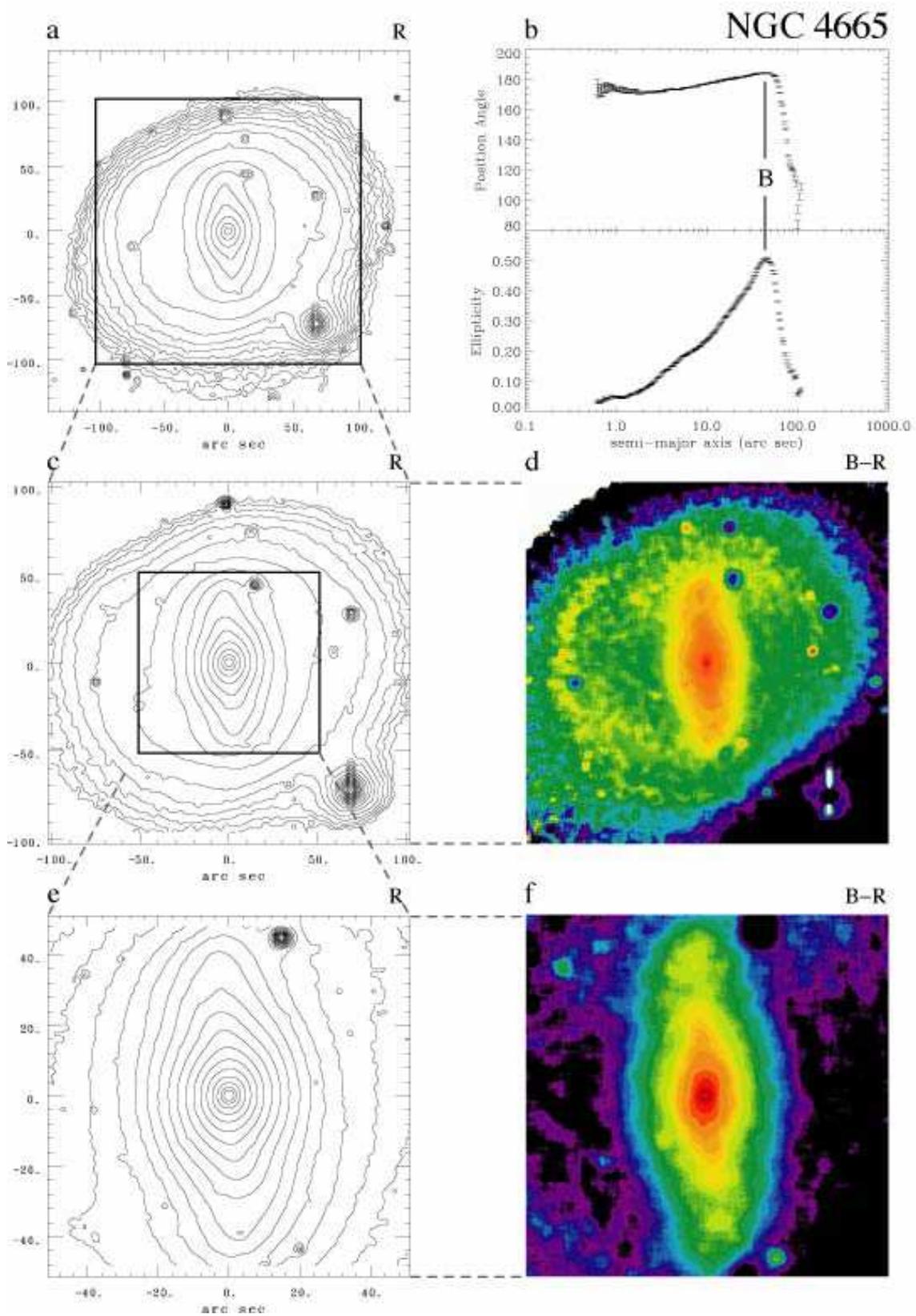}
\end{center}
\caption{NGC 4665: $R$-band isophotes on three scales (a, c, e);
ellipse fits (b, WIYN $R$); \br{} color map on two scales (d, $w =
23$, color range $\ap 0.45$ mag between outer disk [green] and center;
f, same smoothing width, color range = 0.48 mag).
\label{fig:n4665}}
\end{figure}

\begin{figure}
\begin{center}
	\includegraphics[scale=0.85]{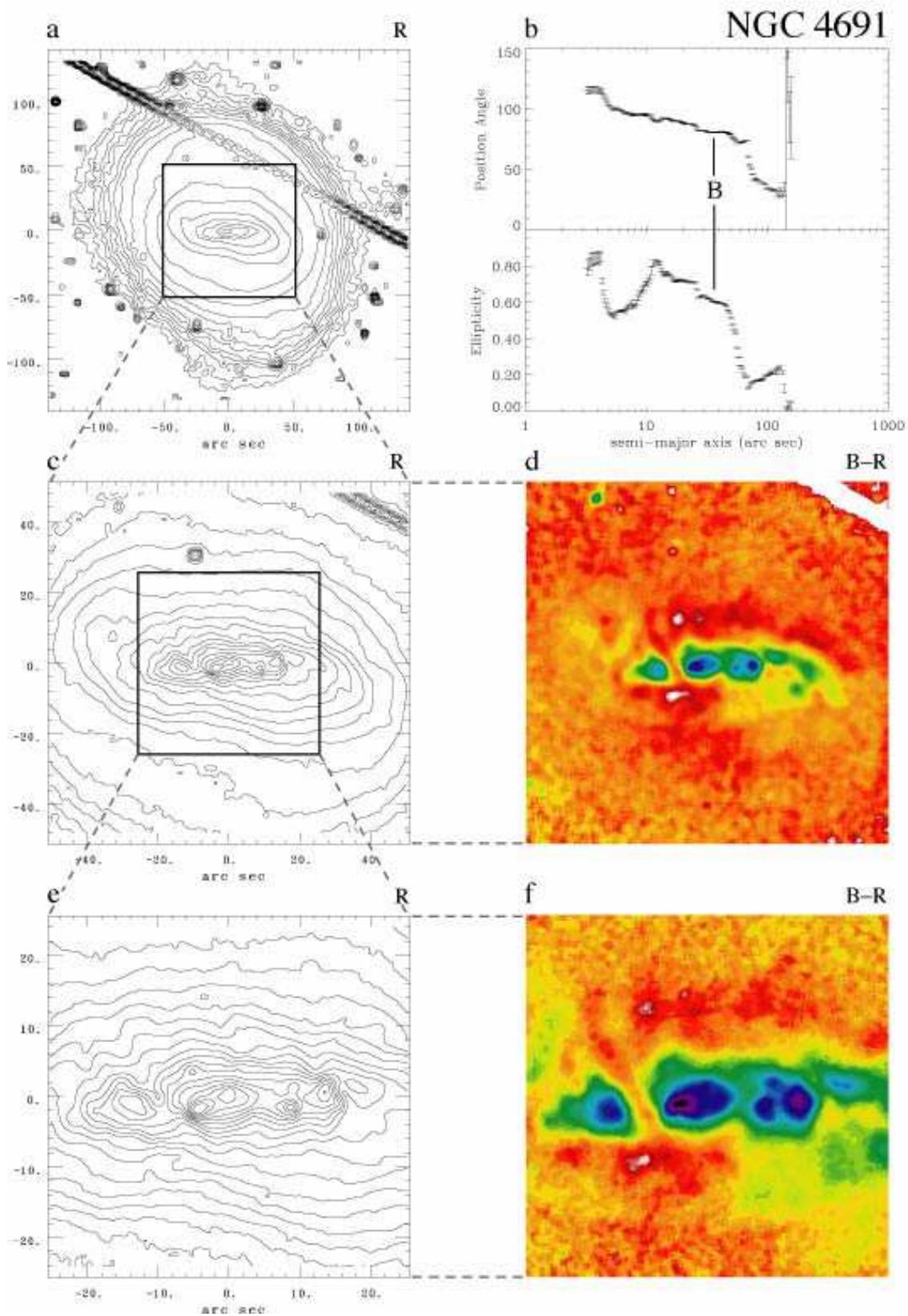}
\end{center}
\caption{NGC 4691: $R$-band isophotes on three scales (a, c, e);
ellipse fits (b, WIYN $R$); \br{} color map on two scales (d, $w =
11$, color range $\ap 0.85$ mag [excluding the white streak in the NW
corner]; f, $w = 5$, color range = 1.02 mag).  The bright streak
crossing the northern part of the galaxy (panels~a, c, and d) is a
satellite or meteor trail.
\label{fig:n4691}}
\end{figure}

\begin{figure}
\begin{center}
	\includegraphics[scale=0.85]{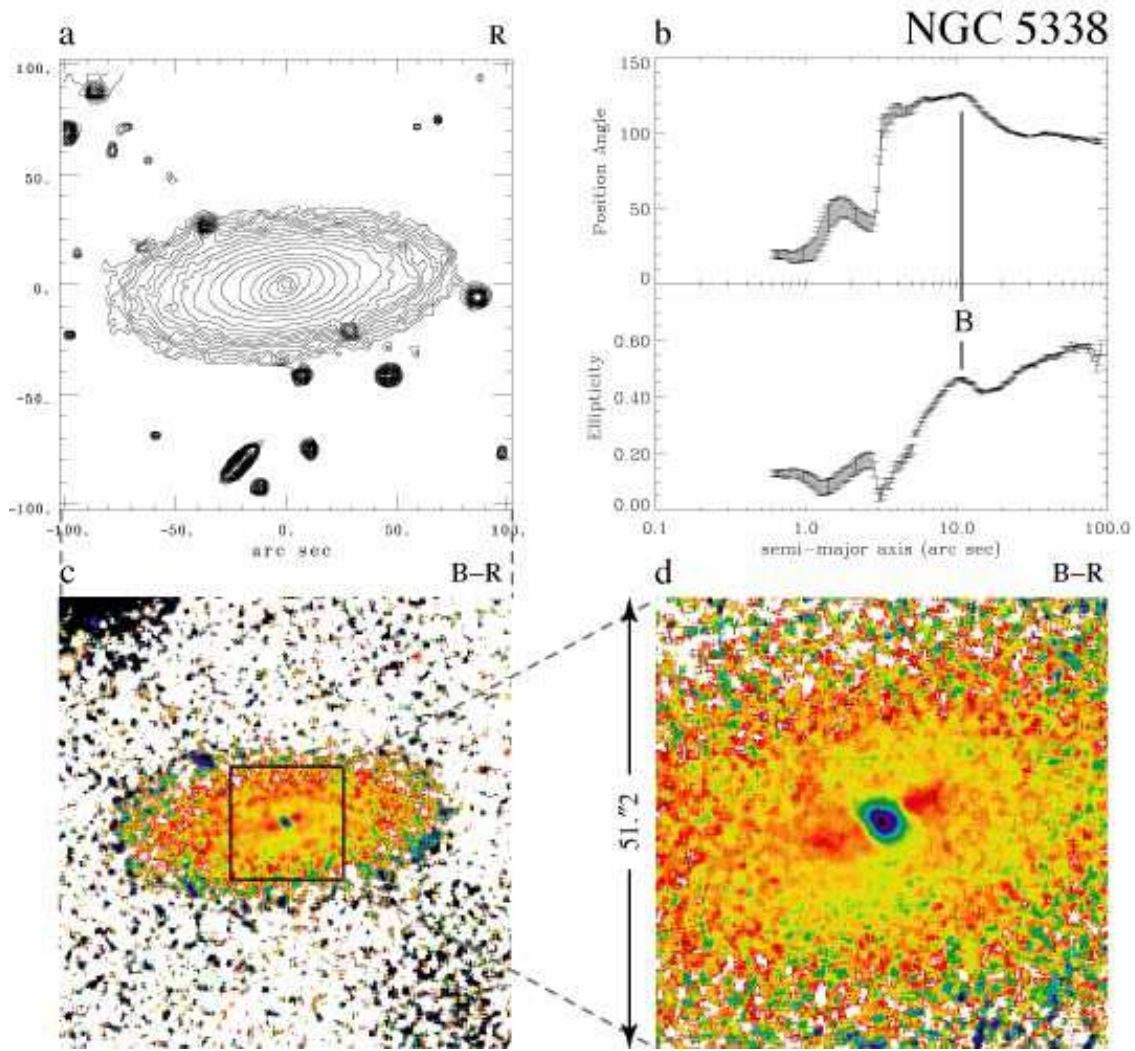}
\end{center}
\caption{NGC 5338: WIYN $R$-band isophotes (a); ellipse fits (b,
WIYN $R$); \br{} color map on two scales (c, $w = 15$, color range
$\ap 0.6$ mag; d, $w = 5$, color range $\ap 2.2$ mag, $\ap 0.9$ mag
between center and dust lanes).
\label{fig:n5338}}
\end{figure}

\begin{figure}
\begin{center}
	\includegraphics[scale=0.85]{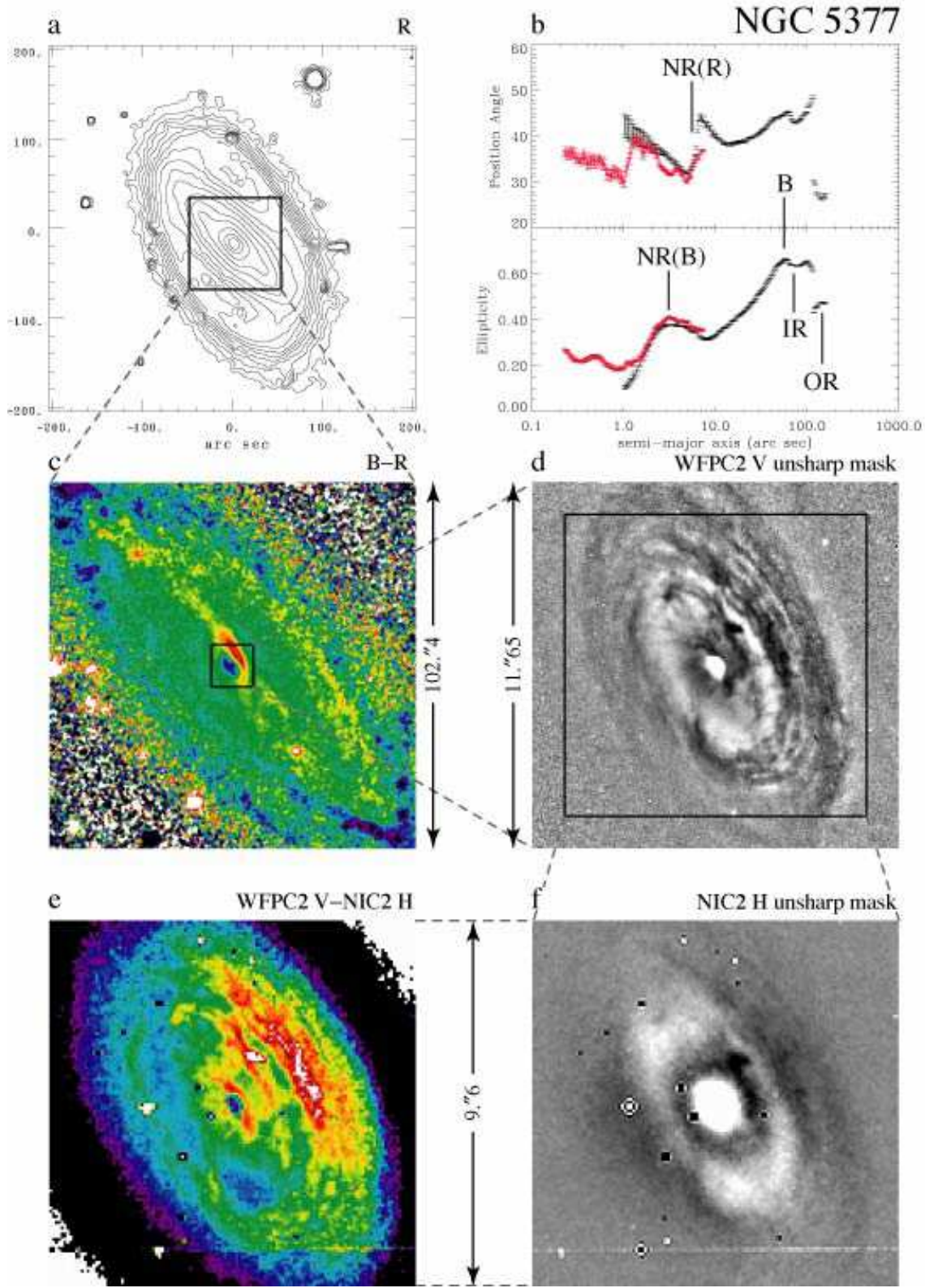}
\end{center}
\caption{NGC 5377: large-scale $R$-band isophotes (a); ellipse fits
(b, WIYN $R$ and NICMOS2 F160W [red]); \br{} color map (c, $w = 5$,
color range = 0.81 mag between blue inner ring and red nuclear ring, =
1.44 mag between inner ring and reddest blobs outside); WFPC2 + NICMOS2
$V\!-\!H$ (F606W$\,-\,$F160W) color map (e, unsmoothed, color range =
1.41 mag); unsharp masks of PC2 F606W image (d, \ds{10}) and NICMOS2
F160W image (f, \ds{10}).
\label{fig:n5377}}
\end{figure}

\begin{figure}
\begin{center}
	\includegraphics[scale=0.85]{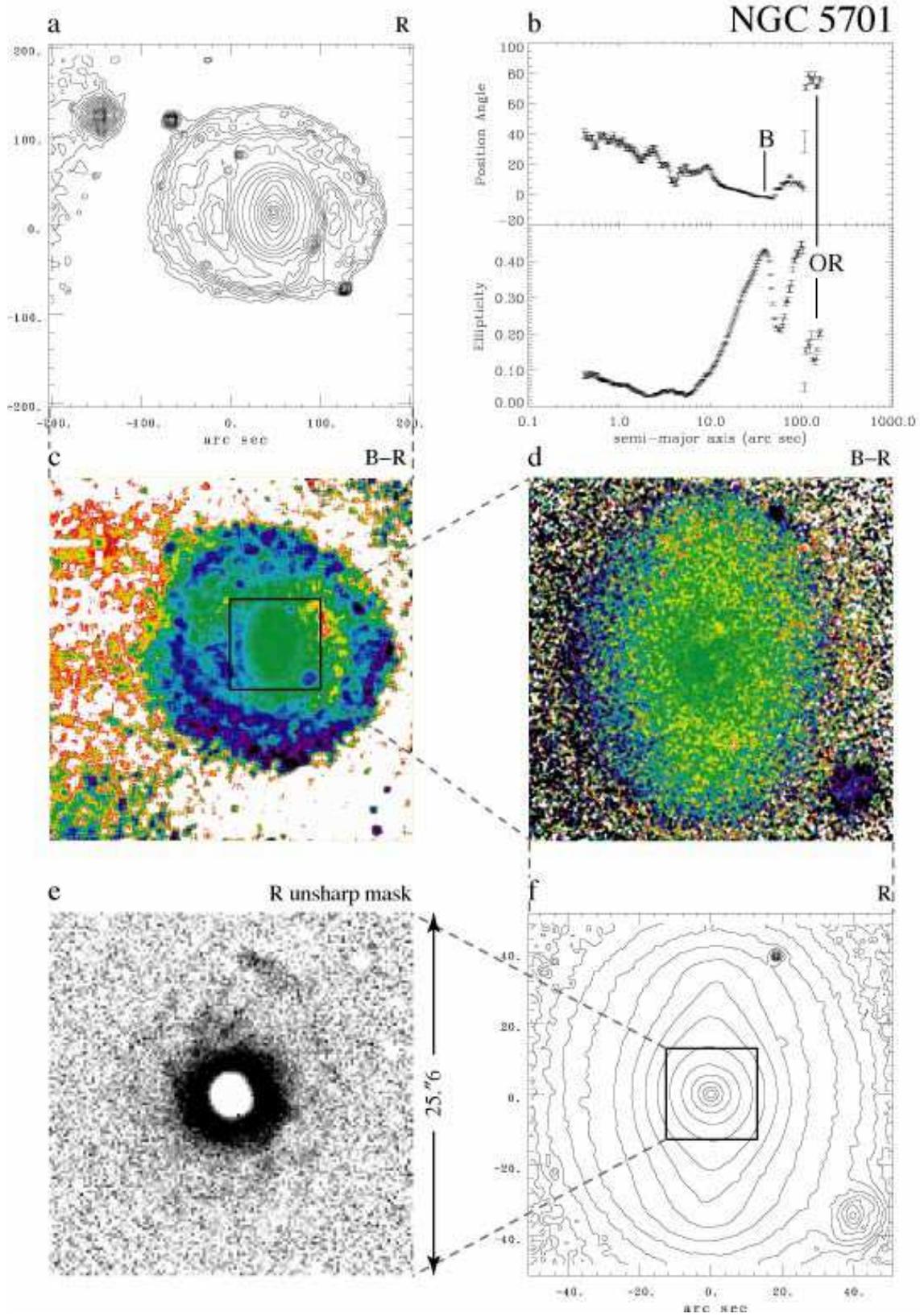}
\end{center}
\caption{NGC 5701: $R$-band isophotes on two scale (a, f); ellipse
fits (b, WIYN $R$); \br{} color maps (c, $w = 41$, color range = 1.44
mag; d, $w = 5$, color range = 1.43 mag); unsharp mask of $R$-band
image (e, \ds{5}).
\label{fig:n5701}}
\end{figure}

\begin{figure}
\begin{center}
	\includegraphics[scale=0.85]{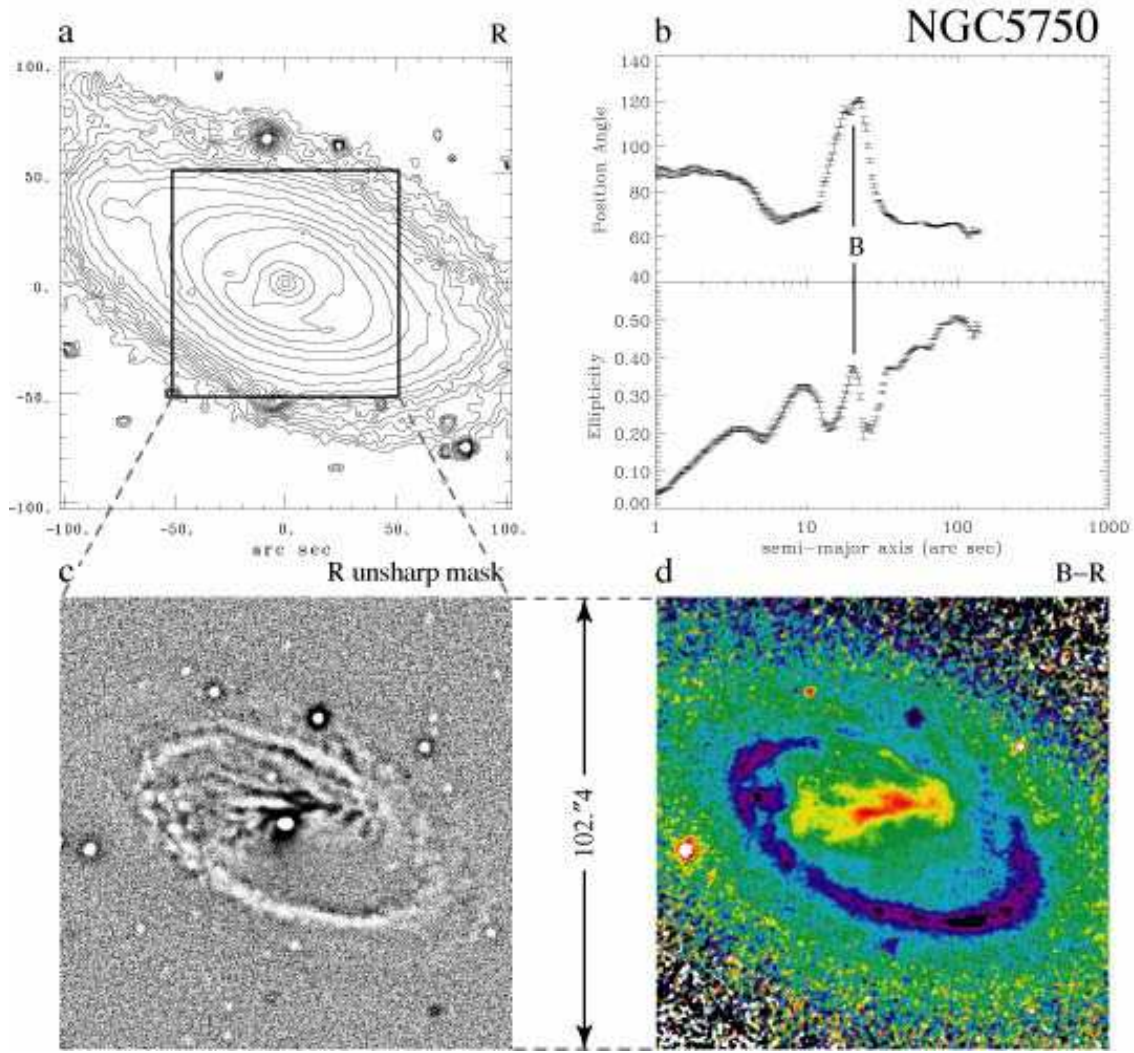}
\end{center}
\caption{NGC 5750: $R$-band isophotes (a); ellipse fits (b, WIYN
$R$); unsharp mask of $R$-band image (c, \ds{5}); \br{} color map (d,
$w = 5$, color range $\ap 0.8$ mag, excluding red star near east
edge).
\label{fig:n5750}}
\end{figure}

\begin{figure}
\begin{center}
	\includegraphics[scale=0.85]{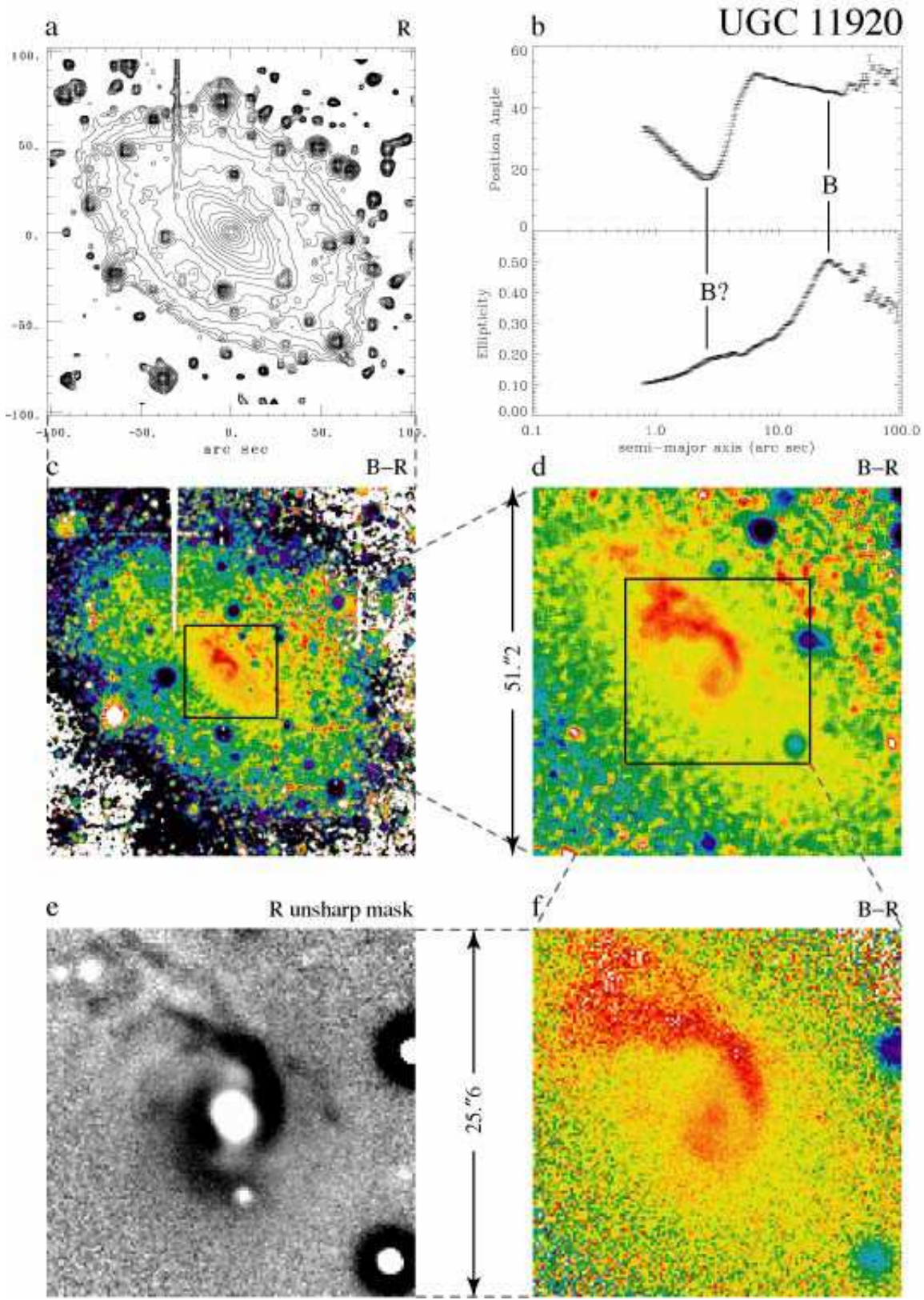}
\end{center}
\caption{UGC 11920: $R$-band isophotes (a, the vertical line
east of galaxy center is spillover from a saturated star); ellipse
fits (b, WIYN $R$ only); \br{} color map (c, $w = 11$,
color range $\ap 0.3$ mag; d, $w = 5$, color range = 1.15
mag; f, unsmoothed, color range $\ap 1.15$ mag); unsharp mask of
$R$-band image (e, \ds{5}).
\label{fig:ugc11920}}
\end{figure}

\begin{figure}
\begin{center}
	\includegraphics[scale=0.85]{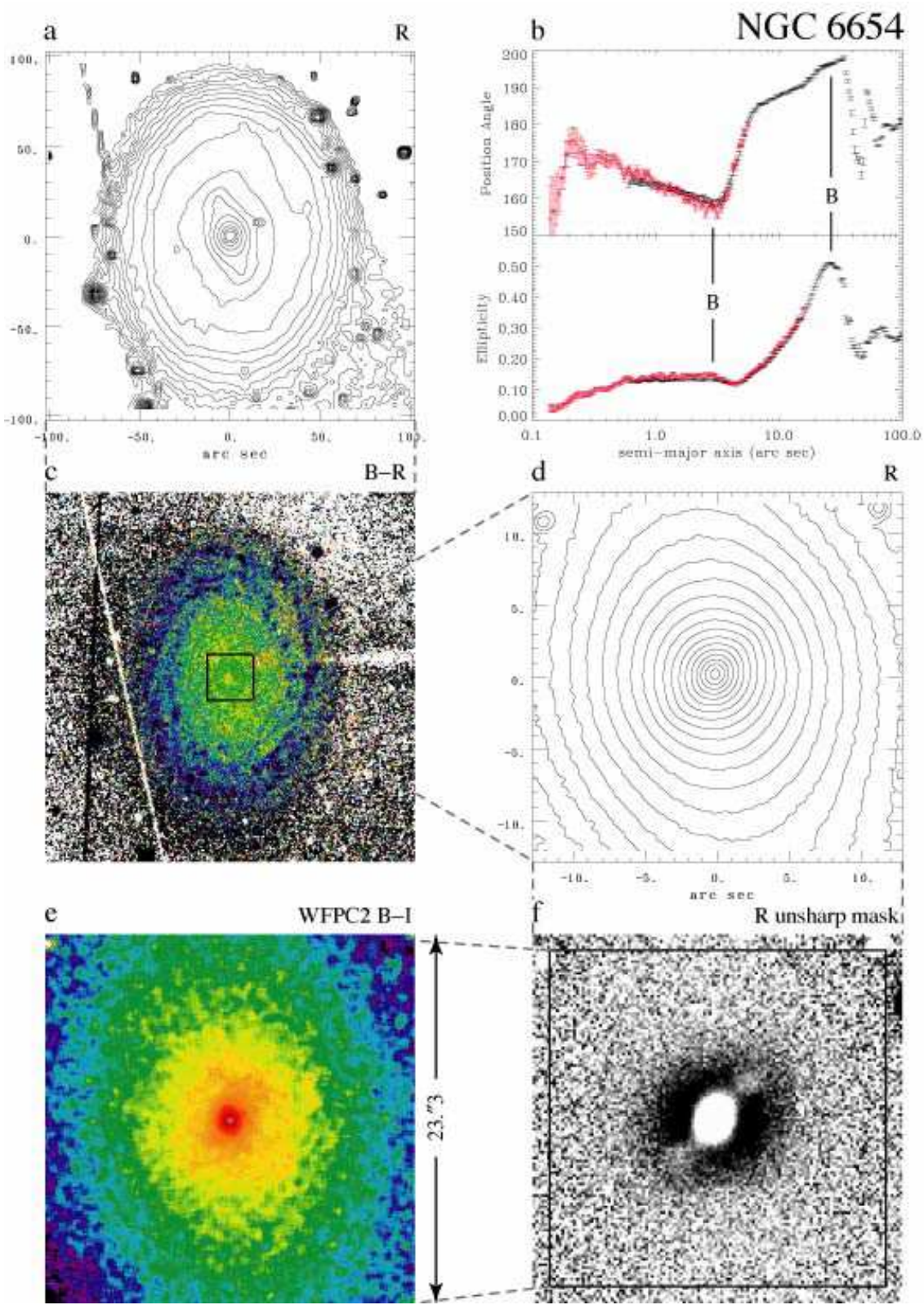}
\end{center}
\caption{NGC 6654: $R$-band isophotes on two scales (a, d); ellipse
fits (b, WIYN $R$ and WFPC2 F814W [red]); \br{} color map (c, $w = 5$,
color range $\ap 0.98$ mag); WFPC2 F450W$\,-\,$F814W color map (e, $w
= 11$, color range = 1.50 mag); unsharp mask of $R$-band image (f,
\ds{5}).  The black and white streaks in the WIYN color map (c) are
due to satellite or meteor trails, one of which can be seen in the
$R$-band image (a).
\label{fig:n6654}}
\end{figure}

\begin{figure}
\begin{center}
	\includegraphics[scale=0.85]{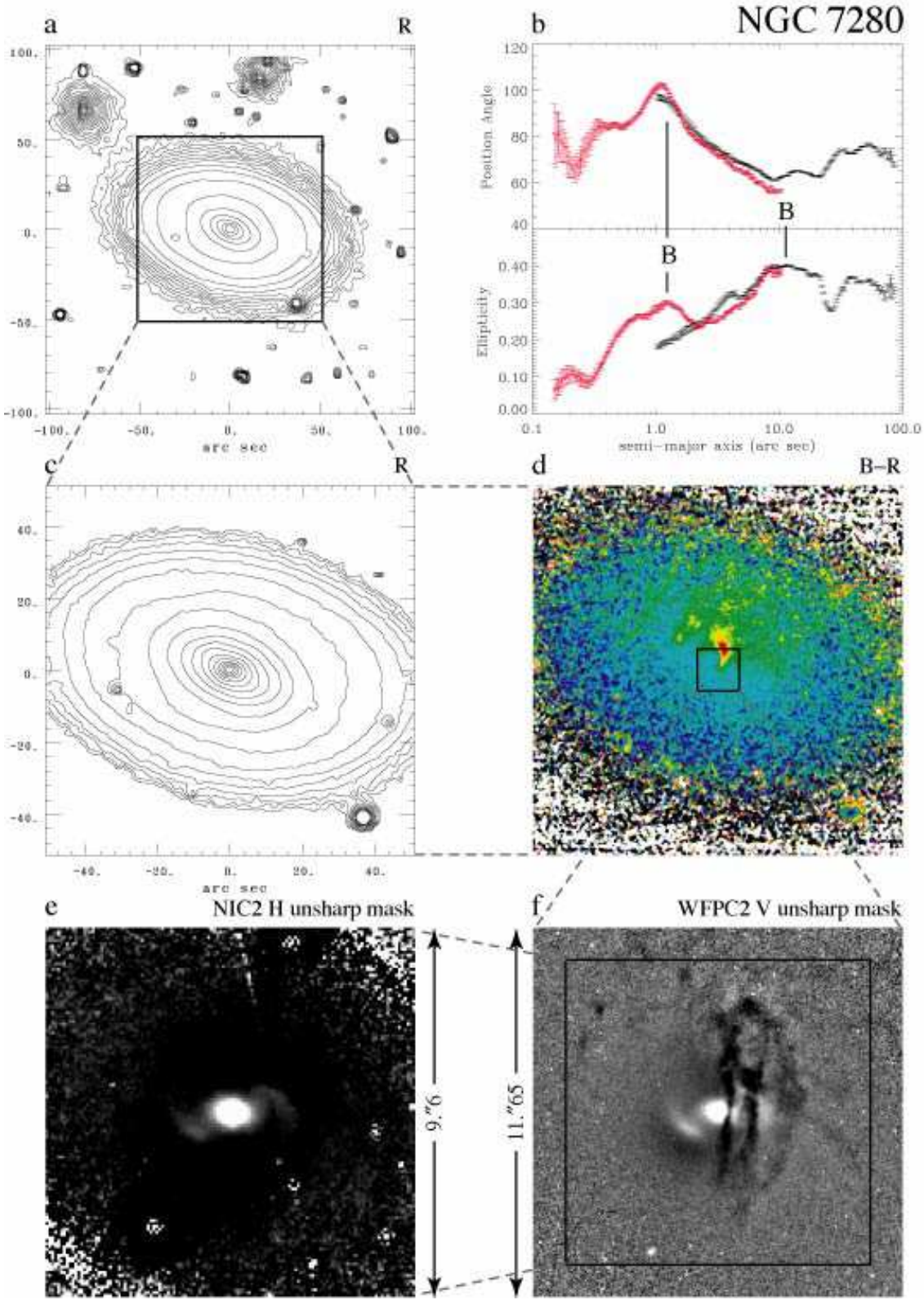}
\end{center}
\caption{NGC 7280:$R$-band isophotes on two scales (a, c); ellipse
fits (b, WIYN $R$ and NICMOS2 F160W [red]); \br{} color map (d, $w =
11$, color range $\ap 0.3$ mag; f, $w = 5$, color range = 1.45 mag);
unsharp masks of NICMOS2 F160W (e, \ds{8}) and WFPC2 F606W (f,
\ds{10}) images.
\label{fig:n7280}}
\end{figure}

Figure~\ref{fig:n3729}: Our WIYN observations of this galaxy were
badly affected by bright sky conditions, and we were unable to
re-observe it.  Only the long $R$-band exposure has usable data, and
there is essentially no discernible signal above the sky outside of
the inner ring.  Consequently, we have used the $B$ and $R$ images
from \citet{tully96}, made available via NED. These were taken with
the Mauna Kea 2.24-m telescope, using a $500 \times 500$ CCD with
0.598\arcsec{} pixels.  We supplement these with ellipse fits to the
central region from the WIYN image, which has higher resolution.

NGC~3729 is believed to be interacting with its ``peculiar'' neighbor
NGC~3718, although \citet{schwarz85} found only tenuous evidence for a
tidal bridge in \hi{} between in the two galaxies, and more recent
\hi{} observations \citep[see][]{cox96} do not show such a bridge. 
There \textit{is} some evidence for distorted optical structure on the
NE side of the galaxy (panel~a of Figure~\ref{fig:n3729}); this blob
of excess light appears quite blue in the color map \citep[as noted
by][]{verheijen01}, so it may be a region of recent star formation.

The color map shows that the inner ring surrounding the bar is blue;
\citet{usui98} noted \ha{} emission from this ring and
\citet{verheijen01} indicate that most of the galaxy's \hi{} is in the
ring as well.  Inside, there are several strong dust lanes.  The
ellipse fits for $a \lesssim 3\arcsec$ are distorted by the presence
of a bright ``second nucleus'' about 2.6\arcsec{} east-northeast of
the galaxy center.  This feature is partially resolved in the WIYN $R$
image (FWHM more than twice the stellar FWHM), so it is probably
\textit{not} a star.  It is noticeably blue in the colormap, so it may
be a luminous, young star cluster.  The combination of this feature
and the strong dust lanes means that we cannot determine whether there
are additional structures inside the bar.

Our determination of the outer disk orientation and inclination is
based on the outer isophotes, and agrees well with the \hi{} kinematic
determinations of \citet{verheijen01}, though the latter might be
affected by noncircular motions in the inner ring.

\subsection{NGC 3941: Double bar, off-plane gas\label{n3941}} 

Figure~\ref{fig:n3941}: This galaxy was recently noted for possessing
ionized gas near the center which counter-rotates with respect to the
stars \citep{fisher94,fisher97}.  In fact, published \hi{} maps
indicate that \textit{all} the gas in this galaxy is counter-rotating. 
The \hi{} synthesis map \citep{vanD89} shows that the neutral gas is
found primarily in two rings.  The outer ring is aligned with the
stellar disk and has approximately the same ellipticity.  The inner
\hi{} ring appears almost circular; if it is intrinsically circular,
then it must be \textit{inclined} $\sim 20\arcdeg$ or more with
respect to the stellar disk.  Both rings share the same sense of
rotation as that of the inner ionized gas observed by Fisher; thus,
\textit{all} the observed gas in this galaxy counter-rotates with
respect to the stars.  We argue below that some or all of the gas may
be off-plane.

Our images show a symmetric stellar disk oriented almost N-S, with a
small primary bar.  Using the isophotes, we estimate the position
angle of this bar to be $\ap 166\arcdeg$; it is probably a typical
``rectangular'' S0 bar \citep[see][]{athan90}, and the extended
``corners'' produced by projection skew the ellipse fits
(Figure~\ref{fig:n3941pa}).  Unsharp masking (\ds{20}, not shown)
indicates weak stellar spiral arms outside the primary bar, with a
counter-clockwise trailing orientation; these may be responsible for
the ellipticity peak outside the bar in the ellipse fits.  Our
inclination for the outer disk is based on the outermost isophotes, at
$a \sim 120\arcsec$.

The ellipse fits and unsharp masks also show a secondary bar (see
Figure~\ref{fig:umaskpa}).  The position angle in the ellipse fits is
only 10\arcdeg{} different from our estimate of the outer disk
orientation --- which would meet our inner disk criteria --- but the
isophotes and unsharp masks indicate a position angle 15\arcdeg{}
further to the east, so we consider this a secondary bar.  The
structure of this feature in the unsharp masks also supports a bar
classification.

The primary bar is redder than the outer disk, with a smooth color
gradient following the isophotes.  This is fairly typical for galaxies
in our sample, including those with no evidence for significant dust
(e.g., NGC~4665), and suggests that the gradient is due to changing
stellar colors.

However, within the bar there are clear indications of spiral dust
lanes with a peculiar orientation (panels~d and f).  While they have
the same sense of rotation --- trailing counter-clockwise --- as the
faint stellar arms outside the bar, the fact that the gas in this
region is counter-rotating presents us with an interesting problem. 
Either some of the spirals in this galaxy are leading, or else the gas
and stars are not coplanar.  The dust lanes in the bulge region are
almost entirely confined to the east side of the galaxy, so they could
be part of an inclined gas disk, as in NGC~2655, NGC~2787, and
NGC~4203.  The \textit{minor}-axis spectrum of \citet{fisher97} shows
rotation in the ionized gas, further indication that the gas and stars
are misaligned.  We suggest that the inner ionized gas and outer \hi{}
are all on inclined orbits, almost perpendicular to the stellar disk. 
In this case, the eastern sides of the gas disks/rings are closer to
us, while the nearer side of the stellar disk is to the west, and both
stellar and dusty spirals are trailing.  Since the inner gas is
probably off-plane, we do not class this as a ``nuclear spiral,''
though the central dust lanes are partly spiral in shape.

\subsection{NGC 3945: Double bar, nuclear ring, inner disk\label{n3945}} 

This excessively complex galaxy --- with two bars, an inner disk,
\textit{and} a stellar nuclear ring --- was discussed in detail by
\citet{erwin99}.  The measurements in Table~\ref{tab:features} are the
same, except for the position angles of the two bars and the
ellipticity of the nuclear ring.  Examination of unsharp masks and of
the isophotes indicates the position angles for both bars that are
different from those given by the ellipse fits.  We also measured a
semi\textit{minor} axis value for the nuclear ring of $\ap
2.8\arcsec$, using the \ds{20} unsharp mask of the WFPC2 F814W image;
this gives an ellipticity of 0.57 for the nuclear ring (the value in
Erwin \& Sparke was from the ellipse fits).  This suggests that the
nuclear ring is intrinsically elliptical, as long as the galaxy's
inclination is $< 67\arcdeg$.

\subsection{NGC 4045: Dust-obscured\label{n4045}} 

Figure~\ref{fig:n4045}: This galaxy is a splendid example of a dusty,
star-forming Sa, similar in some respects to NGC~2273.  The outer disk
has two blue spiral arms which wrap almost 180\arcdeg{} around and
appear to dimple in at the bar ends, forming an \rone{} outer ring;
this causes the ellipticity peak at $a \ap 66\arcsec$.  Numerous
patchy blue spiral arms form a partial inner pseudo-ring around the
bar.  These arms are almost certainly sites of recent or ongoing star
formation; \citet{usui98} traced the ``inner ring'' in \ha{} emission. 
The dust lanes are stronger on the S side, suggesting that side of the
galaxy is nearer.

The bar itself lies almost north-south; unsharp masking and the color
maps show the characteristic leading-edge dust lanes, along with the
associated ``spur'' dust lanes, as in NGC 2273.  The two leading-edge
dust lanes converge on what might be a nuclear dust ring, with some
bright asymmetries near the nucleus, though this is at the limits of
our resolution.

\subsection{NGC 4143: Nuclear spiral, inner disk\label{n4143}} 

Figure~\ref{fig:n4143}: Unsharp masking indicates that the bar in this
galaxy lies at a PA of $\ap 165\arcdeg$, rather than the 156\arcdeg{}
suggested by the ellipse fits.  In the ellipse fits, the bar shows up
as a pronounced position angle twist, but \textit{not} as a distinct
ellipticity peak.  We identify \amax{} with the maximum in position
angle; \amin{} cannot be defined.  Further in, the ellipse fits show
an ellipticity peak at a position angle only 2\arcdeg{} different from
that of the outer disk.  This is then a clear, if not particularly
strong, inner disk candidate.

The \br{} color maps shows one or two possible dust lanes outside the
bar, and several strong dust lanes in the nuclear region.  The nuclear
lanes are complex and asymmetric, but appear to form a spiral pattern,
trailing clockwise, particularly in unsharp masking of the WFPC2
image; \citet{martini01} noted this spiral pattern as well,
classifying it as ``flocculent.''  (By contrast, a weak
counter-clockwise trailing stellar spiral can be seen in unsharp
masks, outside the bar --- this might indicate that the inner gas is
not corotating with the stars.)  A pair of dust lanes south of the
nucleus blur into a single lane in the \br{} color map; these have a
different orientation than the spiral lanes.  In the F606W$\!-\!$F160W
color map, the strongest dust patches are red, while the nucleus
itself is blue \citep[probably an AGN --- ][]{ho97d}.

\subsection{NGC 4203: Off-plane gas\label{n4203}} 

Figure~\ref{fig:n4203}: This weakly barred galaxy harbors one of the
best-documented AGNs in our sample, with broad \ha{} emission, a
power-law X-ray spectrum, and a compact UV pointlike nucleus
\citep[][]{ho97d,iyomoto98,barth98}.  Westerbork \hi{} observations by
\citet{vanD88} show two \hi{} rings, a nearly circular inner ring with
$r \ap 44\arcsec$, probably coplanar with the stellar disk, and an
extended, elliptical outer ring beyond the optical disk; the latter is
probably tilted (by 34\arcdeg{} or 87\arcdeg) with respect to the
optical disk.

The bar is a weak, elliptical distortion aligned with the outer disk;
this alignment means we cannot define $\aten$.  Our value for $\amin$
is much larger than $\amax$; unsharp masking suggests the true end of
the bar is probably just outside $\amax$.

The \br{} color maps show counter-clockwise spiral dust lanes outside
the bar.  Inside the bar, the dust is much stronger, and confined to
the region east of the blue nucleus.  The WFPC2 unsharp masks and
color maps show concentric arcs of dust, centered on the nucleus; this
suggests an inclined gas disk, extending above the stellar disk on the
east side.  Together with the evidence for misalignment between the
outer \hi{} and the optical disk, this leads us to put the galaxy in
our off-plane gas class \citep[it is listed as ``Kinematically
Related'' to polar rings in the catalog of][]{whitmore90}.

Because the dust lanes are quite small-scale and one-sided, we have a
relatively unobstructed view of the central regions, so we do not
consider this a dust-obscured galaxy.  Neither the ellipse fits nor
the unsharp masks show any evidence for an inner bar; an inner disk
would be extremely hard to detect, since the galaxy is close to
face-on.

\subsection{NGC 4245: Nuclear spiral, nuclear ring\label{n4245}} 

Figure~\ref{fig:n4245}: This galaxy is part of a group within the
Coma~I cloud which also includes NGC~4310 and NGC~4314; we adopt a
distance of 12 Mpc for all three galaxies, based on the distances for
galaxies in that cloud given by \citet{forbes96}. 
\nocite{garcia94}Garc\'{\i}a-Barreto, Downes, \& Huchtmeier (1994)
have suggested that this group may be analogous to an extremely poor
cluster, especially since the galaxies in the central part of the
group (projected $r < 130$ kpc, which includes all three of the
galaxies in our sample) are all deficient in \hi.  In fact, NGC~4245
was one of the three most \hi{}-deficient spirals in the group (the
most deficient being NGC~4314), though \citet{gerin94} found the
\textit{molecular} gas content of all three to be typical for their
Hubble types.

Like its better-studied sibling NGC~4314, NGC~4245 harbors a
star-forming nuclear ring.  Leading-edge dust lanes that appear to
flow into the ring are visible in both the unsharp mask and the color
maps.  The nuclear ring shows up in the ellipse fits as a rapid change
in ellipticity and a discontinuity in position angle; we measure the
ring's size, shape, and orientation directly from the WFPC2 image.  In
the unsharp mask of that image, we can see that the ring is made up of
bright knots mixed with spiral dust lanes; the latter lead into a
nuclear spiral which appears to continue all the way into the nucleus. 
The \br{} color map (panel~f) shows that the brightest regions of the
nuclear ring are blue, consistent with their being sites of recent or
ongoing star formation.

We find no evidence for an inner bar within the nuclear ring in the
WFPC2 unsharp mask; the variations in the ellipse fits are almost
certainly due to the dust lanes of the nuclear spiral.  These dust
lanes are weak enough that feel fairly confident that there is no
inner bar in this galaxy, despite not having near-IR images.

\subsection{NGC 4310: Dust-obscured\label{n4310}} 

Figure~\ref{fig:n4310}: This highly inclined galaxy is a member of the
same group as NGC~4245 and NGC~4314 (see discussion of NGC~4245,
above).  The color maps and unsharp masks show strong dust lanes on
the western side of the disk, along with blue regions closer to the
center.  Our identification of the (sole) bar is rather uncertain, and
any central structures could be hidden by the complex of dust lanes.

\subsection{NGC 4314: Double bar, nuclear ring, nuclear spiral\label{n4314}} 

Figure~\ref{fig:n4314}: This galaxy has both a well-studied nuclear
ring and a secondary bar; the availability of both WFPC2 and NICMOS
images, and the large size of the secondary bar, make it idea for
studying the shape of dust lanes within a secondary bar.  See the
notes for NGC~4245 and NGC~4310 (above) for a discussion of this
galaxy's environment, which may be responsible for its extreme \hi{}
deficiency \citep{garcia94}.

As mentioned in Section~\ref{sect:measure}, defining the length of
this galaxy's primary bar is problematic.  Published measurements
agree on lengths of 60--75\arcsec.  The DSS and WFPC2 mosaic images
(panels~a and c) show that the classic, narrow bar, with notably
square ends \citep{athan90}, comes to an end at approximately this
length (roughly at $\amax = 70\arcsec$).  However, the isophotes
remain elongated and parallel with the bar well beyond that radius, as
can be seen in both the images and the ellipse fits, as well as in the
near-IR images of \citet{quillen94} and the ellipse fits of
\citet{friedli96a}.  Consequently, $\aten$ and $\amin$ give much
larger values for the bar size (113\arcsec{} and 127\arcsec,
respectively).  It is not clear why the galaxy has this appearance;
Quillen et al.\ and \citet{patsis97} modeled the bar and placed
corotation at $r = 70\arcsec \pm10\arcsec$.  If that is the case, we
would not expect the stars to have a distribution parallel to the bar
beyond that radius.  One possibility is that we are seeing the bright
inner parts of the two spiral arms which cross over, or trail off
from, the bar ends, forming what may be an incomplete outer ring.  The
WFPC2 mosaic image suggests that these arms extend slightly past the
bar ends on the leading side, so that the isophotes remain roughly
parallel to the bar even though they are no longer bar-shaped. 
Nonetheless, it is curious that the combined stellar distribution
seems parallel to the bar well beyond the bar end itself, particularly
if corotation is close to the bar end.

\citet{benedict96} presented CO imaging of the bar.  Using this and
earlier CO data from \citet{combes92}, they constructed a rotation
curve and identified orbital resonances for the primary bar, assuming
that it ended shortly before corotation.  This rotation curve produced
two inner Lindblad resonances, with the nuclear ring lying between
them and the inner bar ending close to the inner ILR of the primary
bar.

The nuclear ring is visible in the HST images as both a set of bright,
blue knots and several curved dust lanes, and has recently been
studied in detail by \citet{benedict02}, who find cluster ages of 1--5
Myr.  At least some of the star formation sites show up as bright
knots in the F160W unsharp mask (panel~e), which suggests that some of
the near-IR light in the ring is from red supergiants.  Our
measurement of the ring's size and orientation are based on the blue
features in the WFPC2 color map (panel~f).  The position angle we
measure (135\arcdeg) is very close to that determined by
\citet{benedict02} (137\arcdeg) from an ellipse fit to the
distribution of blue star clusters, though they derived a slightly
larger semi-major axis (7.4\arcsec{} versus our 6.8\arcsec).  As
reported by \citet{combes92} and \citet{benedict96}, this ring is
slightly \textit{outside} the main CO ring, but is roughly coincident
with the ring of \ha{} emission \citep{gonzalez97} and 6 cm radio
continuum emission \citep{garcia91}; it also matches the HST \ha$ +
$\nii{} map of \citet{benedict02}.  The strongest dust lanes in the
unsharp mask and the color map do seem to lie just inside the blue
ring of star formation, consistent with the idea that the traces the
molecular gas.

Also clearly visible are two blue arms \textit{outside}, trailing away
from the NW and SE ends of the ring, which resemble features seen
outside the nuclear ring in NGC~2273 (above).  These make up the
elliptical blue region reported by \citet{benedict93}. 
\citet{benedict02} pointed out that these arms are oriented almost
perpendicular to the nuclear ring and to the outer bar, and are
located in between the two ILRs suggested by the CO-derived resonance
curves of \citet{benedict96}; they suggested that these arms might be
associated with $x_{2}$ orbits of the outer bar.  They also found
colors consistent with ages of $\sim 100$--200 Myr for the arms,
indicating a previous episode of star formation, perhaps associated
with the \textit{outer} of the two ILRs.

The inner bar was first noticed by \citet{benedict93}, using WF/PC1
images in the $I$ band.  It is even more prominent in the NICMOS2
F160W image (e.g., the ellipse fits in panel~b and the unsharp mask in
panel~e; also see the bulge-subtracted image in \nocite{ann01}Ann
2001), with an ellipticity peak just inside the nuclear ring and a
position angle of 136\arcdeg{} --- parallel with the nuclear ring
\citep[as noted by][]{ann01}, close to the position angle of the outer
bar, but much different from that of the outer disk.  (The ellipticity
peak outside the ring is due to the blue spiral arms noted above.) 
The presence of this bar has been confirmed by both \citet{ann01} and
\citet{benedict02}, using the NICMOS and WFPC2 images, respectively. 
In contrast to Benedict et al., however, we argue that the inner bar
is probably \textit{not} parallel with the outer bar.  The use of the
F160W image (less affected by dust) provides a better estimate of the
inner bar's orientation; in addition, there is little uncertainty in
the position angle of the \textit{outer} bar (position angles $<
145\arcdeg$ are clearly ruled out).  This distinction is important
because an inner bar parallel to the outer bar can be explained as the
result of $x_{1}$ orbits inside a double ILR in a \textit{single} bar. 
Our measurements suggest that the inner bar actually \textit{trails}
the outer bar by $\ap 10\arcdeg$, and is thus likely to be
independently rotating.  In support of this scenario, \citet{ann01},
who also noted the $\approx 10\arcdeg$ misalignment between outer bar
and nuclear ring, found that agreement between gas flow in
hydrodynamic simulations and the visible dust lanes of this galaxy was
better when the inner bar was independently rotating.

The PC2 unsharp mask and the color map show a wealth of dust lanes
within the nuclear ring.  The lanes are aligned with the secondary
bar, and rotationally symmetric about the center.  Two of the lanes
are clear analogues of the leading-edge lanes in the primary bar, with
the same sense of rotation.  Curiously, they appear to extend past the
midpoint of the inner bar, though there does not appear to be a dust
ring within the inner bar.  For simplicity, we classify these lanes as
a ``nuclear spiral.''

\subsection{NGC 4386: Inner disk\label{n4386}} 

Figure~\ref{fig:n4386}: The bar in this galaxy is almost aligned with
the outer disk, and appears primarily as a peak in the ellipticity,
though it also shows up as a modest red feature in the \br{} color
map.  Inside, there is a weak secondary ellipticity peak aligned with
the outer disk.  This feature, which fits our criteria for an inner
disk, appears as a striking red zone in the color maps, reddest at the
center (a change of $\ap 0.22$ in \br).

Unsharp masking shows two peculiar features: a faint symmetric
brightening just \textit{outside} the inner disk, which might indicate
a faint stellar nuclear ring (panel~d); and a hint of bar-like
structure in the inner disk itself (panel~f).  Our resolution is not
good enough for us to decide if this is really a bar; nonetheless, NGC
4386 is a good candidate for an inner disk which is actually a
secondary bar.

\subsection{NGC 4643: Nuclear spiral, inner disk, off-plane gas?\label{n4643}} 

Figure~\ref{fig:n4643}: This is a strongly barred galaxy with an
unusually thin bar; it had the narrowest bar of the six early-type
barred galaxies studied by \citet{ohta90}.  The stellar velocity field
of \citet{bettoni97} suggested circular rotation in the disk (apart
from bar-related deviations), with $i \ap 40\arcdeg$ and a
line-of-nodes at $\ap 50\arcdeg$ \citep[see also][]{magrelli92}; this
agrees well with our values for the outer disk isophotes. 
\citet{whitmore90} classed this as an ``Object Possibly Related to
Polar Rings,'' noting that an unpublished image showed ``a very faint,
disklike (i.e., about E8) feature \ldots{} aligned along the major
axis extending to about three times the optical radius of the inner
galaxy.''  A similar feature has been reported by David Malin (1998,
private communication).  No sign of this appears in our images, though
this may simply be because of the relatively short exposure.

This galaxy was one of the three strongest double-bar candidates found
by \citet{shaw95}, due to strong position-angle twists within the bar. 
Our ellipse fits show that the twist is due to an inner elliptical
component at the same position angle as the outer disk: we interpret
this feature as an inner disk.  In a \ds{2} unsharp mask, the inner
disk actually shows some signs of being a \textit{ring}, but the
rather low resolution of our images ($R$-band seeing $= 1.3\arcsec$)
means we cannot be certain.

The \br{} color maps show tightly wrapped spiral arms in a fairly
narrow zone just outside the bar; the mix of blue and red arms
suggests both dust and recent star formation.  In the vicinity of the
inner disk we see a dusty nuclear spiral, with the same sense of
rotation as the spiral arms outside the bar.  A separate dust lane,
slightly curved, to the NW of the nucleus could be part of a
\textit{polar} dust lane; the orientation is roughly the same as the
faint optical extensions reported by \citet{whitmore90}.  However, the
\hi{} width of $W_{20} = 351$ \kms{} reported by \citet{wk86} is
consistent with Bettoni \& Galletta's observed stellar $V_{max}$ of
140 \kms{} (at $r = 15\arcsec$), so there is no evidence here for
off-plane gas.

\subsection{NGC 4665\label{n4665}} 

Figure~\ref{fig:n4665}: This galaxy is a relatively featureless SB0,
with a strong bar and only a few asymmetries in the outer disk and a
curious twist inside the bar to mar an otherwise canonically simple
appearance.  \citet{ohta90} performed an extensive analysis of the
surface photometry and decomposed the galaxy into disk, bulge, and bar
components.  It had the highest bar-to-disk luminosity ratio of the
six galaxies in their sample: almost half the light between the bulge
and the ends of the bar was in the extracted bar model.

The ellipse fits show the bar, but no clear signs of any other
structures.  There is a very weak \textit{increase} in ellipticity
near the very center ($a \lesssim 1.5\arcsec$); due to the poor seeing
(1.2\arcsec) we cannot determine what, if anything, this is due to. 
We \textit{do} see the twist in PA along the bar noted by
\nocite{jung97}Jungwiert et al.\ (1997) in their near-IR images
--- amounting to $\approx 12$ degrees between $a = 45\arcsec$ and $a =
1\arcsec$.  This could be a sign of mild triaxiality in the bulge.

Weak stellar spiral arms appear to trail off the ends of the bar, with
the southern arm somewhat stronger than the northern one.  These arms
can also be seen in the processed images of \citet{ohta90}, where
axisymmetric components of the galaxy have been modeled and
subtracted.  Both arms are red features in the color map, which may
indicate some dust content.  More striking is the outer, red arc in
the E side of the galaxy.  Its position and shape are suggestive of a
dusty outer ring, although there is no counterpart in the W side of
the disk.  This color asymmetry is interesting in light of the fact
that this was one of the six \textit{least} asymmetric galaxies in the
60-galaxy $R$-band sample of \citet{rudnick98}, based on $R$-band
images.

Neither the color map nor unsharp masks show any evidence for dust
lanes or rings in the inner regions.  A general inward reddening trend
is visible --- the galaxy center is $\sim 0.4$ magnitudes redder than
the disk just outside the bar --- and the isochromes twist to follow
the isophotes.  This suggests that the isophote twist is \textit{not}
due to leading-edge dust lanes, which would produce a twist in the
opposite sense.  Unlike the isophotes, however, the isochromes remain
elongated into the galaxy center.

\subsection{NGC 4691: Dust-obscured\label{n4691}} 

Figure~\ref{fig:n4691}: This is an unusual and complicated galaxy,
which bears some resemblance to IC~676.  Like that galaxy, it has an
extremely narrow bar with several distinct blue regions of recent star
formation and a dust lane crossing straight through (or over) the bar;
but it lacks the other galaxy's symmetry.

The \hi{} spectra of \citet{huchtmeier85} and \citet{richter87} are
somewhat peculiar and possibly asymmetric, and the velocity widths
reported by Huchtmeier \& Seiradakis ($W_{20} = 167$ \kms{} and
$W_{50} = 76$ \kms) are not consistent with ordered, circular
rotation.  Richter \& Huchtmeier give a smaller value of $W_{20} =
132$ \kms{}, which is only barely consistent with the apparent
inclination (for $i = 42$, it implies $V_{circ} \ap 100$ \kms). 
\citet{young95} reported an extremely narrow CO width ($W_{50} = 35$
\kms), and \nocite{wiklind93}Wiklind, Henkel, \& Sage's (1993) CO
velocity map for $r \lesssim 20\arcsec$ showed essentially \textit{no}
velocity gradient; they argued that the galaxy was face-on.  All of
this suggests that the gas in this galaxy is either not coplanar with
the stars, not in ordered motion, or both.

The outer disk appears somewhat distorted.  The bar itself ends in two
stubby spiral arms which partly wrap around it to form a partial inner
ring; inspection of unsharp masks suggests that the bar may actually
end at $a \sim 42\arcsec$, with the beginnings of the inner ring
keeping the isophotes approximately parallel until $\aten =
55\arcsec$.  Because of the complicated structure within the bar, the
ellipse fits are unreliable for $a \lesssim 25\arcsec$, and the
fitting algorithm fails altogether for $a < 3.5\arcsec$).

The color maps show a total of seven distinct blue regions in the bar,
four of which correspond to the \ha{}-bright knots studied
photometrically and spectroscopically by \citet{garcia95}; the latter
are the four brightest ``nuclei'' in the $R$-band image (panel~e). 
The easternmost blue region is separated from the rest by a narrow
dust lane, which appears to connect with strong dust lanes which run
along the north and south sides of the bar.  There is some resemblance
to the bar-crossing lane in IC~676, but in this galaxy none of the
structure is symmetric about the bar.

\subsection{NGC 5338: Dust-obscured\label{n5338}} 

Figure~\ref{fig:n5338}: This galaxy, among the closest in our sample
($D = 11.5$ Mpc), shows evidence for dust lanes along middle of the
bar.  In the center, there is a rather abrupt transition to an
asymmetric blue region, possibly connecting with a blue spiral arm to
the SE. Although this appears as a change in PA and ellipticity in the
ellipse fits, the unsharp masks show no sign of a secondary bar or
other regular structures.  The innermost regions are complex and
asymmetric, possibly due to star formation.

Although the shape of the outer disk argues for a high inclination ($i
= 68\arcdeg$), the \hi{} $W_{20}$ is only 87 \kms{} (LEDA; 91 \kms{}
in \nocite{wk86}Wardle \& Knapp 1986).  This might indicate that the
\hi{} is not coplanar with the stars.

\subsection{NGC 5377: Nuclear spiral, nuclear ring\label{n5377}} 

Figure~\ref{fig:n5377}: This is another galaxy where the outer
isophotes are dominated by an outer ring, this time a notably blue
one.  It seem clear that the galaxy is highly inclined --- but
\textit{how} highly inclined is difficult to tell.  If we require
$V_{circ} < 500$ \kms, then $i > 21\arcdeg$ \citep[given $W_{20} =
380$ \kms,][]{balkowski83}.  Misalignment between the outer and inner
rings (and the bar) suggests that outer ring may be intrinsically
perpendicular to the bar, and that the line-of-nodes lies somewhere
\textit{between} the apparent position angles of the two rings. 
Unfortunately, this makes determining the inclination more difficult. 
An intrinsically circular outer ring would imply $i = 58\arcdeg$; an
intrinsically circular inner ring means $i = 50\arcdeg$.  We adopt $i
= 55\arcdeg$ and 35\arcdeg{} for the position angle of the
line-of-nodes; this is reasonably close to the outer kinematic
inclination ($\ap 50\arcdeg$) and major axis ($\ap 30\arcdeg$) derived
from VLA \hi{} observations \citep{kornreich01}.

Both the outer and inner rings are quite blue in the \br{} color maps,
and there is evidence for dust on both the inner and outer edges of
the inner ring.  There are also strong leading-edge dust lanes within
the bar, apparently feeding into a nuclear ring.  Determining the size
of the bar is difficult, due to the effect of the inner ring on the
ellipse fits.  We note that the semi-major axis of the inner ring
(measured on the color maps) is less than our value of $\amin$
(78\arcsec), so the true length of the bar is probably somewhere
between 58\arcsec{} and 70\arcsec.

The HST images reveal a dramatic and complex nuclear ring, first
described by \citet{carollo98}.  Examination of unsharp masks and
color maps indicates that the ring consists of three parts.  The
outermost is a red ring of tightly wrapped dust lanes, into which the
leading-edge lanes of the bar flow; this is marked in the ellipse fits
by an abrupt change in position angle (``NR(R)'' annotation in the
figure and in Table~\ref{tab:features}).  The next major feature is a
bright ring, especially visible in the NICMOS2 unsharp mask (panel~f);
this was also noted by Carollo et al.  It appears in the ellipse fits
as an ellipticity maximum, and in both the \br{} and HST color maps as
a blue region on the SE side, which suggests that it may be made of
younger stars.  Finally, there appears to another dust ring right
around the nucleus ($a \ap 0.85\arcsec$, PA $\sim 30\arcdeg$), best
seen in the $V\!-\!H$ color map (panel~e).  Both the outer and inner
dust rings show up in ellipse fits to the F606W image (not shown) as
discontinuities in the position angle.  The nucleus itself appears
quite blue.  In addition, there are spiral dust lanes in between the
two dust rings, threading through the luminous blue ring.

\subsection{NGC 5701: Nuclear spiral\label{n5701}} 

Figure~\ref{fig:n5701}: This galaxy is surrounded by a large, distinct
outer ring, elongated roughly perpendicular to the bar (the size
listed in Table~\ref{tab:features} is more of an upper limit on the
ring's size, since the arms making up the ring wind outward over a
large range in radius).  This coincides with the \ha{} ring seen by
\citet{pogge93}; the blue spiral arms visible in the color map are
thus almost certainly the result of recent star formation.  The
optical galaxy lies within a huge, somewhat lopsided \hi{} disk,
extending to $r \ap 10\arcmin$ \citep{duprie96,kornreich00}.  The
Duprie et al.\ Arecibo velocity map indicates an approximate PA for
the disk line-of-nodes $\sim 45\arcdeg$; Kornreich et al., using VLA
data, find 48\arcdeg.  The low velocity width ($W_{50} = 121$ \kms{}
from Duprie et al.\ and 139 \kms{} from Kornreich et al.)  suggests
that the \hi{} disk is nearly face-on.

The presence of a detached outer ring means that determining the
inclination is difficult.  Assuming that the \hi{} is coplanar with
the stars, and that 100 \kms $ < V_{circ} < 500$ \kms, then $7\arcdeg
< i < 37\arcdeg$.  A more stringent upper limit can be found by
assuming that the isophotes between the bar and the outer ring, which
appear parallel to the bar with $e \ap 0.21$, are those of a
lens, and thus intrinsically parallel with the bar.  An extremely
elliptical inner ring with axis ratio 0.7 (e.g., NGC~1433, slightly
beyond the 3-$\sigma$ limit in \citet{buta95}) would imply $i =
28\arcdeg$, which can be taken as a reasonable upper limit.  We adopt
an intermediate value of 20\arcdeg{}; this implies that the outer ring
is intrinsically perpendicular to the bar, but is fairly round (axis
ratio = 0.85), and that $V_{circ} \ap 175$ \kms.

The bar itself is almost completely dust-free.  There is a gradual
twist in position angle within the bar, noted by \citet{jung97} in
their near-IR images.  As they point out, the low inclination of the
galaxy makes it unlikely that this is a projection effect.  Unsharp
masking shows a faint nuclear spiral extending out to $r \sim
10\arcsec$; this may be responsible for the position angle and
ellipticity variations for $a < 10\arcsec$.  There is no sign of an
inner bar or disk, though the low inclination of the galaxy means an
inner disk would be quite difficult to detect.

\subsection{NGC 5750: Dust-obscured\label{n5750}} 

Figure~\ref{fig:n5750}: We have found no significant observations of
this dusty SB0/a galaxy in the literature, aside from an \hi{}
spectrum in \citet{tifft88}, which gives a velocity width $W_{20} =
411$ \kms.  This seems consistent with the galaxy's rather high
inclination, so the \hi{} is probably coplanar with the stars.  Our
\br{} color map shows a striking blue inner ring surrounding the bar,
with weak evidence for a red ring further out.  The region inside the
inner ring is quite dusty, though virtually all of the dust lanes seem
confined to the northeast side of the galaxy.  The two ellipticity
peaks in the ellipse fits are almost certainly due to the dust; we see
no signs of a secondary bar or inner disk in the unsharp masks, but IR
observations are clearly needed to be certain.

\subsection{NGC 6654: Double bar\label{n6654}} 

Figure~\ref{fig:n6654}: There appear to be almost no observations of
this galaxy in the literature, aside from photographic surface
photometry discussed in \citet{kormendy77a,kormendy77b}.  No \hi{} or
CO emission has been detected \citep{wk86,knapp96}.  The \br{} color
map shows a lacy pattern of tightly wrapped blue spirals in the outer
disk.  These arms are luminous structures in both the $B$ and $R$
images, so they are probably sites of recent or ongoing star
formation; they appear to lie just outside the ``plateau'' region of
Kormendy's surface brightness profiles.  There are no significant
color variations in the bar itself.  Inside the primary bar is a
secondary bar; Figure~\ref{fig:umaskpa} shows that this is one of
several inner bars where the ellipse fits give a misleading position
angle.  There is a suggestion of a color asymmetry in the WFPC2 color
map, but it is rather weak.  Unsharp masking shows no signs of
distinct dust lanes in or around the inner bar.

\subsection{UGC 11920: Dust-obscured\label{ugc11920}} 

Figure~\ref{fig:ugc11920}: We have found no previous observations of
this galaxy in the literature, aside from an \hi{} spectrum in
\citet{kamphuis96}.  The spectrum appears to have $W_{50} \sim
300$--350 \kms, which is consistent with the galaxy's inclination.

The profusion of foreground stars in the image makes measuring the
outer disk's shape and orientation difficult.  The bar is almost
parallel with the apparent major axis of the disk, so we cannot define
an \aten{} length.  Within the bar, the color map shows a strong,
curving dust lane which leads into the central regions; it may be
merging with a dusty nuclear ring, but our resolution is not good
enough to tell.  Interior to this, both the ellipse fits and the
unsharp masks suggest a secondary bar --- but this may simply be the
result of dust lanes.  Although we provide measurements for this
possible secondary bar, we do \textit{not} count UGC 11920 among the
confirmed double-barred galaxies of our sample.

\subsection{NGC 7280: Double bar, off-plane gas\label{n7280}} 

Figure~\ref{fig:n7280}: This galaxy is similar to NGC~3941: it is
double-barred, and there is good evidence for off-plane gas in the
nuclear regions.

Two \hi{} studies suggest the possibility that this galaxy is
interacting with its neighbor UGCA~429, a dwarf irregular 4.2\arcmin{}
ENE of this galaxy with a redshift difference of only 57 \kms. 
\citet{vanD91} found a weak central \hi{} concentration in NGC~7280,
and evidence for a weak extension in the direction of the dwarf
irregular; \citet{duprie96}, on the other hand, found that the \hi{}
in UGCA 429 appeared to extend in the opposite direction from NGC
7280.

Our WIYN images show an elliptical disk with a weak bar almost aligned
with the outer disk; unsharp masking shows an apparent outer ring at
$r \sim 40\arcsec$.  More striking is the color map, which shows
several dust lanes, all in the northern half of the disk.  In addition
to a long, linear dust lane $\ap 23\arcsec$ ENE of the galaxy center,
there are several short lanes in the central regions, including one
arcing right around the nucleus.  The combined F606W WFPC2 image shows
that the last lane is actually composed of two strong, nearly straight
lanes, and a fainter, more curved lane to the west, crossing right
over the nuclear regions (panel~f).  \citet{carollo97} suggested that
these lanes were ``likely rings wrapping several times around the
nucleus and perpendicular to it, reminiscent of the famous polar-ring
galaxy NGC~2685.''  We concur; the asymmetric dust distribution
further out, and the indirect \hi{} evidence for interaction with UGCA
429, are further evidence for this being a polar ring galaxy.  The
recent two-dimensional spectroscopy of \citet{afanasiev00} confirms
this: ionized gas in the inner 10\arcsec{} rotates orthogonally to the
stars.

Although Afanasiev \& Sil'chenko suggest that the innermost elliptical
feature in the isophotes, mostly clearly seen in the NICMOS image, is
an inclined stellar disk, we disagree.  Unsharp masks of both the
WFPC2 and the NICMOS images show the characteristic signs of an inner
bar; the position angle of this bar, based on measurements of the
isophotes and the unsharp masks, is $\ap 115\arcdeg$ (almost 15
degrees away from the position angle given by the ellipse fits). 
Since the region Afanasiev \& Sil'chenko study includes the transition
from an outer bar to an inner bar, their assumption of circular
rotation is probably not justified; they also use position angles of
$\ap 80--100\arcdeg$, rather different from what we find.  Thus, we do
not feel that their kinematics necessarily imply an inclined disk.
 

\subsection{NGC 7743: Nuclear spiral\label{n7743}} 

\begin{figure}
\begin{center}
	\includegraphics[scale=0.85]{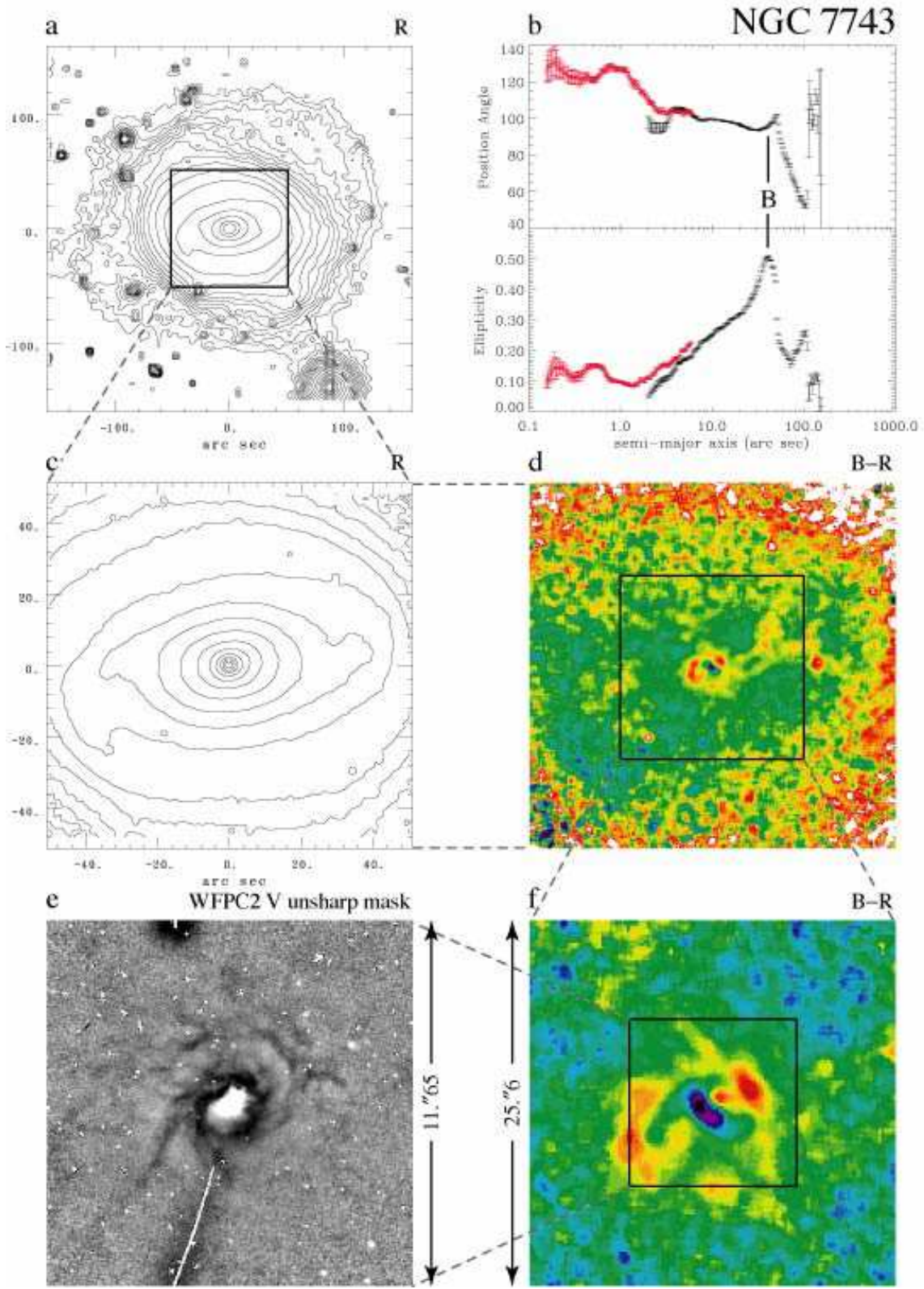}
\end{center}
\caption{NGC 7743: $R$-band isophotes on two scales (a, c; the
nuclear isophotes in c are suppressed to show the bar/spiral isophotes
more clearly); ellipse fits (b, WIYN $R$ and NICMOS2 F160W [red]);
\br{} color map on two scales (d, $w = 11$, color range $\ap 0.3$ mag;
f, $w = 5$, color range = 0.31 mag); unsharp mask (\ds{10}) of WFPC2
F606W image.
\label{fig:n7743}}
\end{figure}

Figure~\ref{fig:n7743}: This is a fairly well-studied Seyfert galaxy
with a dusty nuclear spiral and evidence for star formation in the
nuclear regions (e.g., \ha{} observations by \nocite{pogge87}Pogge \&
Eskridge 1987; extended UV emission in \nocite{barth98}Barth et al.\
1998; near-IR spectrum in \nocite{larkin98}Larkin et al.\ 1998).

The outer isophotes are asymmetric, and there are projections
to the NE and SW, especially noticeable at about 85\arcsec{} from the
center; these can also be seen in the images of \citet{kennicut86}. 
The nature of the galaxy's \hi{} gas is also confusing: while both
\citet{wk86} and \nocite{balkowski83}Balkowski \& Chamaraux (1983)
reported $W_{20} = 310$ \kms{} and Wardle \& Knapp give $W_{50} = 258$
\kms, \citet{lu93} report $W_{50} = 75$ \kms; \citet{duprie96} failed
to detect \textit{any} \hi{} and suggested that previous reports were
confused by two \hi{} clouds to the north (whose redshifts differ from
that of the galaxy, however).

The bar is surrounded by an inner pseudo-ring which does not quite
close, and which is responsible for both the peak in PA at $a \ap
50\arcsec$ and the corresponding peak in ellipticity.  We identify the
bar itself with the minimum in PA at $a \ap 31\arcsec$; the twisting
exterior to the is due to the spiral arms.  Both $\aten$ and $\amin$ 
are affected by the spiral arms and are probably overestimates for the 
bar length.

Within the bar are two parallel dust lanes in the western half, the
southernmost of which could be a leading-edge lane; however, no strong
lanes are visible in the east half.  \nocite{w95}W95 suggested that
this galaxy had a triaxial bulge, on the basis of twisting position
angles between 3 and 10\arcsec.  Since this is the region dominated by
complex, asymmetric dust lanes in the color map (see below), we
suspect most of the twisting is due to dust, not intrinsic triaxiality.

The dust lanes in the central regions ($r \lesssim 8\arcsec$) are
asymmetric and off-center, and do not form an obvious nuclear ring;
the interior is rather blue.  Unsharp masking of the WFPC2 image
(panel e of the figure) resolves this interior into nuclear spiral
dust lanes, some of which can be seen in the PC2-NICMOS color image of
\citet{regan99}.  Although those authors suggested that there was a
secondary bar present, on the basis of what appeared to be straight,
leading-edge dust lanes, our inspection of the HST images, both WFPC2
and NICMOS, shows no sign of a bar or inclined disk.  The ellipticity
peak at $a \ap 0.6\arcsec$ in the NICMOS image is probably due to
strong dust lanes near the center (which can be seen in the figures of
Regan \& Mulchaey).  In particular, there is no hint of any elongated
structure with a length appropriate to the ``straight'' dust lanes
Regan \& Mulchaey identify.  The unsharp mask of the WFPC2 image
(panel~e), which is larger in scale and higher in resolution than
Regan \& Mulchaey's color map, indicates that the dust lanes they
identified are simply parts of an overall nuclear spiral.  We conclude
that there is probably no secondary bar in this galaxy.

\acknowledgments

We are indebted to those people who obtained images for us during
queue time or their own or shared observing runs: Eric Wilcots, Jay
Gallagher, and especially Ted von Hippel, who in addition gave
valuable advice and assistance on this project's first observing run. 
Jay Gallagher provided numerous useful and insightful comments on
various drafts of this paper, the thesis it derives from, and the
early, exploratory versions of the sample.  Witold Maciejewski helped
by maintaining a proper theorist's skepticism of observations; he also
produced images of model double-bar galaxies and helped provide some
of the theoretical motivations for measurement techniques and
analysis.  We also thank Daniel Friedli and Marc Balcells for helpful
comments and criticisms on various drafts of this work.  The work
reported here forms part of the Ph.D.\ thesis of Peter Erwin at the
University of Wisconsin--Madison.

This research has made extensive use of the NASA/IPAC Extragalactic
Database (NED) which is operated by the Jet Propulsion Laboratory,
California Institute of Technology, under contract with the National
Aeronautics and Space Administration.  We also made use of the
Lyon-Meudon Extragalactic Database (LEDA; http://leda.univ-lyon1.fr). 
Based on observations made with the Isaac Newton Group of Telescopes
operated on behalf of the UK Particle Physics and Astronomy Research
Council (PPARC) and the Nederlandse Organisatie voor Wetenschappelijk
Onderzoek (NWO) on the island of Tenerife in the Spanish Observatorio
del Roque de Los Muchachos of the Instituto de Astrof\'{\i}sica de
Canarias.

Finally, this research was supported by NSF grants AST 9320403 and AST
9803114, and by NASA grant AR-0798.01-96A from the Space Telescope
Science Institute, operated by the Association of Universities for
Research in Astronomy, Inc., under NASA contract NAS5-26555.

\clearpage

\begin{deluxetable}{llrrrr}
\tablewidth{0pt}
\tabletypesize{\scriptsize}
\tablecaption{The WIYN Sample\label{tab:galaxies}}
\tablecolumns{6}
\tablehead{
\colhead{Galaxy} & \colhead{Type (RC3)} & \colhead{$R_{25}$ (\arcsec)} & \colhead{$z$ (\kms)} & \colhead{$D$ (Mpc)} & \colhead{Nucleus}}
\startdata
NGC 718 & SAB(s)a & 71 & 1733 & 22.7 & L2\\
NGC 936 & SB(rs)$0^{+}$ & 140 & 1430 & 23.0\tablenotemark{a} & abs.\\
NGC 1022 & (R\arcmin)SB(s)a & 72 & 1453 & 18.5 & SB\tablenotemark{c}\\
NGC 2273 & (R\arcmin)SB(s)a & 97 & 1840 & 27.3 & S2\\
NGC 2655 & SAB(s)0/a & 147 & 1404 & 22.1 & S2\\
NGC 2681 & (R\arcmin)SAB(rs)0/a & 109 & 692 & 17.2\tablenotemark{a} & L1.9\\
NGC 2685 & (R)SB0$^{+}$ pec & 134 & 883 & 14.4 & S2/T2:\\
NGC 2787 & SB(r)$0^{+}$ & 95 & 696 & 7.5\tablenotemark{a} & L1.9\\
NGC 2859 & (R)SB(r)$0^{+}$ & 128 & 1687 & 24.2 & T2:\\
NGC 2880 & SB$0^{-}$ & 62 & 1551 & 21.9\tablenotemark{a} & abs.\tablenotemark{d}\\
NGC 2950 & SB(r)$0^{0}$ & 80 & 1337 & 14.9\tablenotemark{a} & abs.\\
NGC 2962 & (R)SAB(rs)$0^{+}$ & 79 & 1966 & 30.0\tablenotemark{e} & \nodata\\
NGC 3032 & SAB(rs)$0^{0}$ & 60 & 1533 & 22.0\tablenotemark{a} & H II\tablenotemark{f}\\
NGC 3185 & (R)SB(r)a & 71 & 1218 & 17.1 & S2:\\
NGC 3412 & SB(s)$0^{0}$ & 109 & 865 & 11.3\tablenotemark{a} & abs.\\
NGC 3489 & SAB(rs)$0^{+}$ & 106 & 708 & 12.1\tablenotemark{a} & T2/S2\\
IC 676 & (R)SB(r)$0^{+}$ & 74 & 1290 & 19.0 & SB\tablenotemark{g}\\
NGC 3729 & SB(r)a pec & 85 & 1024 & 16.4 & H II\\
NGC 3941 & SB(s)$0^{0}$ & 104 & 928 & 12.2\tablenotemark{a} & S2:\\
NGC 3945 & (R)SB(rs)0$^{+}$ & 157 & 1220 & 19.3 & L2\\
NGC 4045 & SAB(r)a & 81 & 1981 & 26.3 & \nodata\\
NGC 4143 & SAB(s)$0^{0}$ & 69 & 985 & 15.9\tablenotemark{a} & L1.9\\
NGC 4203 & SAB$0^{-}$: & 102 & 1086 & 15.1\tablenotemark{a} & L1.9\\
NGC 4245 & SB(r)0/a: & 87 & 890 & 12.0\tablenotemark{b} & H II\\
NGC 4310 & (R$\arcmin$)SAB$0^{+}$? & 64 & 913 & 12.0\tablenotemark{b} & \nodata\\
NGC 4314 & SB(rs)a: & 125 & 963 & 12.0\tablenotemark{b} & L2\\
NGC 4386 & SAB$0^{0}$: & 74 & 1677 & 27.0\tablenotemark{a} & \nodata\\
NGC 4643 & SB(rs)0/a & 93 & 1319 & 17.8 & T2\\
NGC 4665 & SB(s)0/a & 114 & 785 & 10.4 & abs.\\
NGC 4691 & (R)SB(s)0/a pec & 85 & 1110 & 14.6 & H II\tablenotemark{h}\\
NGC 5338 & SB0: & 76 & 816 & 12.8\tablenotemark{a} & \nodata\\
NGC 5377 & (R)SB(s)a & 112 & 1793 & 26.7 & L2\\
NGC 5701 & (R)SB(rs)0/a & 128 & 1505 & 20.9 & T2:\\
NGC 5750 & SB(r)0/a & 91 & 1687 & 26.2 & L/S2\tablenotemark{i}\\
NGC 6654 & (R$\arcmin$)SB(s)0/a & 79 & 1821 & 28.2 & abs.\\
UGC 11920 & SB0/a & 72 & 1145 & 18.5 & \nodata\\
NGC 7280 & (R)SAB(r)$0^{+}$ & 66 & 1844 & 24.3\tablenotemark{a} & \nodata\\
NGC 7743 & (R)SB(s)$0^{+}$ & 91 & 1710 & 20.7\tablenotemark{a} & S2\\
\enddata
\tablecomments{$R_{25}$ is one-half of $D_{25}$, from RC3; heliocentric redshift $z$
is from NED; distance $D$ is from LEDA (redshift, corrected for Virgo-centric
infall, + $H_{0} = 75$ \kms{} Mpc$^{-1}$), except
as noted.  Nuclear classifications are from \cite{ho97a}, unless otherwise
indicated: L = LINER, S = Seyfert, T = transition between LINER and H II,
H II = H II region, SB = starburst, ``abs.'' = absorption lines only.
Uncertain classifications are indicated by a trailing colon.}
\tablenotetext{a}{\cite{tonry01}}
\tablenotetext{b}{\cite{forbes96}}
\tablenotetext{c}{\cite{ashby95}.}
\tablenotetext{d}{spectra of \cite{munn92} show no emission.}
\tablenotetext{e}{\cite{ajhar01}}
\tablenotetext{f}{\cite{veron86}}
\tablenotetext{g}{\cite{contini98}}
\tablenotetext{h}{\cite{chromey74}, \cite{garcia95}}
\tablenotetext{i}{\cite{veron86}}
\end{deluxetable}

\clearpage

\begin{deluxetable}{lllrr}
\tablewidth{0pt}
\tabletypesize{\footnotesize}
\tablecaption{Observations\label{tab:obs}}
\tablecolumns{5}
\tablehead{
\colhead{Galaxy} & \colhead{Telescope/} & \colhead{Filters}
& \colhead{Seeing (R, \arcsec)} & \colhead{Notes}\\
\colhead{} & \colhead{Instrument} & \colhead{} &\colhead{} & \colhead{}}
\startdata
NGC 718 & WIYN & BR & 1.0 & \\
\\[0.5mm]
NGC 936 & DSS & IIIaJ &  & \\
 & WHT & BVR & 0.85 & \\
 & WFPC2 & 555W &  & \\
\\[0.5mm]
NGC 1022 & WIYN & BR & 1.3 & \\
\\[0.5mm]
NGC 2273 & WIYN & UBRI & 0.65 & \\
 & WFPC2 & 606W,791W &  & \\
 & NIC1/2/3 & 160W,164N,222M &  & \\
\\[0.5mm]
NGC 2655 & WIYN & UBR & 0.8 & non-lin.\ CCD\\
 & WFPC2 & 547M &  & \\
\\[0.5mm]
NGC 2681 & WIYN & BR & 0.8 & \\
 & PC1 & 555W &  & \\
 & NIC3 & 160W,187N &  & \\
 \\[0.5mm]
NGC 2685 & WIYN & BR & 0.65 & \\
 & NIC3 & 160W,187N &  & \\
 \\[0.5mm]
NGC 2787 & WIYN & BR & 0.9 & non-lin.\ CCD\\
 & WFPC2 & 555W,702W,814W &  & \\
 & NIC3 & 160W,187N &  & \\
\\[0.5mm]
NGC 2859 & WIYN & UBR & 0.6 & non-lin.\ CCD\\
\\[0.5mm]
NGC 2880 & WIYN & BR & 0.7 & \\
 & WFPC2 & 450W,555W,814W &  & \\
\\[0.5mm]
NGC 2950 & WIYN & UBR & 0.65 & non-lin.\ CCD\\
 & WFPC2 & 450W,555W,702W,814W &  & \\
\\[0.5mm]
NGC 2962 & WIYN & BR & 1.2 & \\
 & WFCP2 & 814W &  & \\
\\[0.5mm]
NGC 3032 & WIYN & UBR & 0.8 & non-lin.\ CCD\\
 & WFPC2 & 606W &  & \\
 & NIC2 & 160W &  & \\
\\[0.5mm]
NGC 3185 & WIYN & BR & 0.9 & non-lin.\ CCD\\
\\[0.5mm]
NGC 3412 & WIYN & UBR & 1.15 & non-lin.\ CCD\\
 & WFPC2 & 606W &  & \\
\\[0.5mm]
NGC 3489 & WIYN & UBR & 0.85 & \\
 & WFPC2 & 555W,606W,814W &  & \\
\\[0.5mm]
IC 676 & WIYN & BR & 0.95 & \\
\\[0.5mm]
NGC 3729 & WIYN & R & 1.0 & \\
 & MK & BR & 1.3 & \\
\\[0.5mm]
NGC 3941 & WIYN & BR & 0.8 & \\
\\[0.5mm]
NGC 3945 & WIYN & BR & 0.65 & \\
 & WFPC2 & 450W,555W,814W &  & \\
\\[0.5mm]
NGC 4045 & WIYN & BR & 0.85 & \\
\\[0.5mm]
NGC 4143 & WIYN & BR & 0.75 & \\
 & WFPC2 & 218W,606W &  & \\
 & NIC2 & 160W &  & \\
\\[0.5mm]
NGC 4203 & WIYN & BR & 0.95 & \\
 & WFPC2 & 218W,555W,814W &  & \\
\\[0.5mm]
NGC 4245 & WIYN & BR & 0.7 & \\
 & WFPC2 & 606W &  & \\
\\[0.5mm]
NGC 4310 & WIYN & BR & 0.7 & \\
\\[0.5mm]
NGC 4314 & DSS & 103aE &  & \\
 & WFPC2 & 336W,439W,569W,814W &  & \\
 & NIC2 & 160W &  & \\
\\[0.5mm]
NGC 4386 & WIYN & BR & 0.8 & \\
\\[0.5mm]
NGC 4643 & WIYN & BR & 1.3 & non-lin.\ CCD\\
\\[0.5mm]
NGC 4665 & WIYN & BR & 1.2 & \\
\\[0.5mm]
NGC 4691 & WIYN & BR & 1.2 & non-lin.\ CCD\\
\\[0.5mm]
NGC 5338 & WIYN & BR & 0.7 & \\
\\[0.5mm]
NGC 5377 & WIYN & BR & 1.1 & non-lin.\ CCD\\
 & WFPC2 & 606W &  & \\
 & NIC2 & 110W,160W &  & \\
\\[0.5mm]
NGC 5701 & WIYN & VBR & 0.5 & non-lin.\ CCD\\
\\[0.5mm]
NGC 5750 & WIYN & BR & 1.3 & non-lin.\ CCD\\
\\[0.5mm]
NGC 6654 & WIYN & BR & 0.9 & non-lin.\ CCD\\
 & WFPC2 & 450W,814W &  & \\
\\[0.5mm]
UGC 11920 & WIYN & BR & 0.85 & non-lin.\ CCD\\
\\[0.5mm]
NGC 7280 & WIYN & BR & 0.6 & \\
 & WFPC2 & 606W &  & \\
 & NIC2 & 160W &  & \\
\\[0.5mm]
NGC 7743 & WIYN & BR & 0.65 & \\
 & WFPC2 & 547M,606W &  & \\
 & NIC2 & 160W &  & \\
\enddata
\tablecomments{
Image sources: WIYN = Wisconsin-Indiana-Yale-NOAO Telescope, WHT = 
William Herschel Telescope, MK = Mauna Kea 2.24m Telescope (Tully et 
al.\ 1996), DSS = Digitized Sky Survey, PC1 = Planetary Camera of HST
 (pre-refurbishment), WFPC2 = Wide Field/Planetary
 Camera-2 of HST, NIC1/2/3 = NICMOS1, 2, or 3 camera of HST.  Seeing for WIYN,
WHT, or MK observations is FWHM, in arc seconds, of stars in the best
$R$-band image (Moffat profile).  ``non-lin.\ CCD'' indicates
WIYN images were taken with the CCD before the non-linearity was
fixed (see text for details).}
\end{deluxetable}

\clearpage

\begin{deluxetable}{llllcl}
\tablewidth{0pt}
\tabletypesize{\footnotesize}
\tablecaption{Comparison of Bar Length 
Measurements\label{tab:bar-length}}
\tablecolumns{6}
\tablehead{
\colhead{Galaxy} & \colhead{\amax} & \colhead{\aten} & \colhead{\amin}
& \colhead{Other Measurements} & \colhead{Source for Other 
Measurements}}
\startdata
NGC 936  & 41\arcsec & 51 & 60 & 41 & \citet{kormendy83} \\
         &    &    &    & 52 & \citet{kent87} \\
         &    &    &    & 50 & \citet{kent89}\\
\\
NGC 4314 & 70 & 113 & 127 & 71 & \citet{ee85}\\
         &    &     &     & 60 & \citet{quillen94}\\
         &    &     &     & 66 & \citet{shaw95}\\
\\
NGC 4643 & 50 & 62 & 69 & 60 & \citet{ohta90}\\
         &    &    &    & 51 & \citet{shaw95}\\
\\
NGC 4665 & 45 & 65 & 99 & 58 & \citet{ohta90}\\
\enddata
\end{deluxetable}

\begin{deluxetable}{llllll}
\tablewidth{0pt}
\tabletypesize{\footnotesize}
\tablecaption{Model Bar Length 
Measurements\label{tab:bar-length-models}}
\tablecolumns{6}
\tablehead{
\colhead{Component} & \colhead{} & \colhead{\amax} & \colhead{\aten} 
& \colhead{\amin}
& \colhead{Ferrers Bar $a$}}
\startdata
Model 1:  & Inner Bar & 1.82 & 2.11 & 2.11 & 4.2 \\
          & Outer Bar & 3.61 & \nodata & \nodata & 7.0 \\
\\
Model 2:  & Inner Bar & 0.62 & 0.92 & 0.87 & 1.2 \\
          & Outer Bar & 3.60 & \nodata & \nodata & 6.0 \\
\enddata
\end{deluxetable}

\clearpage

\begin{deluxetable}{lrcrlrrr}
\tabletypesize{\footnotesize}
\tablewidth{375pt}
\tablecaption{Observed Structures\label{tab:features}}
\tablecolumns{8}
\tablehead{
\colhead{Galaxy} & \colhead{$i$} & \colhead{Rot.} & \colhead{Notes} & \colhead{Structure} & \colhead{Size} & \colhead{PA} & \colhead{\emax}\\
\colhead{} & \colhead{\arcdeg} & \colhead{} &\colhead{} & \colhead{} & \colhead{\arcsec} &\colhead{\arcdeg} & \colhead{}}
\startdata
NGC 718 & 30 & + & DB, NR & Outer Disk & 71 & 5 & 0.13\\
 & & & & Bar 1 & 20--30 & 152 (158) & 0.23\\
 & & & & NR(B) & $\sim 3.3$ & $\sim 50$ & $\sim 0.15$\\
 & & & & Bar 2 & 1.6--3.6 & $\sim 15$ & 0.19\\
 \\
NGC 936 & 41 & $-$? & NR & Outer Disk & 140 & 130 & 0.24\\
 & & & & Bar & 41--51 & 78 (81) & 0.47\\
 & & & & NR(S) & $\sim 8$ & $\sim 130$ & $\sim 0.23$\\
 \\
NGC 1022 & $\lesssim 24$ & + & dusty & Outer Disk & 72 & $\sim 174$ & 0.08\\
 & & & & Inner Ring & $\sim 30$ & $\sim 60$ & 0.16\\
 & & & & Bar & 19--22 & $\sim 115$ & 0.51\\
 \\
NGC 2273 & 50 & + & NS, NR & [Outer Disk] & 97 & 50 & $\sim 0.34$\\
 & & & & Outer Ring & 145 & 50 & 0.45\\
 & & & & Bar & 14--17 & 116 (109) & 0.43\\
 & & & & NR(SF) & 2.2 & $\sim 30$ & 0.47\\
 \\
NGC 2655 & 26 & $-$? & dusty, OPG & Outer Disk & 147 & $\sim 95$ & 0.10\\
 & & & & Bar(?) & 8.7--63 & 90 & 0.35\\
 \\
NGC 2681 & $\sim 18$ & $-$ & DB, NR & Outer Disk & 109 & 140 & 0.05\\
 & & & & Bar 1 & 50--60 & 30 & 0.23\\
 & & & & NR(B) & 18 & --- & 0.0\\
 & & & & Bar 2 & 18--19 & 73 & 0.33\\
 & & & & Bar 3 & 1.7--3.3 & 20 & 0.26\\
 \\
NGC 2685 & \nodata\tablenotemark{a} & \nodata & ID, OPG & [Outer Disk] & 134 & 38 & $\sim 0.54$\\
 & & & & Outer Ring & 155 & 25 & 0.50\\
 & & & & Bar? & 33--54 & 38 & 0.61\\
 & & & & Disk & 1.8--3.7 & 35 & 0.61\\
 \\
NGC 2787 & 55 & + & dusty, ID, OPG & Outer Disk & 95 & 109 & 0.41\\
 & & & & Bar & 29--36 & 160 (150) & 0.34\\
 & & & & Disk & 18--21 & 113 & 0.34\\
 \\
NGC 2859 & $\sim 25$ & + & DB, NR & [Outer Disk] & 128 & $\sim 90$ & $\sim 0.1$\\
 & & & & Outer Ring & 120 & 83 & 0.23\\
 & & & & Bar 1 & 34--43 & 162 & 0.40\\
 & & & & NR(R) & 7.0 & $\sim 60$ & $\sim 0.2$\\
 & & & & Bar 2 & 4.1--6.2 & 62 & 0.31\\
 \\
NGC 2880 & 52 & $-$? & ID & Outer Disk & 62 & 144 & 0.37\\
 & & & & Bar & 8.0--9.0 & 82 (105) & 0.20\\
 & & & & Disk & 3.0--4.5 & 138 & 0.22\\
 \\
NGC 2950 & 48 & + & DB, NR & Outer Disk & 80 & 120 & 0.32\\
 & & & & Bar 1 & 24--31 & 162 (155) & 0.43\\
 & & & & NR(S) & $\sim 4.2$ & $\sim 120$ & $\sim 0.3$\\
 & & & & Bar 2 & 3.2--3.9 & 85 (92) & 0.36\\
\\
NGC 2962 & 51 & $-$ & DB & Outer Disk & 79 & 10 & 0.35\\
 & & & & Bar 1 & 29--43 & 168 (174) & 0.30\\
 & & & & Bar 2 & 3.5--4.2 & 93 (10) & 0.03\\
 \\
NGC 3032 & 30 & $-$ & NS & Outer Disk & 60 & 95 & 0.13\\
 & & & & Bar? & 7.9--11 & 97 & 0.20\\
 \\
NGC 3185 & 48 & + & dusty & Outer Disk & 71 & 140 & 0.31\\
 & & & & Inner Ring & 39 & 125 & 0.62\\
 & & & & Bar & 31--32 & 114 & 0.58\\
 & & & & NR(R)? & $\sim 4$ & $\sim 140$ & $\sim 0.3$\\
 \\
NGC 3412 & 52 & \nodata & ID & Outer Disk & 109 & 153 & 0.37\\
 & & & & Bar & 15--21 & 100 (115) & 0.26\\
 & & & & Disk & 1.0--6.1 & 154 & 0.33\\
 \\
NGC 3489 & 56 & +? & dusty, NR & Outer Disk & 106 & 71 & 0.42\\
 & & & & Outer Ring & 51 & 71 & 0.51\\
 & & & & Bar? & $\sim 10$ & $\sim 20$ ($\sim 55$) & $\sim 0.17$\\
 & & & & NR(R) & 3.2 & $\sim 60$ & $\sim 0.7$\\
 \\
IC 676 & 44 & $-$? & dusty & Outer Disk & 74 & 15 & 0.27\\
 & & & & Outer Ring & 55 & 8 & 0.32\\
 & & & & Bar & 13--34 & 164 & 0.72\\
 \\
NGC 3729 & $\sim 50$ & + & dusty & Outer Disk & 85 & $\sim 170$ & $\sim 0.35$\\
 & & & & Inner Ring & 51 & 167 & 0.50\\
 & & & & Bar & 23--26 & 26 & 0.66\\
 \\
NGC 3941 & 51 & +? & DB, OPG & Outer Disk & 104 & 10 & $\sim 0.35$\\
 & & & & Bar 1 & 21--32 & 166 (176) & 0.47\\
 & & & & Bar 2 & 3.2--4.4 & 35 (20) & 0.21\\
 \\
NGC 3945 & $\sim 50$ & $-$? & DB, NR, ID & [Outer Disk] & 157 & 158 & $\sim 0.35$\\
 & & & & Outer Ring & 150 & 160 & 0.47\\
 & & & & Bar 1 & 32--39 & 72 (75) & 0.29\\
 & & & & Disk & 10--18 & 158 & 0.36\\
 & & & & NR(S) & 6.5 & 158 & $\sim 0.57$\\
 & & & & Bar 2 & 2.6--3.0 & 90 (115) & 0.11\\
 \\
NGC 4045 & 48 & + & dusty & Outer Disk & 81 & 90 & 0.32\\
 & & & & Outer Ring & 66 & 90 & 0.39\\
 & & & & Bar & 18--20 & $\sim 18$ & 0.30\\
 \\
NGC 4143 & 59 & + & NS, ID & Outer Disk & 69 & 144 & 0.46\\
 & & & & Bar & 17--28 & 163 (156) & $\sim 0.38$\\
 & & & & Disk & 4.2--6.2 & 142 & 0.25\\
 \\
NGC 4203 & 34 & + & OPG & Outer Disk & 102 & 10 & 0.17\\
 & & & & Bar & 13--46 & 9 & 0.24\\
\\
NGC 4245 & 40 & + & NS, NR & Outer Disk & 87 & $\sim 0$ & 0.25\\*
 & & & & Bar & 37--42 & 137 & 0.48\\*
 & & & & NR(SF) & 4.2 & $\sim 145$ & $\sim 0.15$\\
 \\
NGC 4310 & 62 & \nodata & dusty & Outer Disk & 64 & 159 & 0.5\\
 & & & & Bar? & 13--28 & 161 & 0.63\\
 \\
NGC 4314 & 25 & + & DB, NR, NS & Outer Disk & 125 & 65 & 0.09\\
 & & & & Bar 1 & 70--113 & 146 & 0.65\\
 & & & & NR(SF) & $\sim 6.8$ & $\sim 135$ & $\sim 0.24$\\
 & & & & Bar 2 & 4.5--5.6 & 136 & 0.23\\
 \\
NGC 4386 & 48 & \nodata & ID & Outer Disk & 74 & 140 & 0.31\\
 & & & & Bar & 25--36 & 134 & 0.52\\
 & & & & Disk & 2.4--3.2 & 141 & 0.28\\
 \\
NGC 4643 & 38 & +? & NS, ID & Outer Disk & 93 & 55 & 0.20\\
 & & & & Bar & 50--62 & 133 & 0.45\\
 & & & & Disk & 3.5--5.4 & 53 & 0.10\\
 \\
NGC 4665 & 33 & + &  & Outer Disk & 114 & 120 & 0.15\\
 & & & & Bar & 45--65 & 4 & 0.51\\
 \\
NGC 4691 & 42 & \nodata & dusty & Outer Disk & 85 & 30 & 0.24\\
 & & & & Bar & 30--55 & 82 & 0.64\\
 \\
NGC 5338 & 68 & \nodata & dusty & Outer Disk & 76 & 95 & 0.58\\
 & & & & Bar & 11--15 & 125 & 0.46\\
 \\
 \\
 \\
 \\
NGC 5377 & $\sim 55$ & $-$ & NS, NR & [Outer Disk] & 111 & $\sim 35$ & $\sim 0.4$\\
 & & & & Outer Ring & 150 & 27 & 0.47\\
 & & & & Inner Ring & $\sim 70$ & 43 & 0.64\\
 & & & & Bar & 58--78 & 45 & 0.66\\
 & & & & NR(R) & 6.5 & $\sim 30$ & $\sim 0.45$\\
 & & & & NR(B) & $\sim 3$ & $\sim 30$ & 0.41\\
 \\
NGC 5701 & $\sim 20$ & $-$ & NS & [Outer Disk] & 128 & $\sim 45$ & $\sim 0.05$\\
 & & & & Outer Ring & $\sim 125$ & 78 & 0.20\\
 & & & & Bar & 40--58 & 177 & 0.43\\
 \\
NGC 5750 & 62 & $-$ & dusty & Outer Disk & 91 & 65 & 0.50\\
 & & & & Bar & 20--24 & 121 & 0.37\\
 \\
NGC 6654 & 44 & $-$ & DB & Outer Disk & 79 & 0 & 0.27\\
 & & & & Bar 1 & 26--38 & 17 & 0.51\\
 & & & & Bar 2 & 2.7--4.2 & 135 (160) & 0.15\\
 \\
UGC 11920 & 52 & $-$ & dusty & Outer Disk & 72 & 50 & 0.36\\
 & & & & Bar & 26--39 & 45 & 0.51\\
 & & & & Bar 2? & 2.5--3.3 & 0 (17) & 0.19\\
 \\
NGC 7280 & 49 & +? & DB, OPG & Outer Disk & 66 & 73 & 0.33\\
 & & & & Bar 1 & 11--27 & 55 (63) & 0.40\\
 & & & & Bar 2 & 1.2--1.6 & 115 (101) & 0.30\\
\\
NGC 7743 & 28 & $-$ & NS & Outer Disk & 91 & 105 & 0.11\\*
 & & & & Bar 1 & 31--58 & 96 & 0.51\\
\enddata
\tablecomments{For each galaxy, we give disk inclination and rotation,
along with a summary of notable features: DB = double barred;
ID = inner disk (possible secondary bar); NR = nuclear ring;
NS = nuclear spiral; OPG = evidence for off-plane gas or dust,
dusty = too dust-obscured to determine presence or absence of central
structures.  Derivation of disk inclination $i$ is described in the text.
Rotation (+ = counter-clockwise, $-$ = clockwise, \nodata{} = unknown)
assumes that spiral arms trail, if any can be seen.  ``Size'' is the radial
extent of a feature in arc seconds, either semi-major axis or (for outer disks) $R_{25}$.
``[Outer Disk]'' indicates that a distinct outer disk cannot be identified;
we list $R_{25}$ and best estimates for overall disk orientation.}
\tablenotetext{a}{No reliable inclination can be determined for this galaxy.}
\end{deluxetable}

\clearpage

\twocolumn

\clearpage

\onecolumn

\clearpage

\clearpage

 %

\clearpage

\clearpage

\clearpage

\clearpage

\clearpage

\clearpage

\clearpage

\clearpage

\clearpage

\clearpage

\clearpage

\clearpage

\clearpage

\clearpage

\clearpage

\clearpage

\clearpage

\clearpage

\clearpage

\clearpage

\clearpage

\clearpage

\clearpage

\clearpage

\clearpage

\clearpage

\clearpage

\clearpage

\clearpage

\clearpage

\clearpage

\clearpage

\clearpage

\clearpage

\clearpage

\clearpage

\end{document}